\documentclass[onecolumn]{mn2e}
\usepackage{graphicx}
\usepackage{amsbsy}
\usepackage{amsmath}
\usepackage{epsf}
\newcommand{\apj}{ApJ}
\newcommand{\mnras}{MNRAS}

\newcommand{\araa}{ARAA}

\newcommand{\aap}{A\&A}
\newcommand{\leteq}{\begin{subequations}}
\newcommand{\beq}{\end{subequations}}
\newcommand{\simlt}{\la}
\newcommand{\lesssim}{\la}
\newcommand{\simgt}{\ga}
\newcommand{\gtrsim}{\ga}

\newcommand{\bc}{\begin{center}}
\newcommand{\lambdad}{\tilde{\lambda}}
\newcommand{\ec}{\end{center}}
\newcommand{\cc}{\rm{cm}^{-3}}
\newcommand{\htwo}{\mbox{\sc Hii}}
\newcommand{\ccb}{\rm{cm}^{3}}

\newcommand{\vnx}{v_{{\rm n},x}}
\newcommand{\vny}{v_{{\rm n},y}}
\newcommand{\vnz}{v_{{\rm n},z}}
\newcommand{\vix}{v_{{\rm i},x}}
\newcommand{\viy}{v_{{\rm i},y}}
\newcommand{\viz}{v_{{\rm i},z}}
\newcommand{\Bz}{B_z}
\newcommand{\By}{B_y}
\newcommand{\Bo}{B_0}

\newcommand{\vaio}{v_{\rm{A,i,0}}}
\newcommand{\vano}{v_{\rm{A,n,0}}}
\newcommand{\vmso}{v_{\rm{ms,n,0}}}
\newcommand{\vaiod}{\tilde{v}_{\rm{A,i,0}}}
\newcommand{\vanod}{\tilde{v}_{\rm{A,n,0}}}
\newcommand{\vmsod}{\tilde{v}_{\rm{ms,n,0}}}
\newcommand{\rhon}{\rho_{\rm n}}
\newcommand{\rhono}{\rho_{\rm n,0}}
\newcommand{\tni}{\tau_{\rm{ni}}}
\newcommand{\tnio}{\tau_{{\rm{ni,0}}}}
\newcommand{\tniod}{\tilde{\tau}_{{\rm{ni,0}}}}
\newcommand{\tinod}{\tilde{\tau}_{{\rm{in,0}}}}

\newcommand{\tin}{\tau_{\rm{in}}}
\newcommand{\tino}{\tau_{{\rm{in,0}}}}

\newcommand{\xe}{x_{\rm e}}

\newcommand{\xxio}{x_{\rm i,0}}
\newcommand{\nn}{n_{\rm{n}}}
\newcommand{\nno}{n_{\rm{n,0}}}
\newcommand{\nio}{n_{\rm{i,0}}}

\newcommand{\kB}{k_{\rm B}}
\newcommand{\me}{m_{\rm e}}

\newcommand{\mn}{\mu m_{\rm H}}
\newcommand{\mi}{m_{\rm{i}}}
\newcommand{\wpe}{\omega_{\rm{p,e}}}
\newcommand{\HII}{\rm{H}_{2}}
\newcommand{\mH}{m_{\HII}}

\newcommand{\Alf}{Alfv\'{e}n}
\newcommand{\AD}{{ }_{\rm{AD}}}
\newcommand{\freefall}{{ }_{\rm{ff}}}

\newcommand{\tad}{\tau_{\AD}}
\newcommand{\tff}{\tau_{\freefall}}
\newcommand{\tffo}{\tau_{{}_{\rm{ff,0}}}}

\newcommand{\nni}{n_{\rm i}}
\newcommand{\nne}{n_{\rm e}}

\newcommand{\vff}{\nu_{\freefall}}
\newcommand{\vffo}{\nu_{\rm{ff},0}}
\newcommand{\HCO}{\rm{HCO}^+}
\newcommand{\Na}{\rm{Na}^+}
\newcommand{\Mg}{\rm{Mg}^+}

\newcommand{\bm}{\boldmath}
\newcommand{\vvec}{\mbox{\bm $v$}}

\newcommand{\Bvec}{\mbox{\bm $B$}}
\newcommand{\Bveco}{\mbox{\bm $B$}_{0}}

\newcommand{\kvec}{\mbox{\bm $k$}}

\newcommand{\rvec}{\mbox{\bm $r$}}
\newcommand{\evec}{\mbox{\bm $e$}}

\newcommand{\ehat}{\hat{\evec}}

\newcommand{\vnvec}{\vvec_{\rm n}}
\newcommand{\vivec}{\vvec_{\rm i}}

\newcommand{\del}{\nabla}
\newcommand{\cross}{\mbox{\bm $\times$}}
\newcommand{\rhoi}{\rho_{\rm i}}
\newcommand{\rhoio}{\rho_{\rm i,0}}
\newcommand{\rhoe}{\rho_{\rm e}}

\newcommand{\czero}{C_{\rm{a},0}}
\newcommand{\lamJ}{\lambdad_{\rm{J,th}}}
\newcommand{\lamJdim}{\lambda_{\rm{J,th}}}
\newcommand{\lami}{\lambdad_{\rm{A,i}}}
\newcommand{\lamidim}{\lambda_{\rm{A,i}}}
\newcommand{\lamn}{\lambdad_{\rm{A,n}}}
\newcommand{\lamndim}{\lambda_{\rm{A,n}}}
\newcommand{\lammsn}{\lambdad_{\rm{ms,n}}}
\newcommand{\lammsndim}{\lambda_{\rm{ms,n}}}
\newcommand{\lamcs}{\lambdad_{\rm{s,n}}}
\newcommand{\lamcsdim}{\lambda_{\rm{s,n}}}
\newcommand{\Asnfac}{{\cal A}_{\rm{s,n}}}
\newcommand{\Ssmfac}{{\cal S}_{\rm{n}}}
\newcommand{\thetamax}{\theta_{\rm{max}}}
\newcommand{\lamms}{\lambdad_{\rm{J,mag}}}
\newcommand{\lammsdim}{\lambda_{\rm{J,mag}}}
\newcommand{\vph}{\tilde{v}_{\phi}}
\newcommand{\vphd}{{v}_{\phi}}
\newcommand{\damp}{\rm{d}}
\newcommand{\growth}{\rm{gr}}
\newcommand{\tdamp}{\tilde{\tau}_{{}_{\damp}}}
\newcommand{\tdampd}{\tau_{{}_{\damp}}}
\newcommand{\tgrowth}{\tilde{\tau}_{{}_{\growth}}}
\newcommand{\tgrowthd}{\tau_{{}_{\growth}}}
\newcommand{\diffi}{\tilde{\cal D}_{\rm{a,i}}}
\newcommand{\diffn}{\tilde{\cal D}_{\rm{a,n}}}
\newcommand{\diff}{\tilde{\cal D}_{\rm{a}}}
\newcommand{\diffncs}{\tilde{\cal D}_{\rm{P}}}

\newcommand{\sigin}{{\langle \sigma w \rangle}_{\rm{i}\HII}}
\newcommand{\zcr}{\zeta_{{}_{\rm{CR}}}}
\newcommand{\zcrd}{\tilde{\zeta}_{{}_{\rm{CR}}}}
\newcommand{\alphdr}{\alpha_{\rm{dr}}}
\newcommand{\alphdrd}{\tilde{\alpha}_{\rm{m,dr}}}

\newcommand{\rhoiod}{\tilde{\rho}_{\rm{i,0}}}

\newcommand{\omegad}{\tilde{\omega}}
\newcommand{\rhond}{\tilde{\rho}_{\rm n}}
\newcommand{\rhoid}{\tilde{\rho}_{\rm i}}
\newcommand{\vnxd}{\tilde{v}_{{\rm n},x}}
\newcommand{\vnyd}{\tilde{v}_{{\rm n},y}}
\newcommand{\vnzd}{\tilde{v}_{{\rm n},z}}
\newcommand{\vizd}{\tilde{v}_{{\rm i},z}}
\newcommand{\viyd}{\tilde{v}_{{\rm i},y}}
\newcommand{\vixd}{\tilde{v}_{{\rm i},x}}
\newcommand{\vims}{\tilde{v}_{\rm i,ms}}
\newcommand{\Bxd}{\tilde{B}_x}
\newcommand{\Byd}{\tilde{B}_y}
\newcommand{\Bzd}{\tilde{B}_z}
\newcommand{\kd}{\tilde{k}}

\begin{document}

\title[MHD Waves in Weakly Ionised Media]{Hydromagnetic Waves in Weakly Ionised Media. I. Basic 
Theory, and Application to Interstellar Molecular Clouds }
    
\author[Mouschovias, Ciolek, and Morton]{Telemachos Ch. Mouschovias,
\thanks{E-mail: tchm@astro.illinois.edu} 
Glenn E. Ciolek,
\thanks{Current affiliation: New York Center for Astrobiology, and 
Department of Physics, Applied Physics, and Astronomy, Rensselaer 
Polytechnic Institute, 110 8th Street, Troy, NY 12180. 
E-mail: cioleg@rpi.edu} 
and Scott A. Morton \\
Departments of Physics and Astronomy, University of Illinois at Urbana-Champaign, 
1002 W. Green Street, Urbana, IL 61801 \\}

\maketitle

\label{firstpage}
\begin{abstract}

We present a comprehensive study of MHD waves and instabilities in a weakly ionised
system, such as an interstellar molecular cloud. We determine all the critical 
wavelengths of perturbations across which the sustainable wave modes can change radically 
(and so can their decay rates), and various instabilities are present or absent. Hence, 
these critical wavelengths are essential for understanding the effects of MHD waves 
(or turbulence) on the structure and evolution of molecular clouds. Depending on the 
angle of propagation relative to the zeroth-order magnetic field and the physical 
parameters of a model cloud, there are wavelength ranges in which no wave can be 
sustained as such. Yet, for other directions of propagation or different properties 
of a model cloud, there may always exist some wave mode(s) at all wavelengths (smaller 
than the size of the model cloud). For a typical model cloud, {\em magnetically-driven 
ambipolar diffusion} leads to removal of any support against gravity that most short-wavelength
waves (or turbulence) may have had, and {\em gravitationally-driven ambipolar diffusion} 
sets in and leads to cloud fragmentation into stellar-size masses, as first suggested 
by Mouschovias more than three decades ago -- a single-stage fragmentation theory 
of star formation, distinct from the then prevailing hierarchical fragmentation picture.
The phase velocities, 
decay times, and eigenvectors (e.g., the densities and velocities of neutral particles 
and the plasma, and the three components of the magnetic field) are determined as 
functions of the wavelength of the disturbances in a mathematically transparent way and 
are explained physically. Comparison of the results with those of nonlinear analytical or 
numerical calculations is also presented where appropriate, excellent agreement is found, 
and confidence in the analytical, linear approach is gained to explore phenomena difficult 
to study through numerical simulations. Mode splitting (or bifurcation) and mode merging,
which are impossible in single-fluid systems for linear perturbations (hence, the term 
``normal mode'' and the principle of superposition), occur naturally in multifluid systems 
(as do transitions between wave modes without bifurcation) and have profound consequences 
in the evolution of such systems.
\end{abstract}

\section{Introduction -- Background}

     A typical molecular cloud which has not yet given birth to stars is a
cold ($T \simeq 10~\rm{K}$) but complex, partially ionised system, in which
self-gravitational and magnetic forces are of comparable magnitude, with
thermal-pressure forces becoming important at high densities ($\gtrsim 3 \times
10^8~\cc$) or along magnetic field lines. Mouschovias (1976) showed that, 
barring external disturbances, if the magnetic field were to be frozen in the 
matter, interstellar clouds that have not yet given birth to stars 
would remain in magnetohydrostatic (MHS) equilibrium states.
However, ambipolar diffusion (the relative motion of neutral particles and charged 
particles attached to magnetic field lines) is an unavoidable process in partially 
ionised media. It reveals itself in two distinct ways, depending on whether it is
magnetically or gravitationally driven (see discussion in \S~4). The two kinds of 
ambipolar diffusion acting together initiate fragmentation and star formation in 
molecular clouds (Mouschovias 1987a). 

In this fragmentation theory, the evolutionary (or fragmentation, or
core formation) timescale is the gravitationally-driven ambipolar-diffusion timescale, 
$\tad$. This does not mean, however, that it takes a time equal to $\tad$ to form 
stars. The star-formation timescale can be a fraction or a multiple of $\tad$, 
depending on the mass-to-flux ratio of the parent cloud and the degree to which
hydromagnetic waves (HM) contribute to the support of the cloud: the closer 
to its critical value the mass-to-flux ratio is and/or the greater the contribution 
of HM waves to cloud support, the faster the evolution and the 
shorter the star-formation timescale (e.g., see Mouschovias 1987a; Fiedler \& Mouschovias 
1993, Fig. 9a; Ciolek \& Basu 2001; Tassis \& Mouschovias 2004, Fig. 4). \footnote{A 
number of authors {\it assume} that molecular clouds are highly magnetically 
supercritical (e.g., Mac Low \& Klessen 2004 and references therein; Lunttila {\it et al.} 
2008, 2009). Zeeman observations (e.g., Crutcher 1999) are sometimes used as 
the observational justification of that assumption. However, geometrical 
corrections are ignored in those assumptions. The corrections are necessary 
because (1) only the line-of-sight component of the magnetic field is measured, 
and (2) the measured column density of a cloud flattened along the magnetic 
field lines is statistically greater than that needed for the calculation of 
the cloud's mass-to-flux ratio (e.g., see Shu {\it et al.} 1999). However, even 
if the Zeeman observations were to be taken at face value, without any 
geometrical correction at all, they still do not reveal highly supercritical 
molecular clouds. In order to obtain a magnetically supercritical molecular 
cloud model of mass $\approx 1840 \, M_{\odot}$ and mean density 
$\approx 100 \, \cc$, Lunttila {\it et al.} assume a magnetic field of only  
$0.69 \, \mu {\rm G}$ or $0.34 \, \mu {\rm G}$. These values are smaller by a factor
of $\simeq \, 10$ than even the observed strength of the magnetic field in the 
general interstellar medium $\approx 6 \, \mu {\rm G}$ (Heiles \& Crutcher 2005). 
There is no conceivable physical mechanism that can possibly increase the 
density of a forming cloud by $2 - 4$ orders of magnitude while 
{\it de}creasing its magnetic field strength, especially by that large a 
factor ($\simeq \, 10$). Hence, the main assumption of many turbulence simulations, 
i.e., that molecular clouds are highly supercritical, has neither an observational 
basis nor any theoretical justification.} The {\em collapse retardation factor}, 
$\vff = \tff/\tni$ (where $\tff$ is the free-fall time 
and $\tni$ the neutral-ion collision time), is the factor by which magnetic 
forces slow down the contraction relative to free-fall. This was a new theory of 
fragmentation (or core formation), initiated by the decay, due to ambipolar diffusion, 
of relatively small-wavelength perturbations (Mouschovias 1987a, 1991a).

Magnetic braking operates on a timescale shorter than the ambipolar-diffusion 
timescale and even the free-fall timescale, and keeps a cloud (or fragment) 
essentially corotating with the background up to densities $\simeq 10^4 - 10^6~\cc$ 
and thus resolves the angular momentum problem of star formation. More 
specifically, the entire range of periods of binary stars from 10 hr to 
100 yr was shown to be accounted for by this {\em self-initiated} mode of 
star formation (Mouschovias 1977). Even single stars and
planetary systems become dynamically possible (Mouschovias 1978, 1983). 

Star formation, whether self-initiated or triggerred (e.g., by a spiral density
wave, or the expansion of an $\htwo$ region or a supernova remnant; see review
by Woodward 1978) is an inherently nonlinear process. Ambipolar-diffusion$-$initiated
star formation has been studied analytically (Mouschovias 1979, 1991a, b) and 
numerically using adaptive grid techniques in axisymmetric geometry up to 
densities $\sim 10^{10}~\cc$, by which isothermality begins to break down 
(Fiedler \& Mouschovias 1992, 1993; Ciolek \& Mouschovias 1993, 1994, 1995; 
Basu \& Mouschovias 1994, 1995a, b). More recently, these calculations were 
extended into the opaque phases of star formation (Desch \& Mouschovias 2001; 
Tassis \& Mouschovias 2007a, b, c; Kunz \& Mouschovias 2009, 2010). The key 
conclusions of the earlier analytical calculations have been verified and 
numerous new, specific, quantitative predictions have been made, many of which 
have been confirmed by observations (e.g., see Crutcher {\it et al.} 1994; 
Ciolek \& Basu 2000; Chiang {\it et al.} 2008; reviews by Mouschovias 1995, 1996). 
One may nevertheless take a step back from the relatively complicated numerical 
calculations and ask which results, if any, of the nonlinear simulations can be 
recovered with a {\em linear} analysis. With one's confidence increased in the 
validity of the linear approach (within some self-evident limits), one may then 
make new predictions concerning phenomena that cannot be or have not been included
yet in the nonlinear calculations.

The propagation, dissipation, and growth of perturbations in a physical
system depends both on the nature of the perturbations and the properties of
the system. Hydromagnetic waves (or MHD turbulence) seem to play a significant 
role in molecular clouds on lengthscales typically greater than $\sim 0.1~\rm{pc}$.
They have been shown to account quantitatively for the observed supersonic but
sub{\Alf}ic spectral linewidths: an observational almost-scatter diagram of
linewidth versus size is converted into an almost perfect straight line if
plotted in accordance with a theoretical prediction by Mouschovias (1987a),
which relates the linewidth, the size, and the magnetic field strength of an
observed object (see Mouschovias \& Psaltis 1995, Figs. 1 and 2, and update by 
Mouschovias {\it et al.} 2006, Figs. 1 and 2). 

Using a linear analysis as a first step in understanding nonlinear
phenomena is not, of course, a new idea. Jeans (1928) used it to
obtain his famous instability criterion for the collapse of a cloud against
thermal-pressure forces. Hardly any astrophysical system exists whose
stability with respect to small-amplitude disturbances has not yet been
studied by using at least an idealized, mathematically tractable model of the
physical system. A magnetically supported molecular cloud, however, defies a
simple linear analysis. First, no realistic equilibrium states have been
obtained by analytical means. Second, to study the role of ambipolar diffusion
in star formation, one must use at least the two-fluid magnetohydrodynamic
(MHD) equations governing the motions of the neutral particles and the plasma
(ions and electrons). In fact, as shown by Ciolek \& Mouschovias
(1993), charged (and even neutral) grains play a very significant role in the
ambipolar-diffusion$-$initiated protostar formation. One then has to use at least 
the four-fluid (neutral molecules, plasma, negatively-charged and neutral grains) 
MHD equations even for a linear analysis to be realistic and 
relevant to typical molecular clouds. In this paper we use the two-fluid MHD 
equations to study the propagation, dissipation, and growth of HM waves in an 
idealized model molecular cloud. In a subsequent paper we consider the effects 
of the grain fluid(s).

Langer (1978) studied the stability of a model molecular cloud (infinite
in extent and uniform in density and magnetic field) with respect to 
small-amplitude, adiabatic perturbations in the presence of ambipolar diffusion. 
For propagation along the magnetic field lines, he recovered, as one would expect
intuitively, the Jeans dispersion relation and instability in the absence of
the magnetic field. He then investigated the wave propagation perpendicular to
the field lines. He showed that the Jeans instability is still present, that
the critical wavenumber for instability is independent of the magnetic field
strength, but that the growth rate depends on both the field strength and the
degree of ionisation. Aside from two spurious curves in his Figure 1, which
exhibits the growth rate and decay time of some modes, our results for
propagation of the low-frequency modes perpendicular to the magnetic field are
in agreement with Langer's -- he ignored the high-frequency ion modes. Yet even
in this, previously studied case, we offer new analytical expressions and new
physical insight and interpretation of the results. Moreover, we present not
only the eigenvalues (frequencies or, equivalently, phase velocities) as
functions of wavelength but the eigenvectors as well (i.e., material
velocities, densities, magnetic-field components, etc.). We also study
propagation at arbitrary angles with respect to the magnetic field, and we
offer a thorough discussion of the wave modes, not just the 
ambipolar-diffusion--induced instability.

Pudritz (1990) revisited Langer's problem (with the minor difference of
considering isothermal perturbations) but introduced a new effect: he assumed
that there exists a power-law spectrum of small-amplitude waves, and then he
studied the effect that this spectrum has on the ambipolar-diffusion--induced,
Jeans-like instability. \footnote{The plasma force equation in Pudritz's (1990) 
paper (eqs. [2.2b], [3.4], [A4], and [A10]) and in Langer's (1978) paper, 
eq. (6), contains an error. The thermal-pressure force should be multiplied 
by a factor of 2 to account for the presence of electrons. Although this omission 
does not affect Langer's results because he did not consider the ion modes,
it does introduce errors in some of the ion modes considered by Pudritz.} 
He concluded that the slope of the spectrum (considered as a function of 
wavelength) has an important effect on the growth rate of the instability; the 
steeper the spectrum, the greater the growth rate. (The growth rate of 
gravitationally-driven ambipolar diffusion, however, cannot
possibly exceed the free-fall rate.) 

Several other papers have appeared in print since 1990, studying different
aspects of weakly ionised systems, focusing usually on the stability of certain
MHD modes or shocks, especially as it may relate to the formation of structures 
in molecular clouds and/or on the effect of the grain fluid(s) on the allowable wave 
modes or shocks (e.g., Wardle 1990; Balsara 1996; Zweibel 1998; Kamaya \& Nishi 
1998, 2000; Mamun \& Shukla 2001; Cramer et al. 2001; Falle \& Hartquist 2002; 
Tytarenko et al. 2002; Zweibel 2002; Ciolek et al. 2004; Lim et al. 2005; Oishi 
\& Mac Low 2006; Roberge \& Ciolek 2007; van Loo et al. 2008; Li \& Houde 2008). 
In this paper we present a general theory of the propagation, dissipation and growth of 
MHD waves in partially ionised media in three dimensions, with emphasis on mathematical 
transparency of the formulation and analytical solution of the problem, the physical 
understanding and interpretation of all modes, including their eigenvectors, 
the many critical wavelengths that exist and which separate regimes dominated by
different waves or instabilities, and on specific features relevant to the 
evolution of molecular clouds. As mentioned above, even when a particular result 
agrees with previous work, we offer new insight into its physical understanding.
       
In \S~2 we present the equations governing the behaviour of a weakly
ionised, magnetic, self-gravitating interstellar cloud. The equations are
linearised, Fourier-analyzed, and put in dimensionless form. The free
parameters of the problem are identified, their physical meaning explained,
and their typical values given. The different hydromagnetic modes and their
dependence on wavelength for different directions of propagation relative to
the magnetic field are calculated and explained physically in \S~3. Analytical
expressions for the phase velocities, damping timescales, growth timescales,
including critical or cutoff wavelengths, are also obtained. A physical discussion 
of the eigenvectors is an integral part of this presentation. Section 4
summarizes some of the results and their relevance to the formation of protostellar
fragments (or cores) and to other observable phenomena. It also gives in two Tables 
all the critical wavelengths and the ranges of wavelengths in which different modes 
can exist in molecular clouds, for propagation parallel, perpendicular, and at
arbitrary angles with respect to the magnetic field.

\vspace{-3ex}
\section{FORMULATION OF THE PROBLEM}
\subsection{Basic Equations}

     We consider a weakly ionised medium (e.g., an interstellar molecular
cloud) consisting of neutral particles ($\HII$ with a $20\%$ helium abundance
by number; subscript n), electrons, and singly-charged positive ions (subscript i). 
For specificity we assume that the ions are molecular ions (such as
$\HCO$); for the densities of interest in this paper ($\sim 10^3 \, \cc$), this 
is sufficient since atomic ions (such as $\Na$ or $\Mg$) are less abundant (a result
of depletion of metals in dense clouds) and, in any case, they have masses comparable 
to that of $\HCO$ (for more detailed treatments of the chemistry, 
see Ciolek \& Mouschovias 1995, 1998, or the appendix of Mouschovias \& Ciolek 1999). 
Interstellar grains, which have been shown to have significant effects on the
formation and contraction of protostellar cores (Ciolek \& Mouschovias 1993,
1994, 1995) and in the opaque phase of star formation (Tassis \& Mouschovias 
2007a, b, c; Kunz \& Mouschovias 2009, 2010) are neglected in this analysis; 
they are accounted for in a subsequent paper. 

The magnetohydrodynamic (MHD) equations governing the evolution of the above
two-fluid system are

\leteq
\begin{eqnarray}
\label{contineqa}
\frac{\partial \rhon}{\partial t} +\del \cdot \left(\rhon \vnvec \right) &=& 0 , \\
\label{ioncontineqa}
\frac{\partial \rhoi}{\partial t} +\del \cdot \left(\rhoi \vivec \right) &=&
\mi \zcr \frac{\rhon}{\mn} - \frac{\alphdr}{\mi} \rhoi^2, \\
\label{neutforceqa}
\rhon \left[\frac{d}{d t}\right]_{\rm n}\vnvec &=& - \del P_{\rm n} - \rhon \del
\psi -\frac{\rhon}{\tni}\left(\vnvec - \vivec
\right), \\
\label{ionforceqa}
\rhoi \left[\frac{d}{d t}\right]_{\rm i}\vivec &=&
\frac{\left(\del \cross \Bvec \right) \cross \Bvec}{4 \pi} -
\frac{\rhoi}{\tin} \left(\vivec - \vnvec \right), \\
\label{heateqa}
\frac{\kB}{\mn} \left\{\frac{\rhon}{\gamma-1}
\left[\frac{d}{dt}\right]_{\rm n}T - T \left[\frac{d}{dt}\right]_{\rm n} \rhon \right\}
&=& \Gamma_{\rm n} - \Lambda_{\rm n}, \\
\label{inducteqa}
\frac{ \partial \Bvec}{\partial t} &=& \del \cross \left(\vivec \cross \Bvec
\right), \\
\label{gaseqa}
P_{\rm n} &=& \rhon \frac{\kB T}{\mn}, \\
\label{poissoneqa}
\del^{2} \psi &=& 4 \pi G \rhon, \\
\label{gausseqa}
\del \cdot \Bvec &=& 0,
\end{eqnarray}
\beq
where $\rho_\alpha$ and $\vvec_\alpha$ are, respectively, the density and 
velocity of species $\alpha$,
\begin{eqnarray}
\label{ddtdefeq}
\left[\frac{d}{dt}\right]_{\alpha} = \frac{\partial}{\partial t} +
\vvec_{\alpha} \cdot \del 
\end{eqnarray}
is the time-derivative comoving with species $\alpha$, $\psi$ the
gravitational potential, $P_{\rm n}$ the neutral pressure, $\Bvec$ the
magnetic field, $T$ the temperature,  $\Gamma_{\rm n}$ and $\Lambda_{\rm n}$
the heating and cooling rates (per unit volume) of the neutral gas, and
\leteq
\begin{eqnarray}
\label{tnieq}
\tni &=& 1.4 \frac{\mi+\mH}{\mi} \frac{1}{\nni \sigin}, \\
\label{tineq}
\tin &=& 1.4 \frac{\mi+\mH}{\mn} \frac{1}{\nn \sigin},
\end{eqnarray}
\beq
the neutral-ion and ion-neutral mean collision (i.e., momentum-exchange)
times. The quantity $G$ is the universal gravitational constant, and $\kB$ is 
Boltzmann's constant; $\mu$ is the mean mass of a neutral particle in units of the 
atomic-hydrogen mass and is equal to 2.33 for a $\HII$ gas with a $20\%$ helium
abundance by number. The quantities $\zcr$ and $\alphdr$ in the ion mass
continuity equation (\ref{ioncontineqa}) are, respectively, the cosmic-ray
ionisation rate and the coefficient for dissociative recombination of
molecular ions and electrons (in $\rm{cm}^3~\rm{s}^{-1}$). In writing equation
(\ref{ioncontineqa}), we have used the condition of local charge neutrality 
$e(\nni - \nne)=0$, where $\nni$ and $\nne$ are, respectively, the number 
densities of ions and electrons. We may use the assumption of local charge 
neutrality because the various HM modes of interest here have frequencies much
smaller than the electron plasma frequency $\wpe = (4 \pi \nne e^2/\me)^{1/2}
= (4 \pi \xe \nn e^2/\me)^{1/2}= 5.64 \times 10^2~
(x_{\rm e}/10^{-7})^{1/2}(\nn/10^3~\cc)^{1/2}~\rm{s}^{-1}$ (where $\xe = \nne/\nn$
is the abundance of electrons relative to the neutrals, and is equal to the 
degree of ionisation for weakly-ionised systems); hence, any excess
charge density is quickly shielded by the mobile electrons, so that
$\nne=\nni$ for timescales $\simgt \wpe^{-1}$. Since we neglect the effects of
grains in this paper, capture of ions onto grains is not included on
the right-hand side of equation (\ref{ioncontineqa}) as a sink term for ions.

Heat conduction and viscosity are not important for the densities and
lengthscales of interest ($\nn \simeq
10^2-10^5~\cc$, $L \simeq 10^{-3}-10^{1}~\rm{pc}$), and are therefore ignored
as possible sources of heating/cooling in the model clouds. (Note that the
left-hand side of equation [\ref{heateqa}] is equal to 
$\rhon T [d/dt]_{\rm n} S$, where $S$ is the entropy per gram of matter.)

Because $\rhon \gg \rhoe,~\rhoi$, we include only the neutral density as a
source term in Poisson's equation (\ref{poissoneqa}). Similarly, the
gravitational and thermal-pressure forces (per unit volume) on the plasma
(ions and electrons) have been neglected in the plasma force equation
(\ref{ionforceqa}). One can easily show that, for the physical conditions in
typical molecular clouds, they are completely negligible in comparison to the
magnetic force exerted on the plasma, except in a direction almost exactly
parallel to the magnetic field. Ignoring them parallel to the magnetic field
implies that we are neglecting the ion acoustic waves and the (extremely 
long-wavelength) Jeans instability in the ions.

The quantity $\sigin$ in equations (\ref{tnieq}) and (\ref{tineq}) is the
elastic collision rate between ions and neutrals. For $\HCO-\HII$ collisions,
$\sigin=1.69\times 10^{-9}~\ccb~\rm{s}^{-1}$
(McDaniel \& Mason 1973). The factor 1.4 in equations (\ref{tnieq}) and (\ref{tineq})
accounts for the inertial effect of He on the motion of the neutrals (for
a discussion, see \S~2.1 of Mouschovias \& Ciolek 1999).

\vspace{-4ex}
\subsection{Linear System}

To investigate the propagation, dissipation, and growth of HM waves
in molecular clouds, we follow the original analysis by Jeans (1928; see also
Spitzer 1978, \S~13.3a; and Binney \& Tremaine 1987, \S~5.1), and assume that
the zeroth-order state is uniform, static (i.e., $\vvec_\alpha=0$), and in 
equilibrium. 
\footnote{It is well known that the assumption that the gravitational potential
is uniform in the zeroth-order state is not consistent with Poisson's equation
(\ref{poissoneqa}) (e.g., see Spitzer 1978; Binney \& Tremaine 1987). However,
it is not well known that, for an infinite uniform system, no such inconsistency
exists; i.e., there is no net gravitational force on any fluid element, hence 
this state is a true, albeit unstable, equilibrium state.}
We consider only adiabatic perturbations; therefore, the net heating rate 
($\Gamma_{\rm n}-\Lambda_{\rm n}$) on the right-hand side of equation 
(\ref{heateqa}) vanishes.

We write any scalar quantity or component of a vector  $q_{\rm{tot}}(\rvec,t)$
in the form $q_{\rm{tot}}(\rvec,t)=q_{0}+ q(\rvec,t)$, where $q_0$ refers to
the zeroth-order state, and the first-order quantity $q$ satisfies the condition
$|q| \ll |q_0|$. We thus obtain from equations
(\ref{contineqa}) - (\ref{gausseqa}) the linearised system
\leteq
\begin{eqnarray}
\label{contineqb}
\frac{\partial \rhon}{\partial t} &=& - \rhono \left(\del \cdot \vnvec\right), \\
\label{ioncontineqb}
\frac{\partial \rhoi}{\partial t} &=& - \rhoio \left(\del \cdot \vivec\right)
+ \frac{\rhoio}{\mn} \xxio \alphdr \rhon - 2 \frac{\rhoio}{\mi} \alphdr
\rhoi, \\
\label{neutforceqb}
\rhono \frac{\partial \vnvec}{\partial t} & = & - \del P_{\rm n} - \rhono \del
\psi - \frac{\rhono}{\tnio}\left(\vnvec-\vivec\right), \\
\label{ionforceqb}
\rhoio \frac{\partial \vivec}{\partial t} &=&
\frac{\left(\del \cross \Bvec \right) \cross \Bveco}{4 \pi}
-\frac{\rhoio}{\tino}\left(\vivec - \vnvec \right), \\
\label{heateqb}
\frac{1}{T_0}\frac{\partial T}{\partial t} &=& \left(\gamma - 1\right)
\frac{1}{\rhono}\frac{\partial \rhon}{\partial t} , \\
\label{inducteqb}
\frac{\partial \Bvec}{\partial t} &=& \del \cross \left(\vivec \cross
\Bveco\right), \\
\label{gaseqb}
\frac{P_{\rm n}}{P_{\rm{n,0}}} &=& \frac{\rhon}{\rhono} + \frac{T}{T_0}, \\
\label{poissoneqb}
\del^2 \psi &=& 4 \pi G \rhon, \\
\label{gausseqb}
\del \cdot \Bvec &=& 0 .
\end{eqnarray}
\beq
Equation (\ref{ioncontineqb}) has been simplified by using the relation 
\begin{eqnarray}
\label{zeroorderchargeeq}
\zcr \frac{\rhono}{\mn} = \alphdr \left(\frac{\rhoio}{\mi}\right)^{2} ,
\end{eqnarray}
which expresses equilibrium of the ion density in the zeroth-order state, as a
result of balance between the rate of creation of ions from ionisation of
neutral matter by high-energy ($E \gtrsim 100~\rm{MeV}$) cosmic rays and the
rate of destruction of ions by electron$-$molecular-ion dissociative
recombinations. This relation allowed us to replace $\zcr$ by
$\alphdr(\rhoio/\mi)^2(\mn/\rhono)$ in equation (\ref{ioncontineqb}).  The
quantity $\xxio \equiv \nio/\nno$ in equation (\ref{ioncontineqb}) is the
degree of ionisation (where $\nio$ and $\nno$ are the number densities of ions
and neutrals in the unperturbed state). For an ideal gas (with only 
translational degrees of freedom), $\gamma=5/3$ in equation (\ref{heateqb}).

We seek plane-wave solutions of the form $q(\rvec, t)=\overline{q}\exp
(i\kvec\cdot \rvec - i \omega t)$, where $\kvec$ is the propagation vector,
$\omega$ the frequency, and $\overline{q}$ the amplitude (in general, complex)
of the perturbation. Equations (\ref{contineqb}) - (\ref{gausseqb}) reduce to
\leteq
\begin{eqnarray}
\label{contineqc}
\omega \rhon &=& \rhono \kvec \cdot \vnvec, \\
\label{ioncontineqc}
\omega \rhoi &=& \rhoio \kvec \cdot \vivec + i \frac{\rhoio}{\mn} \xxio
\alphdr \rhon - i2 \frac{\rhoio}{\mi} \alphdr \rhoi, \\
\label{neutforceqc}
\omega \vnvec &=&
\left(\czero^2 k - \frac{1}{\tffo^2 k}\right)\frac{\rhon}{\rhono}
\frac{\kvec}{k} - \frac{i}{\tnio} \vnvec + \frac{i}{\tnio} \vivec , \\
\label{ionforceqc}
\omega \vivec &=&
-\frac{\left(\kvec
\cross \Bvec\right)\cross \Bveco}{4 \pi \rhoio} + \frac{i}{\tino}\vnvec -
\frac{i}{\tino}\vivec, \\
\label{inducteqc}
\omega \Bvec &=& - \kvec \cross \left(\vivec \cross \Bveco\right), \\
\label{gausseqc}
\kvec \cdot \Bvec &=&0,
\end{eqnarray}
\beq
where
\begin{eqnarray}
\label{czerodefeq}
\czero \equiv \left(\frac{\gamma P_{\rm{n,0}}}{\rhono}\right)^{1/2}
=\left(\frac{\gamma \kB T_{0}}{\mn}\right)^{1/2}
\end{eqnarray}
is the adiabatic speed of sound in the neutrals, and 
\begin{eqnarray}
\label{tffdefeq}
\tffo \equiv \left(4 \pi G \rhono\right)^{-1/2}
\end{eqnarray}
is the (one-dimensional) neutral free-fall timescale. The quantities $T$,
$P_{\rm n}$, and $\psi$ have been eliminated by using equations
(\ref{heateqb}), (\ref{gaseqb}), and (\ref{poissoneqb}), respectively.

\subsection{The Dimensionless Problem}

We put equations (\ref{contineqc}) - (\ref{gausseqc}) in dimensionless  form by
adopting $\rhono$, $\Bo$, $\tffo$, and $\czero$ as units of density, magnetic
field strength, time, and speed, respectively. The implied unit of length is
$\czero \tffo$, which is proportional to the one-dimensional thermal Jeans
lengthscale $\lambda_{\rm{J,th}}$ (see \S~3.1.1). For convenience, we adopt a
cartesian coordinate system such that the propagation vector $\kvec$ is in the
$x$-direction and the zeroth-order magnetic field $\Bveco$ is in the
$(x,z)$-plane at an angle $\theta$ with respect to $\kvec$ (see Fig. 1). Then
the three unit vectors are 
\leteq
\begin{eqnarray}
\label{xdefeq}
\ehat_{x} &\equiv& \frac{\kvec}{k}, \\
\label{ydefeq}
\ehat_{y} &\equiv& \frac{\Bveco \cross \kvec}{|\Bveco \cross \kvec|}, \\
\label{zdefeq}
\ehat_{z} &\equiv& \ehat_{x} \cross \ehat_{y}  .
\end{eqnarray}
\beq
%
\begin{center}
\begin{figure}
\includegraphics[width=70mm]{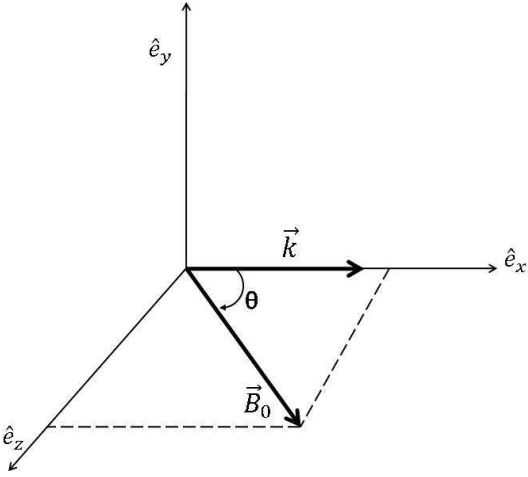}
\caption{{\it Coordinate system used in analysing the linearised hydromagnetic
equations.} The $x$-axis is aligned with the propagation vector $\kvec$,
and the magnetic field $\Bveco$ is in the $(x,z)$-plane at an angle $\theta$
with respect to $\kvec$ (see eqs. [\ref{xdefeq}] - [\ref{zdefeq}]).}
\end{figure}
\end{center}
%
One may write $\Bveco = \ehat_{x} \Bo \cos \theta + \ehat_{z} \Bo \sin \theta$,
and the dimensionless form of equations (\ref{contineqc}) - (\ref{gausseqc}) can
be written in component form as
\leteq
\begin{eqnarray}
\label{contineqd}
\omegad \rhond &=& \kd \vnxd, \\
\label{ioncontineqd}
\omegad \rhoid &=& \rhoiod \kd \vixd + i \rhoiod^2 \alphdrd \rhond - 
i2 \rhoiod \alphdrd \rhoid , \\
\label{xneutforceq}
\omegad \vnxd &=& \left(\kd - \frac{1}{\kd}\right) \rhond -
\frac{i}{\tniod}\vnxd + \frac{i}{\tniod}\vixd, \\
\label{xionforceq}
\omegad \vixd &=&
\frac{i}{\tinod}\vnxd - \frac{i}{\tinod}\vixd + \vaiod^2 \kd \Bzd \sin \theta,
\\
\label{xinducteq}
\omegad \Bxd &=& 0, \\ \nonumber \\
\label{yneutforceq}
\omegad \vnyd &=& - \frac{i}{\tniod} \vnyd + \frac{i}{\tniod} \viyd, \\
\label{yionforceq}
\omegad \viyd &=& \frac{i}{\tinod} \vnyd - \frac{i}{\tinod} \viyd - \vaiod^2
\kd \Byd \cos \theta  , \\
\label{yinducteq}
\omegad \Byd &=& - \kd \viyd \cos \theta    , \\ \nonumber \\
\label{zneutforceq}
\omegad \vnzd &=& - \frac{i}{\tniod}\vnzd + \frac{i}{\tniod}\vizd, \\
\label{zionforceq}
\omegad \vizd &=& \frac{i}{\tinod} \vnzd - \frac{i}{\tinod} \vizd - \vaiod^2
\kd \Bzd \cos \theta    , \\
\label{zinducteq}
\omegad \Bzd &=& \kd \vixd \sin \theta - \kd \vizd \cos \theta  , \\
\label{gausseqd}
\kd \Bxd &=& 0 .
\end{eqnarray}
\beq
Note that equation (\ref{gausseqd}) is redundant in that it gives the same
information as equation (\ref{xinducteq}), namely, that there cannot be a
nonvanishing component of the perturbed magnetic field in the direction of
propagation.

The dimensionless free parameters appearing in equations (\ref{contineqd}) -
(\ref{gausseqd}) are given by
\leteq
\begin{eqnarray}
\label{tnioddefeq}
\tniod &\equiv& \frac{\tnio}{\tffo} = 0.506
\left(\frac{10^{-7}}{\xxio}\right)
\left(\frac{10^3~{\cc}}{\nno}\right)^{1/2}, \\
\label{tinoddefeq}
\tinod &\equiv& \frac{\tino}{\tffo} = 6.29 \times 10^{-7}
\left(\frac{10^3{~\cc}}{\nno}\right)^{1/2}, \\
\label{vaioddefeq}
\vaiod &\equiv& \frac{\vaio}{\czero} = 5.00 \times 10^3
\left(\frac{\Bo}{30~\mu \rm{G}}\right)
\left(\frac{10^{-7}}{\xxio}\right)^{1/2}
\left(\frac{10~{\rm K}}{T}\right)^{1/2}
\left(\frac{10^3~{\cc}}{\nno}\right)^{1/2}, \\
\label{alphdrddefeq}
\alphdrd &\equiv& \frac{\alphdr}{\mi} \rhono \tffo = 1.41 \times 10^9 
\left(\frac{\alphdr}{10^{-6}~\rm{cm}^3~\rm{s}^{-1}}\right)\left(\frac{\nno}
{10^3~\cc}\right)^{1/2} \left(\frac{29~{\rm{amu}}}{\mi}\right); 
\end{eqnarray}
they represent, respectively, the neutral-ion collision time, the ion-neutral
collision time, the ion {\Alf} speed $\vaio=\Bo/(4 \pi\rhoio)^{1/2}$, and 
the electron$-$molecular-ion dissociative recombination rate per unit ion 
mass. In evaluating the numerical constants in equations
(\ref{tnioddefeq}) - (\ref{alphdrddefeq}) we have used $\mu=2.33$ and
$\gamma=5/3$; we have also normalized the ion mass to that of $\HCO$ (= 29 amu).
For any given ion mass $\mi$ and mean mass per neutral particle $\mu$ 
(in units of $m_{\rm H}$), the {\em ion mass fraction} $\rhoiod$ and the cosmic-ray
ionisation rate $\zcrd$ are not free parameters in the problem; the former is 
determined by the ratio $\tinod/\tniod$ and the latter by the product 
${\rhoiod}^2 \alphdrd$, where $\alphdrd$ is the dimensionless 
dissociative-recombination coefficient for molecular ions.
\beq

We note that $\tniod = 1/\vffo$, where $\vffo$ is the {\em collapse
retardation factor}, which is a parameter that measures the effectiveness with
which magnetic forces are transmitted to the neutrals via neutral-ion
collisions (Mouschovias 1982), and appears naturally in the timescale for the
formation of protostellar cores by ambipolar diffusion (e.g., see reviews by
Mouschovias 1987a, \S~2.2.5; 1987b, \S~3.4; 1991b, \S~2.3.1; and 
discussions in Fiedler \& Mouschovias 1992, 1993; Ciolek \& Mouschovias 1993,
1994, 1995; Basu \& Mouschovias 1994, 1995). It is essentially the factor by
which ambipolar diffusion in a magnetically supported cloud retards the 
formation and contraction of a protostellar fragment (or core) relative to free 
fall up to the stage at which the mass-to-flux ratio exceeds the critical value 
for collapse. It is discussed further in \S~3.2.1. 

Equations (\ref{contineqd}) - (\ref{zinducteq}) govern the behaviour of 
small-amplitude disturbances in a weakly ionised cloud; they (without eq. 
[\ref{xinducteq}]) form a $10 \times 10$ homogeneous system. In general, the 
dispersion relation $\omegad(\kd)$ can be obtained by setting the determinant 
of the coefficients equal to zero. To each root (eigenvalue) of the dispersion
relation there corresponds an eigenvector (or ``mode"), whose components are
the dependent variables appearing in equations
(\ref{contineqd}) - (\ref{zinducteq}). (Note that, once the dependent variables
in eqs. [\ref{contineqd}] - [\ref{zinducteq}] are known, one may use eqs.
[\ref{heateqb}], [\ref{gaseqb}], and [\ref{poissoneqb}] to solve for the
perturbed quantities $T$, $P_{\rm n}$, and $\psi$, respectively.) Since, in
general, $\omegad$ is complex, modes with $\rm{Im}\{\omegad\}<0$ decay and
those with $\rm{Im}\{\omegad\} > 0$ grow exponentially in time. In what
follows we investigate the propagation, dissipation, and growth of the
allowable HM modes in typical interstellar molecular clouds.

\vspace{-4ex}
\section{SOLUTION, PHYSICAL INTERPRETATION, AND APPLICATIONS}

For specificity, we consider a representative molecular cloud of density $\nno=2
\times 10^3~\cc$, magnetic field strength $\Bo=30~\mu\rm{G}$, temperature
$T=10~\rm{K}$, dissociative recombination rate $\alphdr=10^{-6}~\rm{cm}^3~{\rm
s}^{-1}$, and cosmic-ray ionisation rate $\zcr= 5 \times
10^{-17}~\rm{s}^{-1}$, implying a degree of ionisation $\xxio = 1.58 \times
10^{-7}$ (and, hence, ion mass fraction $\rhoiod = 1.97 \times 10^{-6}$). The
unit of time $\tffo$ (see eq. [\ref{tffdefeq}]) for this model
is equal to $3.92\times 10^{5}~\rm{yr}$, and the unit of speed is 
$\czero=0.243~\rm{km}~\rm{s}^{-1}$ (see eq.
[\ref{czerodefeq}]). Hence, the unit of length is $\czero \tffo = 9.72 \times
10^{-2}~\rm{pc}$. The four (dimensionless) free parameters of the problem (see 
eqs. [\ref{tnioddefeq}] - [\ref{alphdrddefeq}]) are: the ion-neutral
collision time $\tinod=4.45\times 10^{-7}$, the neutral-ion
collision time $\tniod=0.226$, the {\Alf} speed in the ions
$\vaiod=2.81\times 10^3$, and the dissociative-recombination
coefficient $\alphdrd=1.99 \times 10^9$. (These imply a  dimensionless
cosmic-ray ionisation rate $\zcrd=6.19 \times 10^{-4}$.)

In the following subsections we present the solutions for propagation along
($\theta=0^\circ$), perpendicular ($\theta=90^\circ$), and at intermediate
angles ($\theta=45^\circ$, $10^\circ$, and $80^\circ$) with respect to the
unperturbed magnetic field $\Bveco$ (see Fig. 1). We use the velocity vector
$\vvec$ in relation to $\kvec$ as defining the polarisation of each wave mode.
Modes that have only $\vnxd$, $\vixd \neq 0$ are said to be {\em longitudinally
polarised}; it follows from equation (\ref{contineqd}) that these modes are
compressible, i.e., $\rhond \neq 0$. Modes that have $\vnxd$, $\vixd =0$ are
said to be {\em transversely polarised}; they are incompressible, i.e.,
$\rhond =0$. Note in what follows that, since the thermal pressure in the
plasma has been neglected, no ion sound waves are present.

\subsection{Propagation Along $\Bveco$ ($\kvec \parallel \Bveco$)}

For $\theta=0^\circ$, equations [\ref{contineqd}] - [\ref{zinducteq}] become
uncoupled in the three mutually orthogonal directions $\ehat_x$, $\ehat_y$,
and $\ehat_z$. The modes polarised in the $x$-direction are given by (see eqs.
[\ref{contineqd}] - [\ref{xionforceq}]) 
\leteq
\begin{eqnarray}
\label{contineqf}
\omegad \rhond &=& \kd \vnxd, \\
\label{ioncontineqf}
\omegad \rhoid &=& \rhoiod \kd \vixd + i \rhoiod^2 \alphdrd \rhond - 
i2 \rhoiod \alphdrd \rhoid , \\
\label{xneutforceqb}
\omegad \vnxd &=& \left(\kd - \frac{1}{\kd}\right) \rhond -
\frac{i}{\tniod}\vnxd + \frac{i}{\tniod}\vixd, \\
\label{xionforceqb}
\omegad \vixd &=&
\frac{i}{\tinod}\vnxd - \frac{i}{\tinod}\vixd .
\end{eqnarray}
\beq
Equations (\ref{yneutforceq}) - (\ref{yinducteq}) yield for the modes with
motions in the $y$-direction
\leteq
\begin{eqnarray}
\label{yneutforceqb}
\omegad \vnyd &=& - \frac{i}{\tniod} \vnyd + \frac{i}{\tniod} \viyd, \\
\label{yionforceqb}
\omegad \viyd &=& \frac{i}{\tinod} \vnyd - \frac{i}{\tinod} \viyd - \vaiod^2
\kd \Byd, \\
\label{yinducteqb}
\omegad \Byd &=& - \kd \viyd,
\end{eqnarray}
\beq
while equations (\ref{zneutforceq}) - (\ref{zinducteq}) for the modes polarised
in the $z$-direction become
\leteq
\begin{eqnarray}
\label{zneutforceqb}
\omegad \vnzd &=& - \frac{i}{\tniod}\vnzd + \frac{i}{\tniod}\vizd, \\
\label{zionforceqb}
\omegad \vizd &=& \frac{i}{\tinod} \vnzd - \frac{i}{\tinod} \vizd - \vaiod^2
\kd \Bzd, \\
\label{zinducteqb}
\omegad \Bzd &=& - \kd \vizd .
\end{eqnarray}
\beq

\subsubsection{Longitudinal Modes for $\kvec \parallel \Bveco$: Eigenfrequencies 
and Eigenvectors}

From equations (\ref{contineqf}) - (\ref{xionforceqb}) the dispersion relation 
for the longitudinal modes is
\begin{eqnarray}
\label{longdispeqa}
\left(\omegad + 2 i \rhoiod \alphdrd \right)\left[\omegad^3 + i
\left(\frac{1}{\tinod} + \frac{1}{\tniod}\right) \omegad^2 -
\left(\kd^2-1\right) \omegad - \frac{i}{\tinod}\left(\kd^2-1\right)\right]=0.
\end{eqnarray}
The four eigenvalues and eigenvectors  as functions of dimensionless
wavelength $\lambdad$ ($= 2 \pi/\kd$) are obtained from direct numerical 
solution of the eigensystem (\ref{contineqf}) -(\ref{xionforceqb})
and displayed in Figures 2 and 3. Figure $2a$ shows the absolute value of the phase 
velocity $\vph$ ($= \omegad_{\rm r}/\kd$, where $\omegad_{\rm r}=\rm{Re}\{\omegad\}$). 
The damping (or dissipation) timescale $\tdamp$ ($=1/\omegad_{\rm i}$, $\omegad_{\rm i} 
< 0$, where $\omegad_{\rm i}=\rm{Im}\{\omegad\}$) is exhibited in Figure $2b$;
growth times $\tgrowth$ ($=1/\omegad_{\rm i}$, $\omegad_{\rm i} > 0$) are
shown in Figure $2c$. The absolute values of the $\vnxd$-components of the 
eigenvectors for the various modes are shown in Figure $3a$, while those of 
$\vixd$, $\rhond$, and $\rhoid$ are displayed in Figures $3b$, $3c$, and $3d$,
respectively. Note that, in Figures $3a$ - $3d$, all the eigenvectors have been 
normalised to unity; i.e., they satisfy the condition 
$\left( |\rhond|^2 + |\rhoid|^2 + |\vnxd|^2 + |\vixd|^2 \right)^{1/2}=1$.
\begin{figure}
\includegraphics[width=180mm]{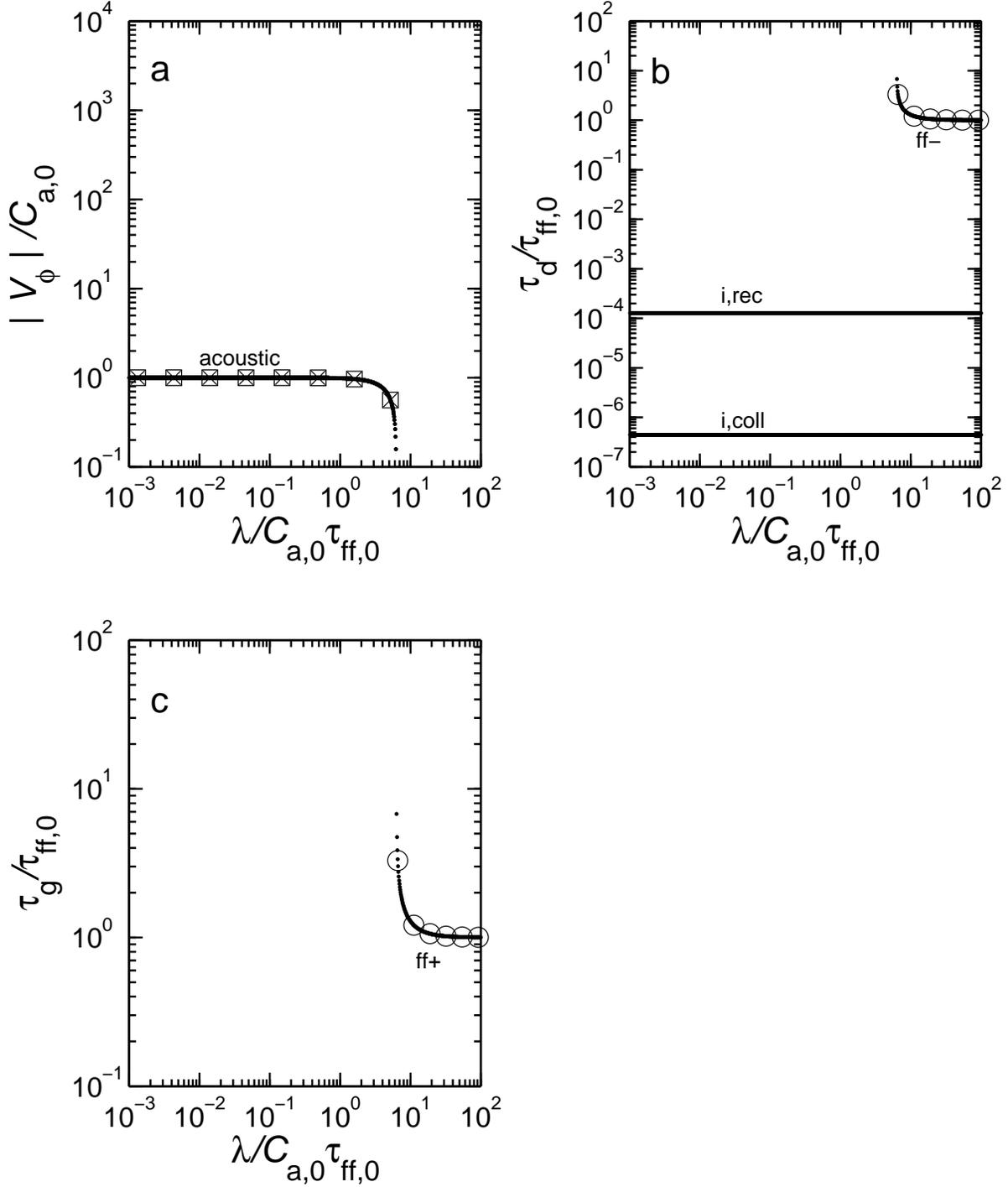}
\caption{{\it Eigenvalues of longitudinal modes as functions of wavelength,
normalised to $\czero \tffo= 9.72 \times 10^{-2}~\rm{pc}$, at an angle of
propagation $\theta=0^\circ$ with respect to $\Bveco$}. Because of degeneracy,
there are a total of four different modes at each wavelength. Each curve is 
identified by a label (a mnemonic for the mode it represents) as explained in 
Table 1. ({\it a}) Absolute value of the phase velocity $\vphd$, normalised to $\czero=0.243~\rm{km}~\rm{s}^{-1}$. Also overplotted as boxes with interior 
$\times$'s are values obtained from eq. (\ref{thermaljeansvphi}). 
({\it b}) Damping timescale $\tdampd$, in units of $\tffo= 3.92\times
10^{5}~\rm{yr}$. Also shown (open circles) are values calculated from
eq. (\ref{tffgrowtheq}). ({\it c}) Growth timescale $\tgrowthd$, normalised
to $\tffo$. Also plotted (open circles) are values calculated using eq. 
(\ref{tffgrowtheq}).}
\end{figure}
\begin{figure}
\includegraphics[width=180mm]{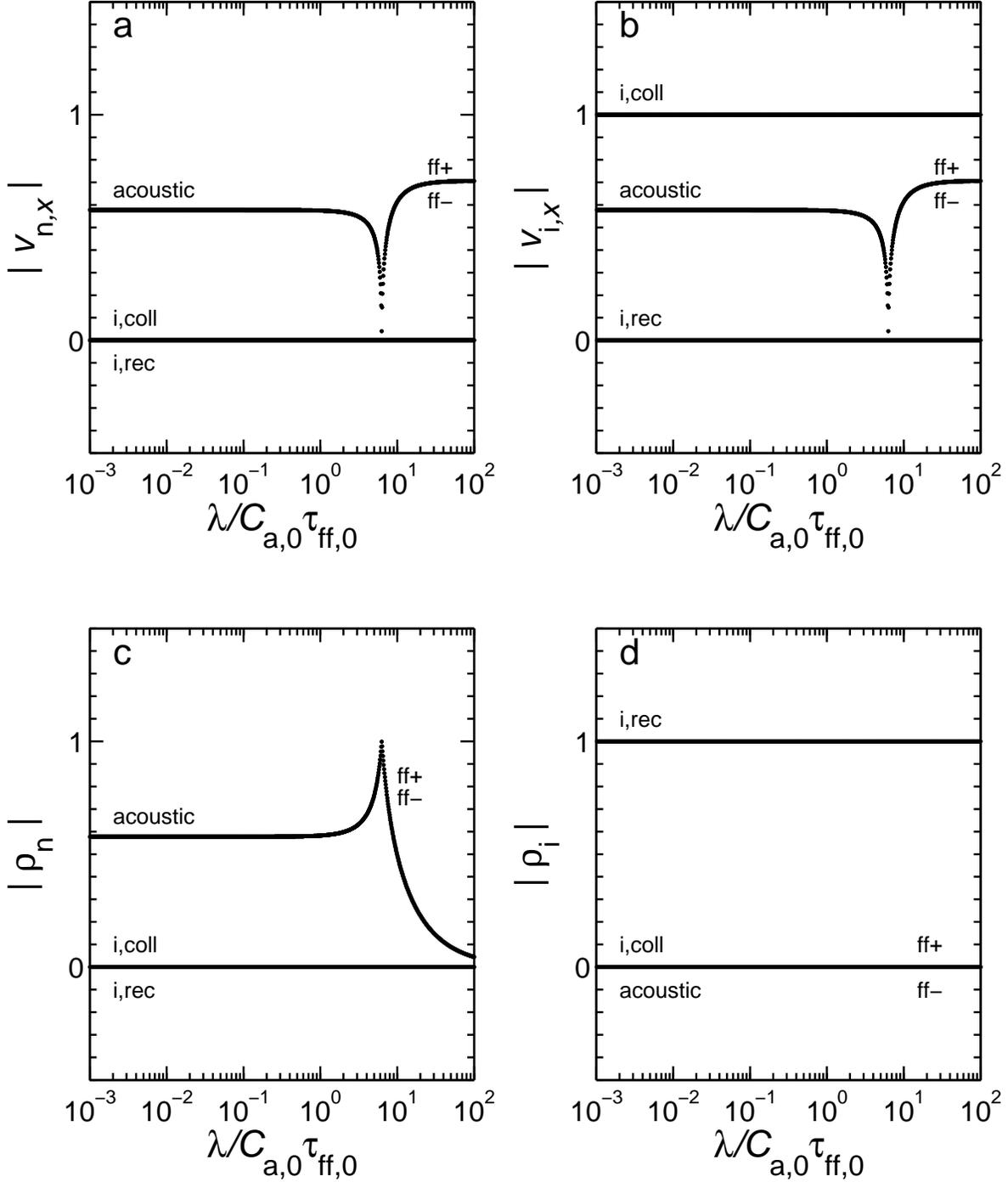}
\caption{{\it Magnitudes of eigenvectors of longitudinal modes for
$\theta=0^\circ$ as functions of wavelength, normalised as in Fig. 2.} As in
Fig. 2, there are a total of four modes. All eigenvectors have been
normalised to unity. Each curve is identified by a label (a mnemonic for the mode 
it represents) as explained in Table 1. ({\it a}) Longitudinal
($x$-) component of the neutral velocity, $|\vnx|$. ({\it b}) Longitudinal ($x$-) 
component of the ion velocity, $|\vix|$. ({\it c}) Neutral density $|\rhon|$. 
({\it d}) Ion density $|\rhoi|$.}
\end{figure}
The first mode for this system of equations is a nonpropagating, {\em ion
collisional-decay mode}, i.e., the ions are streaming through a sea of fixed
neutrals. Their motion decays because of ion-neutral collisions. It is
characterized by
$\vnxd=0=\rhond$. Solving equations (\ref{contineqf}) - (\ref{xionforceqb})
under these conditions (or, equivalently, solving eq. [\ref{longdispeqa}] in
the limit $|\omegad| \gg 1/\tniod$), one finds 
\begin{eqnarray}
\label{iondampeq}
\omegad=-\frac{i}{\tinod}.
\end{eqnarray}
Hence, $\vph=0$ and $\tdamp=\tinod = 4.45 \times 10^{-7}$ (see Fig. 2$b$, 
line labeled ``i,coll''). This mode is
independent of wavelength (see Figs. $2b$, $3a - 3d$, lines labeled ``i,coll").

The second mode is one in which density enhancements in the ions rapidly decay
by dissociative recombinations of molecular ions and electrons; because the
degree of ionisation is so small ($\xxio =1.58 \times 10^{-7}$), this mode
does not involve any motion of the neutrals and leaves the neutral density 
essentially unchanged (see Figs. $3a$ and $3c$, lines labeled
``i,rec"). Solving equation (\ref{ioncontineqf}) with $\rhond=0$ and
$\vixd=0$, one finds that $\vph=0$ and 
\begin{eqnarray}
\label{ionrecmode}
\tdamp = \left(2 \alphdrd \rhoiod\right)^{-1}~,
\end{eqnarray}
which is equal to $1.28\times 10^{-4}$ for the model described here. 
It is again the case that this mode is independent of $\kd$ (see Figs. $2b$
and $3a$ - $3d$, lines labeled ``i,rec").

The remaining two longitudinal modes are low-frequency modes, with $|\omegad|
\ll 1/\tinod$. Because the inertia of the ions {\em along} the magnetic field
is small ($\rhoio \ll \rhono$), the neutrals are able to sweep up the ions,
and, as a result, $\vixd \simeq \vnxd$ (see Figs. 3$a$ and 3$b$). For these 
conditions, equation (\ref{longdispeqa}) yields the {\em thermal Jeans modes} 
\begin{eqnarray}
\label{jeanmodeeq}
\omegad= \pm \kd \left[1 - \left(\frac{1}{\kd^{2}}\right)\right]^{1/2}
\end{eqnarray}
(e.g., see Chandrasekhar 1961, Ch. XIII; or Spitzer 1978, \S~13.3a).
Therefore, for $\kd>1$, 
\begin{eqnarray}
\label{thermaljeansvphi}
\vph = \pm \left[1-\left(\frac{\lambdad}{2 \pi}\right)^2\right]^{1/2}; 
\end{eqnarray}
i.e., the two acoustic waves have the same phase velocity, modified by gravity, but
propagate in opposite directions (along the field lines). In the limit $\lambdad \ll
1$, $\vph=\pm 1$ and $\tdamp=\infty$; i.e., these modes are undamped sound
waves (recall that the unit of speed is $\czero$). At longer wavelengths,
gravitational forces become increasingly more important and the phase velocity
becomes less than unity (see Fig. $2a$). The waves are {\em gravitationally
suppressed} (i.e., $\vph=0$) at wavelengths greater than the thermal Jeans
wavelength
\begin{eqnarray}
\label{lamJdefeq}
\lamJ=2\pi
\end{eqnarray}
($=2 \pi \czero \tffo=0.611~\rm{pc}$, dimensionally). For $\lambdad > \lamJ$
(i.e, $\kd < 1$), it follows from equation (\ref{jeanmodeeq}) that each of the
Jeans modes splits into two separate, conjugate modes. One is a gravitational
growth (or fragmentation) mode, with timescale 
\begin{eqnarray}
\label{tffgrowtheq}
\tgrowth=\left[1 - \left(\frac{\lamJ}{\lambdad}\right)^2\right]^{-1/2} 
\end{eqnarray}
(see Fig. $2c$). This is the classical Jeans instability. As $\lambdad
\rightarrow \infty$, $\tgrowth \rightarrow 1$; dimensionally, this is just the
free-fall timescale, $\tffo$. The corresponding eigenvector is labeled as
``ff+" in Figures $3a$ - $3d$. The other mode is one of exponential decay,
with damping timescale $\tdamp$ also given by equation (\ref{tffgrowtheq})
(see eq. [\ref{jeanmodeeq}]); it is the curve labeled by ``ff$-$" in Figure $2b$.
The eigenvector, also labeled by ``ff$-$", is shown in Figures $3a$ - $3d$. This
mode is one in which an initial density enhancement causes expansive motion,
opposed by gravity, at such a rate that the density enhancement decreases to
zero at the same time that the velocity vanishes. Hence, this is a
monotonically decaying mode; no wave motion is involved. It is
similar to the well-known classical cosmological problem of an expanding 
``flat" universe. Note that, as $\lambdad \rightarrow \infty$, 
$\tdamp \rightarrow 1$. 

     In order to better understand the neutral thermal (Jeans) modes, we
examine more closely the eigenvectors (and their features shown in Figures
$3a$ - $3d$). We substitute equation (\ref{jeanmodeeq}) in equation
(\ref{contineqf}), and we use the normalisation condition 
$|\rhond|^2 + |\vnxd|^2 + |\vixd|^2 = 1$ and the fact
that $|\vnxd| = |\vixd|$ to find that
\leteq
\begin{eqnarray}
\label{rhonkjeanseq}
|\rhond|^2 &=& \frac{\kd^2}{3 \kd^2 - 2}
\end{eqnarray}
and
\begin{eqnarray}
\label{vnxkjeanseq}
|\vnxd|^2 &=& \frac{\kd^2 - 1}{3 \kd^2 - 2}.
\end{eqnarray}
\beq
From these equations, it follows that
\newcounter{temp}
\begin{list}
{(\alph{temp})}{\usecounter{temp}}
\item{as $\kd \rightarrow \infty$ ($\lambdad \rightarrow 0$), $|\rhond| = 
     |\vnxd| = |\vixd| \rightarrow 1/\sqrt{3}) = 0.577$, as seen in Figures 
     $3a$ - $3c$ (curves labeled ``acoustic");}
\item{as $\kd \rightarrow 1$ ($\lambdad \rightarrow$ 2$\pi$), $|\rhond| 
     \rightarrow 1$ but $|\vnxd| = |\vixd| \rightarrow 0$, as also seen in
     Figures $3a$ - $3c$;}
\item{for $\kd < \left(2/3\right)^{1/2}$ (i.e., $\lambdad > 
     \left(3/2\right)^{1/2} 2\pi = 7.70$), 
     $~\rhond$ becomes imaginary (its absolute value is shown in Fig. $3c$);}
\item{for $\left(2/3\right)^{1/2} < \kd < 1$ (i.e., $2\pi < \lambdad < 7.70$),
     $~\vnxd = \vixd$ is imaginary (Figs. $3a$ and $3b$ show the absolute values 
     of these velocities).}
\end{list}

     For the convenience of the reader, Table 1 contains a list of all
abbreviations (and their meaning) used to label the curves in all the figures
of this paper.
\begin{table}
\caption{\sc Glossary of Labels in Figures}
\label{firsttable}
\begin{tabular}{ll} \hline\hline \\
\mbox{Label \hspace{7em}}  & Meaning \\
\hline
acoustic & neutral acoustic wave \\
i,coll & ion collisional-decay mode \\
i,rec & ion dissociative-recombination mode \\
ff+ & Jeans free-fall mode \\
ff$-$ & conjugate Jeans (``cosmological") mode \\
i,A & ion {\Alf} wave \\
n,A & neutral {\Alf} wave \\
AD & magnetically-driven ambipolar-diffusion mode \\
n,coll & neutral collisional-decay mode \\
i,ms & ion magnetosonic wave \\
n,ms & neutral magnetosonic wave \\
PD & neutral pressure-driven diffusion mode \\
AD,fr &  neutral gravitationally-driven AD fragmentation mode \\
i,fast & ion fast wave \\
n,slow & neutral slow wave \\
n,fast & neutral fast wave \\
\hline
\hline
\end{tabular}
\end{table}

\subsubsection{Transverse Modes for $\kvec \parallel \Bveco$: Eigenfrequencies 
and Eigenvectors}

Comparing the systems of equations for the modes with motions only in the $y$ and
only in the $z$ directions, (\ref{yneutforceqb}) - (\ref{yinducteqb}) and
(\ref{zneutforceqb}) - (\ref{zinducteqb}), we note that they are identical. 
Hence, for propagation along the field ($\theta=0^\circ$), $\omegad(\kd)$ is
degenerate for these, transverse modes. Figure $4a$ displays the absolute
value of their phase velocity, obtained by solving the dispersion relation
\begin{eqnarray}
\label{transvdispeq}
\omegad^3 + i \left(\frac{1}{\tniod} + \frac{1}{\tinod}\right)\omegad^2 -
\left(\vaiod \kd\right)^2 \omegad - i \frac{\left(\vaiod
\kd\right)^2}{\tniod} = 0.
\end{eqnarray}
\begin{figure}
\includegraphics[width=180mm]{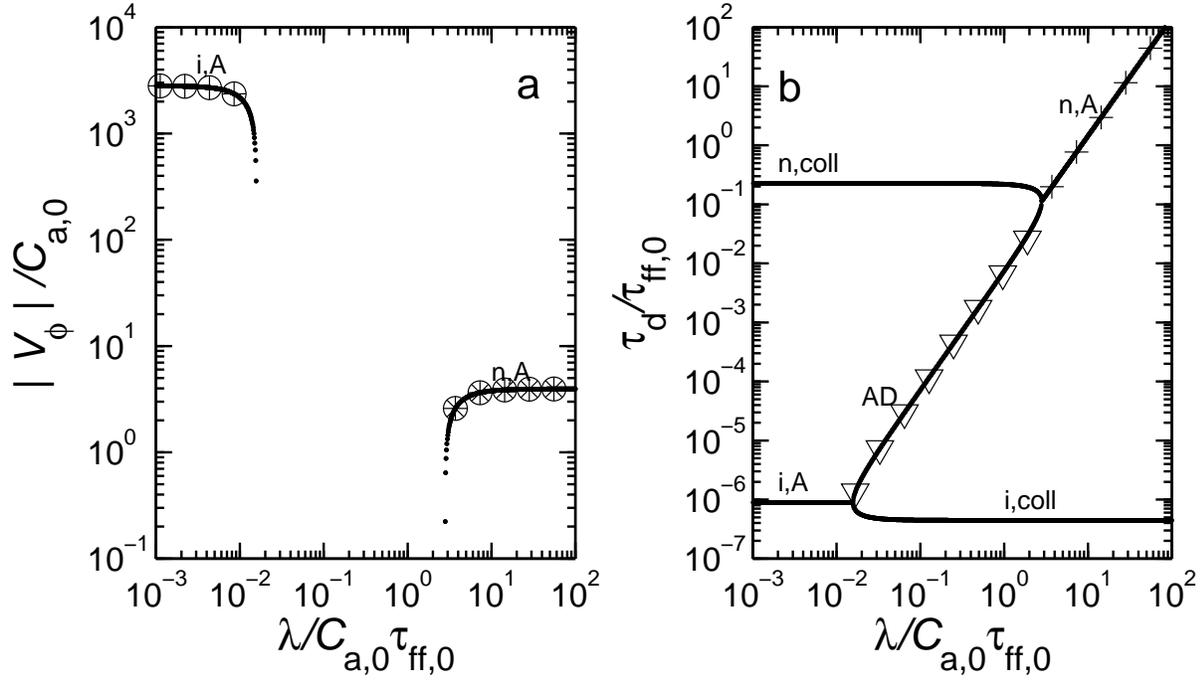}
\caption{{\it Eigenvalues of transverse modes as functions of wavelength, at an
angle of propagation $\theta=0^\circ$ with respect to $\Bveco$}. All
normalisations are as in Fig. 2. At each wavelength there are four different modes 
in all: two with motions in the $y$ direction, and two with motions in the
$z$ direction. ({\it a}) Absolute value of phase velocity, $|\vphd|$. 
Phase velocities resulting from eqs. (\ref{vaieq}) and (\ref{vaneq}) are
also displayed (circles with interior crosses, and circles with interior
asterisks, respectively). ({\it b}) Damping timescale $\tdampd$. Also shown 
are values (downward-facing triangles) calculated from eq. 
(\ref{iADdiffeq}), and values (crosses) calculated from eq. (\ref{nADdiffeq}).}
\end{figure}
(Note that, because of the degeneracy, there are four different modes.) Damping 
timescales are shown in Figure $4b$ as functions of $\lambdad$. None of the 
modes are unstable. The absolute values of the eigenvectors are exhibited in 
Figure 5: $|\vnyd|$ and $|\vnzd|$ in Figure $5a$, $|\viyd|$ and $|\vizd|$ 
in Figure $5b$, and $|\Byd|$ and $|\Bzd|$ 
in Figures $5c$ and $5d$. Note that the $|\Byd|$ (and $|\Bzd|$) axis in 
Figure $5d$ is logarithmic in order to show the behaviour of the modes 
at small wavelengths.

\begin{figure}
\includegraphics[width=180mm]{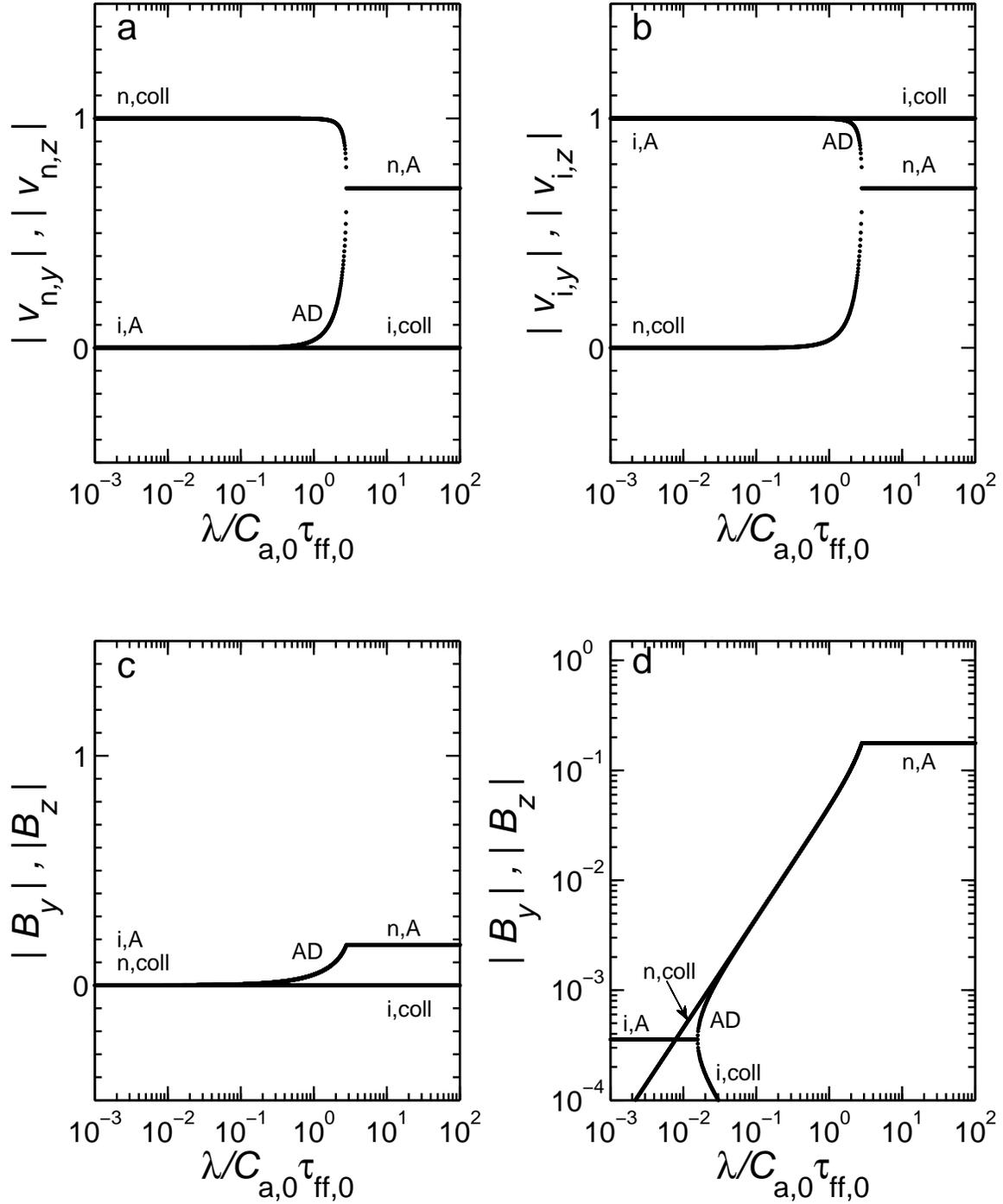}
\caption{{\it Magnitudes of eigenvectors of transverse modes for $\theta=0^\circ$ 
as functions of wavelength.} All labels and normalisations are as in Fig. 3 and 
Table 1.  As in Fig. 4, there are four modes at each wavelength. ({\it a}) 
Transverse ($y$- and $z$-) components of the neutral velocities, $|\vny|$,
and $|\vnz|$. ({\it b}) Transverse ($y$- and $z$-) components of the ion 
velocities, $|\viy|$ and $|\viz|$. ({\it c, d}) Transverse ($y$- and $z$-)
components of the magnetic field, $|\By|$ and $|\Bz|$ shown on two 
different vertical scales, which bring out different important features 
(see text).}
\end{figure}

From the dispersion relation (\ref{transvdispeq}) and Figures $4a$
and $5a$, it is evident that small-wavelength, high-frequency 
($|\omegad| \simgt 1/\tinod$) ion modes propagate
with $\vnyd$, $\vnzd \simeq 0$. Solving equations
(\ref{yneutforceqb}) - (\ref{yinducteqb}) (or, equivalently, eqs.
[\ref{zneutforceqb}] - [\ref{zinducteqb}]) in these limits yields
\begin{eqnarray}
\label{ionalfveneq}
\omegad =
\pm \vaiod \kd \left[1- \left(\frac{1}{2\vaiod \tinod \kd}\right)^2
\right]^{1/2} -\frac{i}{2 \tinod} .
\end{eqnarray}
(In deriving eq. [\ref{ionalfveneq}] we have used the fact that $1/\tinod \gg
1/\tniod$; see eqs. [\ref{tnioddefeq}] and [\ref{tinoddefeq}].) For $\lambdad$
less than the {\em ion {\Alf} cutoff wavelength}
\begin{eqnarray}
\label{ioncutoffeq}
\lami \equiv 4 \pi \vaiod \tinod,
\end{eqnarray}
waves propagate with
\begin{eqnarray}
\label{vaieq}
\vph = \pm \vaiod \left[1-\left(\frac{\lambdad}{\lami}\right)^2\right]^{1/2}~;
\end{eqnarray}
$\lami=1.57 \times 10^{-2}$ (i.e., $\lamidim = 1.53 \times 10^{-3}~\rm{pc}$)
for the model cloud parameters specified at the beginning of \S~3. Hence, for
$\lambdad \ll \lami$, the waves are {\Alf} waves, with $\vph = \pm \vaiod =
2.81 \times 10^3$ (see Fig. $4a$, curve labeled ``i,A"). There are four waves
in all. The two polarised in the $y$-direction are {\em normal, shear 
{\Alf} waves}, and the two polarised in the $z$-direction are {\em modified 
{\Alf} waves}. (At $\theta > 0$ the latter set of waves are {\em fast} waves.) 
All the waves are damped on the timescale $\tdamp = 2 \tinod$ (see eq. 
[\ref{ionalfveneq}] and Fig. $4b$, curve labeled ``i,A") because of
collisions with the neutrals. It is noteworthy that the damping time is longer
by a factor of 2 than that ($\tinod$) referring to the dissipation (momentum
exchange) of ion streaming motion relative to the neutrals. This is so
because, although a typical ion indeed loses memory of the collective (wave)
motion on a timescale $\tinod$, {\em half of the wave energy is stored as
potential energy in the magnetic field. Therefore, it takes twice as long for
collisions to damp the wave than it takes them to damp ion streaming.} 

{\em At $\lambdad = \lami$ the ion {\Alf} waves are critically damped.} For
$\lambdad \geq \lami$, the ion-neutral collision frequency $1/\tinod$ is
greater than the wave (angular) frequency $\omegad$, and the waves can no
longer propagate ($\vph=0$). At $\lambdad = \lami$ there is a bifurcation in
the ion modes (see Fig. $4b$). Two of the modes (one polarised in the
$y$-direction and the other in the $z$-direction), corresponding to the
negative root in equation (\ref{ionalfveneq}), become ion collisional-decay
modes (discussed earlier in \S~3.1.1; curves labeled ``i,coll" in Figs. $4b$
and $5a$ - $5d$), with
\begin{eqnarray}
\label{negrooteq}
\tdamp =\frac{2 \tinod}{1 +
\left[1-\left(\lami/\lambdad\right)^2\right]^{1/2}} ,~~~~\lambdad \geq \lami .
\end{eqnarray}
In the limit $\lambdad \gg \lami$, $\Byd$, $\Bzd \rightarrow 0$ (see Figs.
$5c$ and $5d$), and $\tdamp \rightarrow \tinod$, just as in equation
(\ref{iondampeq}).
Thus, as $\lambdad$ increases, magnetic restoring forces on the ions become
negligible, and the motion of the ions simply decays on a timescale $\tinod$
because of collisions with the neutrals (see Fig. $4b$, curve labeled
``i,coll"). The remaining two ion modes (again,
one polarised in the $y$-direction and the other in the $z$-direction),
corresponding to the positive roots of the dispersion relation (eq.
[\ref{ionalfveneq}]), are {\em magnetically-driven ambipolar-diffusion modes}
(see Fig. $4b$, curve labeled ``AD"), in which the ions (and electrons) 
diffuse quasistatically (i.e., with negligible acceleration; this is 
equivalent to having $|\omegad| \ll 1/\tinod$ in eqs.
[\ref{yionforceqb}] and [\ref{zionforceqb}]) relative to the stationary
neutrals. The damping timescale for these modes is 
\begin{eqnarray}
\label{posrooteq}
\tdamp= \frac{2 \tinod}{1-\left[1-\left(\lami/\lambdad\right)^2\right]^{1/2}},
\end{eqnarray}
%
which is the curve labeled as ``AD" in Figure $4b$. In the limit $\lambdad \gg 
\lami$, $\tdamp$ becomes equal to the ambipolar-diffusion timescale,
\leteq
\begin{eqnarray}
\label{iADdiffeq}
\tau_{{}_{\rm a}}=\frac{\lambdad^2}{4 \pi^2 \diffi},
\end{eqnarray}
%
where
\begin{eqnarray}
\label{diffidefeq}
\diffi \equiv \frac{\lami^2}{16 \pi^2 \tinod}= \vaiod^2 \tinod
\end{eqnarray}
\beq
is the {\em ion ambipolar-diffusion coefficient}. For the values of the free
parameters cited at the beginning of \S~3, $\diffi=3.52$. 

In Figures $4b$ and $5$ it is also evident that there exists two small-wavelength, 
low-frequency ($|\omegad| \ll 1/\tinod$) neutral modes. In these
modes the plasma and magnetic field lines are essentially stationary (i.e., $\viyd
\simeq 0$, $\vizd \simeq 0$ and $\Byd \simeq 0$, $\Bzd \simeq 0$). Under these
constraints, equations (\ref{yneutforceqb}) - (\ref{yinducteqb}) (and,
similarly, eqs. [\ref{zneutforceqb}] - [\ref{zinducteqb}]) yield
\begin{eqnarray}
\label{neutdampeq}
\omegad=-\frac{i}{\tniod}.
\end{eqnarray}
These modes are the {\em neutral collisional-decay modes} (see curves labeled
by ``n,coll" in Figs. $4$ and $5$), in which the motion of the neutrals in
the $y$ and $z$ directions decays due to collisions with ions that are held
fixed in space by the magnetic field; $\vph=0$, and $\tdamp
= \tniod =0.226$ for these modes (see Fig. $4b$).

At longer wavelengths ($\sim \lamn$, see eq. [\ref{lamndefeq}] below),
collisions between the neutrals and the ions cause the two fluids to begin to
move together. Hence, magnetic forces on the ions are more readily transmitted
to the neutrals. For large enough $\lambdad$, the neutrals can sustain a HM
wave. {\em At the point} $\lambdad=\lamn$, {\em the ion ambipolar-diffusion modes 
and the neutral collisional-decay modes merge} (see Fig. $4b$); waves can propagate 
at longer wavelengths. In the limit $|\omegad| \ll 1/\tinod$, the 
dispersion relation (\ref{transvdispeq}) has the solution 
\leteq
\begin{eqnarray}
\label{neutalfveneqa}
\omegad &=&
\pm \vaiod \left(\frac{\tinod}{\tniod}\right)^{1/2} \kd \left[1 -
\frac{\left(\vaiod \tinod \kd \right)^2}{4 (\tinod/\tniod)}\right]^{1/2}
-\frac{i}{2} \vaiod^2 \tinod \kd^2 , \\
\label{neutalfveneqb}
&=&
\pm \vanod \kd \left[1-
\frac{1}{4}\left(\vanod \tniod \kd \right)^2\right]^{1/2} 
-\frac{i}{2} \vanod^2 \tniod \kd^2 ,  
\end{eqnarray}
\beq
where $\vanod$ [$=\vano/\czero$, $\vano \equiv \Bo/\left(4 \pi \rhono \right)^{1/2}$]
is the dimensionless {\Alf} speed in the neutrals. In equation
(\ref{neutalfveneqb}) we have eliminated $\tinod$ and $\vaiod$ in favor of
$\tniod$ and $\vanod$ (see eqs. [\ref{tnieq}] and [\ref{tineq}], and recall
that $\rhono = \mn \nno$). It follows from equation (\ref{neutalfveneqb}) that
$\vph > 0$ for all $\lambdad > \lamn$, where 
\begin{eqnarray}
\label{lamndefeq}
\lamn = \pi \left(\frac{\tinod}{\tniod}\right)^{1/2} \vaiod \tniod = \pi\vanod \tniod 
\end{eqnarray}
is the {\em neutral {\Alf} cutoff wavelength}. (This is referred to simply as
the {\em {\Alf} lengthscale} in Mouschovias 1987a, 1991a.) In the typical model
cloud, $\lamn=2.80$, which means that the dimensional {\Alf} cutoff wavelength
is $\lamndim = 0.273~\rm{pc}$. For $\lambdad > \lamn$, equations
(\ref{neutalfveneqb}) and (\ref{lamndefeq}) yield 
\begin{eqnarray}
\label{vaneq}
\vph=\pm \vanod \left[1 - \left(\frac{\lamn}{\lambdad}\right)^2\right]^{1/2}~. 
\end{eqnarray}
In the limit $\lambdad \gg \lamn$, $\vph \rightarrow \pm \vanod$ ($=\pm 3.94$, since
$|\vano| =0.957~\rm{km}~\rm{s}^{-1}$); these modes are {\Alf} waves in the
neutrals (see curves labeled ``n,A" in Figs. $4$ and $5$). The two (oppositely
propagating) waves polarised in the $y$-direction are normal {\Alf}
waves. The two waves polarised in the $z$-direction are modified {\Alf} waves;
at $\theta > 0$ they are {\em fast} waves. As seen from equation
(\ref{neutalfveneqb}), these modes damp on the ambipolar-diffusion timescale  
\leteq
\begin{eqnarray}
\label{nADdiffeq}
\tdamp= 2 \tau_{{}_{\rm a}}= 2 \frac{\lambdad^2}{4 \pi^2 \diffn},
\end{eqnarray}
where
\begin{eqnarray}
\label{diffndefeq}
\diffn \equiv \frac{\lamn^2}{\pi^2 \tniod} = \vanod^2 \tniod
\end{eqnarray}
\beq
is the {\em neutral ambipolar-diffusion coefficient}. Comparing equations
(\ref{diffidefeq}) and (\ref{diffndefeq}), we note that $\diffn = \diffi$
($=3.52$ for this typical model cloud). We may therefore denote the 
ambipolar-diffusion coefficient simply as $\diff$, without the subscript i or n. 
One should bear in mind, however, that the expression for $\tdamp$ for the neutral 
{\Alf} waves contains an extra factor of 2 compared to the expression for $\tdamp$
for the ion ambipolar-diffusion mode (compare eqs. [\ref{nADdiffeq}] and 
[\ref{iADdiffeq}], and make use of eqs. [\ref{diffndefeq}] and [\ref{diffidefeq}]). 
As explained in the case of the ion {\Alf} waves, it takes twice as long to damp 
a wave than it takes to damp streaming motion (or diffusion). The existence of 
$\lamndim =  \pi v_{\rm{A,n}} \tni$ for {\Alf} waves in the neutrals was first 
shown by Kulsrud \& Pearce (1969), who studied the excitation and propagation of 
HM waves in the intercloud medium due to cosmic-ray streaming (see, also, Parker 
1967). Mouschovias (1987a, 1991a, b) discussed the importance of the lengthscale 
$\lamn$ and the thermal Jeans lengthscale $\lamJ$ in the formation of protostellar 
fragments (or cores) in self-gravitating molecular clouds. He proposed that 
fragmentation is initiated by the {\em decay} of HM waves due to magnetically-driven 
ambipolar diffusion and the almost simultaneous onset of a Jeans-like instability, 
due to gravitationally-driven ambipolar diffusion (see discussion in \S~4 below). 

\subsubsection{Transverse Modes for $\kvec \parallel \Bveco$: Further Discussion 
of Eigenvectors}

     More insight in the physics of the transverse modes can be gained by
understanding analytically certain key features of the eigenvectors shown in 
Figures $5a$ - $5d$. In the case of the ion {\Alf} mode, we may ignore the 
motion of and the dissipation due to the neutrals at short wavelengths and we
may use equations (\ref{yionforceqb}), (\ref{yinducteqb}), and the dispersion
relation $\omegad \simeq \mp \vaiod \kd$ (see eq. [23]), to find that
\begin{eqnarray}
\label{vybythetazeroeq}
\frac{\viyd}{\Byd} \simeq \mp \vaiod.
\end{eqnarray}
(The $\mp$ sign on the right-hand side of eq. [\ref{vybythetazeroeq}] refers
to propagation in the $\pm${\em x}-direction.) Since these modes are
incompressible (i.e., $\rhond = 0 = \rhoid$), we may use the normalisation
condition
\begin{eqnarray}
\label{normalizeqb}
|\viyd|^2 + |\Byd|^2 = 1
\end{eqnarray}
to find that
\leteq
\begin{eqnarray}
\label{viythetazeroeq}
\viyd \simeq \frac{\vaiod}{\left(\vaiod^2 + 1\right)^{1/2}} \simeq 1,
\end{eqnarray}
and
\begin{eqnarray}
\label{bythetazeroeq}
\Byd \simeq \mp \frac{1}{\left(\vaiod^2 + 1\right)^{1/2}} \simeq
\mp \frac{1}{\vaiod}.
\end{eqnarray}
\beq

Since $\vaiod = 2.81 \times 10^3$, it follows from the last equation that
$|\Byd| \simeq 3.56 \times 10^{-4}$. This is in agreement with the curve labeled
``i,A" in Figure $5d$, which is the same as Figure $5c$ but with a logarithmic
scale for $|\Byd|$ so as to show this small but finite value of $|\Byd|$ at small
$\lambdad$. Similarly, the result $\viyd \simeq 1$ is as shown in Figure $5b$,
curve labeled ``i,A".

     As $\lambdad$ increases, $|\viyd|$ remains large and $|\vnyd|$ (and $|\Byd|$) 
small even for $\lambdad > \lami$ (see Figs. $5a$ - $5d$, curves labeled ``i,A"); 
the ion ambipolar-diffusion mode maintains a significant $|\viyd|$ although ion 
{\Alf} waves do not exist for $\lambda > \lami$. As $\lambdad$ increases toward 
the neutral {\Alf} cutoff wavelength $\lamn$ (see eq. [\ref{lamndefeq}]), 
$|\viyd|$ decreases and $|\vnyd|$ increases because the ions begin to couple to 
(and induce motions in) the neutrals via collisions. At exactly $\lambdad = 
\lamn$, the two velocities become equal. For $\lambdad > \lamn$, {\Alf} waves can 
be sustained by the neutrals, and the magnitudes of all three quantities $\viyd$, 
$\vnyd$, and $\Byd$ are significant. In the long-wavelength limit, the ions are well 
coupled to the neutrals ($\viyd = \vnyd$). Using the normalisation condition
\begin{eqnarray}
\label{normalizeqc}
|\vnyd|^2 + |\viyd|^2 + |\Byd|^2 = 2|\vnyd|^2 + |\Byd|^2 = 1
\end{eqnarray}
and equations (\ref{yionforceqb}) - (\ref{yinducteqb}), we now find that
\leteq
\begin{eqnarray}
\label{vnybythetazeroeq}
\frac{\vnyd}{\Byd} = \mp \vanod,
\end{eqnarray}
\begin{eqnarray}
\label{vnythetazeroeq}
\vnyd = \viyd = \frac{\vanod}{\left(1 + 2 \vanod^2 \right)^{1/2}}
\end{eqnarray}
and
\begin{eqnarray}
\label{bythetazeroeq}
\Byd = \mp \frac{1}{\left(1 + 2 \vanod^2 \right)^{1/2}}.
\end{eqnarray}
\beq
Since for our typical model cloud $\vanod$ = 3.94, it 
follows that $\vnyd = \viyd = 0.696$ and $\Byd$ = 0.177, in agreement with
Figures $5a$ - $5d$, curves labeled ``n,A".

As $\lambdad$ increases across $\lami$, the ion {\Alf} mode disappears 
(damps) and bifurcates into the (ion) ambipolar-diffusion mode and the ion
collisional-decay mode, as discussed in relation to Figure $4b$; hence the
placement of the labels ``i,A", ``AD", and ``i,coll" on these curves
in Figures $5a$ - $5d$.

     It is also clear from Figures $5a$ - $5d$ that the eigenvectors for both
the neutral and the ion collisional-decay modes behave exactly as expected on
the basis of our discussion of these modes in relation to Figure $4b$.

\subsection{Propagation Perpendicular to $\Bveco$ ($\kvec \perp \Bveco$)}

For $\theta=90^\circ$, equations (\ref{contineqd}) - (\ref{zinducteq}) again
decouple into three independent subsystems. The subsystem involving
material motions in the $x$-direction is 
\leteq
\begin{eqnarray}
\label{contineqg}
\omegad \rhond &=& \kd \vnxd, \\
\label{ioncontineqg}
\omegad \rhoid &=& \rhoiod \kd \vixd + i \rhoiod^2 \alphdrd \rhond - 
i2 \rhoiod \alphdrd \rhoid , \\
\label{xneutforceqc}
\omegad \vnxd &=& \left(\kd - \frac{1}{\kd}\right) \rhond -
\frac{i}{\tniod}\vnxd + \frac{i}{\tniod}\vixd, \\
\label{xionforceqc}
\omegad \vixd &=&
\frac{i}{\tinod}\vnxd - \frac{i}{\tinod}\vixd + \vaiod^2 \kd \Bzd, \\
\label{zinducteqc}
\omegad \Bzd &=& \kd \vixd.
\end{eqnarray}
\beq
Motions in the $y$-direction are governed by
\leteq
\begin{eqnarray}
\label{yneutforceqc}
\omegad \vnyd &=& - \frac{i}{\tniod} \vnyd + \frac{i}{\tniod} \viyd, \\
\label{yionforceqc}
\omegad \viyd &=& \frac{i}{\tinod} \vnyd - \frac{i}{\tinod} \viyd, \\ 
\label{yinducteqc}
\omegad \Byd &=& 0.
\end{eqnarray}
\beq
Similarly, the equations for the $z$-components of the neutral and ion
velocities are
\leteq
\begin{eqnarray}
\label{zneutforceqc}
\omegad \vnzd &=& - \frac{i}{\tniod}\vnzd + \frac{i}{\tniod}\vizd, \\
\label{zionforceqc}
\omegad \vizd &=& \frac{i}{\tinod} \vnzd - \frac{i}{\tinod} \vizd.
\end{eqnarray}
\beq

\subsubsection{Longitudinal Modes for $\kvec \perp \Bveco$: Eigenfrequencies}

The dispersion relation is easily obtained from equations
(\ref{contineqg}) - (\ref{zinducteqc}):
\begin{eqnarray}
\label{longdispeq}
\left(\omegad + 2 i \rhoiod \alphdrd \right) \left\{\omegad^4 +
i\left(\frac{1}{\tinod} + \frac{1}{\tniod}\right)\omegad^3 -
\left[\left(\vaiod^2 + 1\right)\kd^2 -1\right]\omegad^2
\right. \nonumber \\
\left. \hspace{6em} - i \left[\frac{1}{\tniod}\vaiod^2 \kd^2 +
\frac{1}{\tinod}\left(\kd^2-1\right)\right] \omegad +
\left(\kd^2-1\right)\vaiod^2 \kd^2 \right\} = 0.
\end{eqnarray}
Because it is a fifth-order polynomial, there are five longitudinal modes in
all. The phase velocities, damping timescales, and growth timescales for the
longitudinal modes are displayed in Figures $6a$, $6b$, and $6c$,
respectively; eigenvectors are shown as functions of $\lambdad$ in Figure
$7a$ - $7e$.
\begin{figure}
\includegraphics[width=180mm]{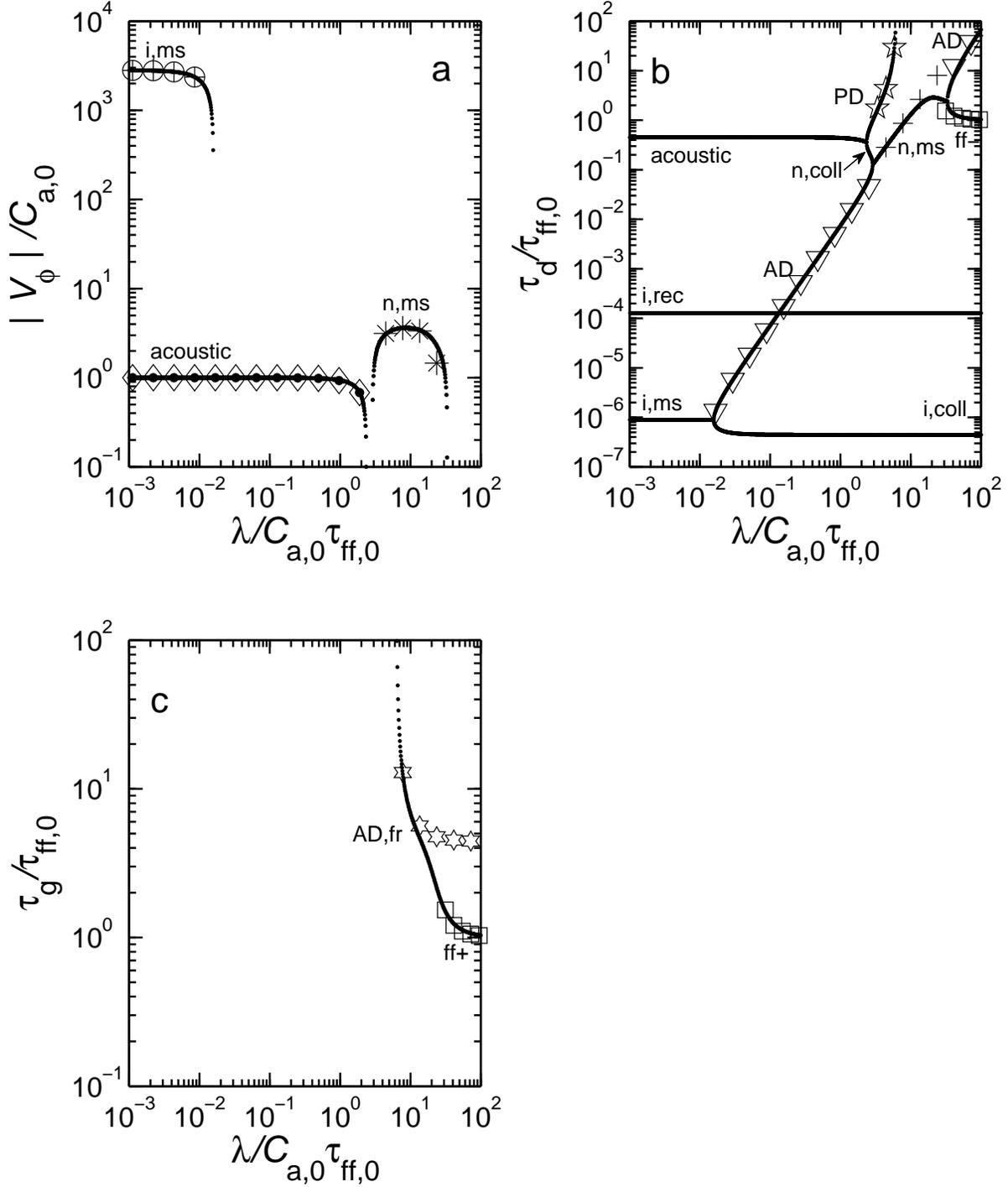}
\caption{{\it Eigenvalues of longitudinal modes as functions of wavelength,
at an angle of propagation $\theta=90^\circ$ with respect to $\Bveco$}. All
normalisations are as in Fig. 2. There are five different modes at each
wavelength.  ({\it a}) Absolute value of phase velocity, $|\vphd|$. Phase 
speeds derived from eq. (\ref{vaieq}) are displayed as circles with 
interior crosses, while speeds obtained from eq. (\ref{acousticeq}) are 
displayed as diamonds with interior black circles. Phase speeds determined 
from eq. (\ref{nmswaveeq}) are displayed as asterisks. ({\it b}) Damping 
timescales $\tdampd$. Also shown are values (downward-facing triangles) 
calculated from eq. (\ref{iADdiffeq}), as well as values (plus signs) from
eq. (\ref{nADdiffeq}). Results (five-pointed stars) from
eq. (\ref{diffcseq}) are also depicted, and so are those (boxes) calculated
by using eq. (\ref{cosmodeeq}). ({\it c}) Growth timescale $\tgrowthd$. 
Values predicted (six-pointed stars) from the ambipolar-diffusion 
fragmentation timescale (eq.  [\ref{tadtimescaleeq}]) are also displayed, 
along with values (boxes) calculated from the magnetic Jeans instability 
mode, eq. (\ref{magjeanmodeeq}).}
\end{figure}
\begin{figure}
\includegraphics[width=180mm]{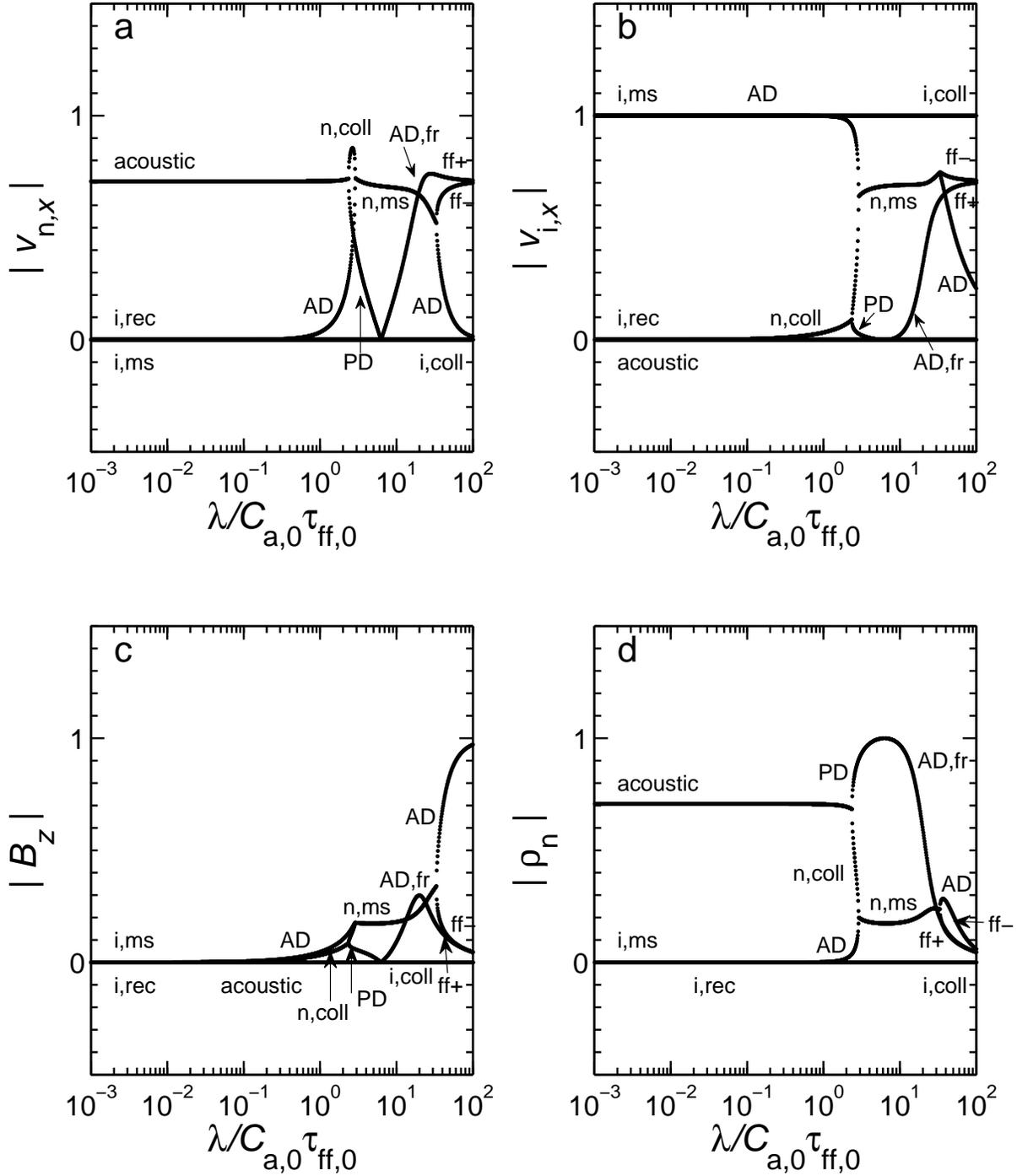}
\caption{{\it Magnitudes of eigenvectors of longitudinal modes for
$\theta=90^\circ$ as functions of wavelength.} All labels and normalisations
are as in Fig. 3 and Table 1.  As in Fig. 6, there are five modes.
({\it a}) Longitudinal ($x$-) component of the neutral velocity, $|\vnx|$. ({\it
b}) Longitudinal ($x$-) component of the ion velocity, $|\vix|$. ({\it c})
Magnetic field component $|\Bz|$. ({\it d}) Neutral density $|\rhon|$. 
}
\end{figure}
\addtocounter{figure}{-1}
\begin{figure}
\includegraphics[width=180mm]{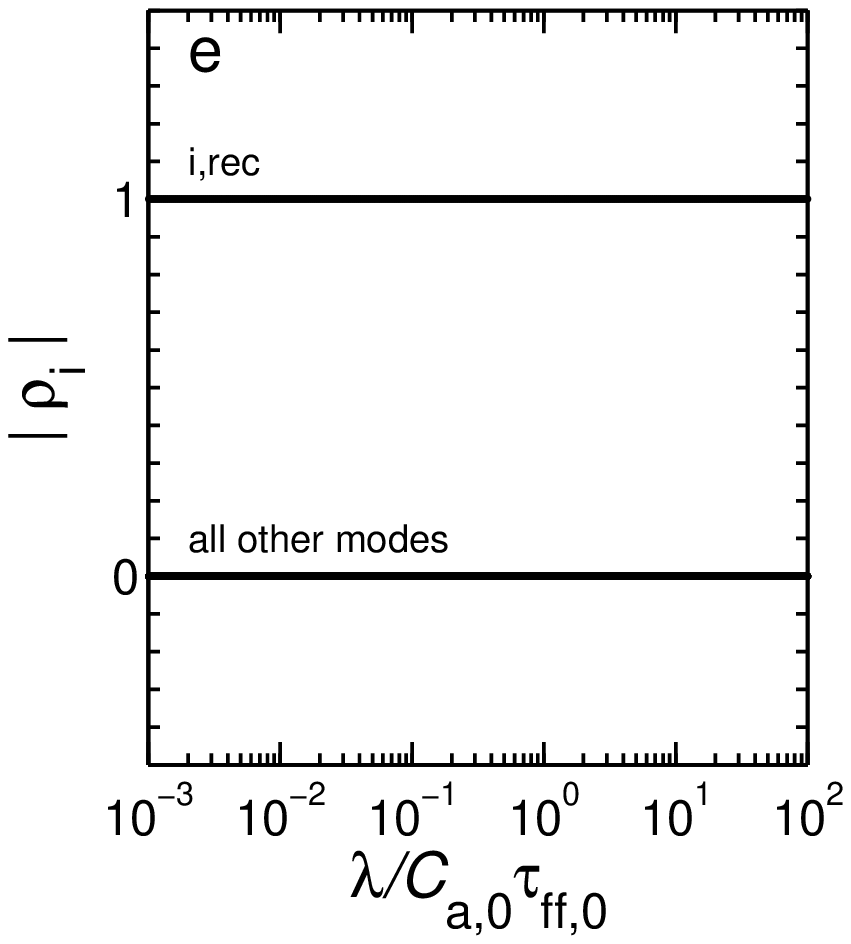}
\caption{{\it Cont.} ({\it e}) Ion density $|\rhoi|$.}
\end{figure}

     At small wavelengths there are again two high-frequency ($|\omegad|
\sim 1/\tinod$) ion modes, with $\vnxd \simeq 0$ (they are degenerate with
respect to the direction of propagation). In this limit, the solution of the
dispersion relation is identical with that given by equation
(\ref{ionalfveneq}). The modes it represents in this case are {\em ion
magnetosonic modes}. Because the speed of sound of the ions ($< \czero$, since
$\mi > \mn$) is negligible compared to the ion {\Alf} speed (see eq.
[\ref{vaioddefeq}]), the phase velocity of the waves for $\lambdad < \lami$
is again given by equation (\ref{vaieq}) and the characteristic decay time
is $\tdamp = 2 \tinod$ (see Figs. $6a$ and $6b$, curves labeled ``i,ms").
For $\lambdad \geq \lami$ ($=1.57 \times 10^{-2}$ for the typical model 
cloud) the ion magnetosonic waves cannot propagate because of frequent 
ion-neutral collisions. Instead, each mode bifurcates (see Fig. $6b$), just
as in the case of ion {\Alf} waves described in \S~3.1.2. One of the two 
resulting modes is an ion collisional decay mode ($\tdamp=\tinod$; curve 
labeled ``i,coll" in Fig. $6b$) and the other is a magnetically-driven ion
ambipolar-diffusion mode, curve labeled ``AD" in Figure $6b$ 
($\tdamp = \lambdad^2/4 \pi^2\diff$ for $\lambdad \gg \lami$; see 
discussion preceding eq. [\ref{diffidefeq}]).

The third mode is an ion dissociative-recombination decay mode, as discussed
in \S~3.1.1, in which density enhancements in the ions decay rapidly
[$\tdamp=\left(2 \rhoiod \alphdrd\right)^{-1}= 1.28 \times 10^{-4}$ for the
typical model cloud parameters] because of dissociative recombinations between
molecular ions and electrons (see Fig. $6b$, curve labeled ``i,rec"). The
only nonvanishing component of the eigenvector for this mode is $\rhoiod$
(see Figs. $7e$ and $7a$ - $7d$).

There also exist two small-wavelength, low-frequency ($|\omegad| \ll
1/\tinod$) wave modes in the neutrals. For these modes, $\vixd \simeq 0$ and
$\Bzd \simeq 0$ (curves labeled ``acoustic'' in Figs. 7b and 7c), i.e., 
the neutrals oscillate in the $x$-direction in an effectively fixed 
background of ions and magnetic field. Under these circumstances, the 
solution of the dispersion relation (\ref{longdispeq}) is 
\begin{eqnarray}
\label{soundeq}
\omegad =
\pm \kd \left[1 - \left(1+
\frac{1}{4\tniod^2}\right)\frac{1}{\kd^2}\right]^{1/2} -\frac{i}{2 \tniod} . 
\end{eqnarray}
The phase velocity can be written as 
\begin{eqnarray}
\label{acousticeq}
\vph = \pm \left[1 - \left(\frac{\lambdad}{\lamcs}\right)^2\right]^{1/2} , \, \, \, \, \,
{\rm for} \, \, \lambdad < \lamcs,
\end{eqnarray}
where
\begin{eqnarray}
\label{lamcsdefeq}
\lamcs \equiv \frac{4 \pi \tniod}{\left[1 + (2 \tniod)^2\right]^{1/2}}
\end{eqnarray}
is the {\em acoustic-wave cutoff wavelength}; $\lamcs =2.59$ in the typical
model cloud. In dimensional form,
$\lamcsdim \simeq 4 \pi \czero \tnio/[1 + (2\tniod)^2]^{1/2}=0.252~\rm{pc}$. 
(Note that for $2\tniod \ll 1$, $\lamcs \simeq 4\pi\tniod$; in this
limit $\lamcsdim = 4 \pi \czero \tnio = 0.277$ pc.) 
These are sound waves in the neutrals modified by gravity and neutral-ion 
collisions. For $\lambdad \ll \lamcs$, $\vph = \pm 1$ ($=\pm \czero$, 
dimensionally), hence, the waves are pure sound waves (see Fig. $6a$, curve 
labeled ``acoustic"). They decay on a timescale $\tdamp=2 \tniod$ (= 0.452; 
see curve labeled ``acoustic" in Fig. $6b$).  For $\lambdad \geq \lamcs$, the 
waves are damped because of collisions with the ions, and $\vph=0$, much 
like the damping of the ion {\Alf} and magnetosonic modes for 
$\lambdad \geq \lami$. For $\lambdad > \lamcs$, this modified neutral sound
wave bifurcates (see Fig. $6b$) into a {\em pressure-driven diffusion mode}
(curve labeled ``PD") and a {\em collisional-decay mode} (curve labeled 
``n,coll"). We examine these modes in that order.

{\em The pressure-driven diffusion mode} corresponds to the positive root of
equation (\ref{soundeq}). The neutrals are diffusing quasistatically (i.e.,
with $|\omegad| \ll 1/\tniod$) through a background of effectively stationary
ions and magnetic field. The eigenvector for this mode is labeled ``PD" in
Figs. $7a$ - $7e$. For $\lamcs \leq \lambdad \leq \lamJ$, the damping timescale
is
\leteq
\begin{eqnarray}
\label{diffcseq}
\tdamp &=& \frac{2\tniod}{1 - \left\{1 - 4
\tniod^2\left[\left(\lamJ/\lambdad\right)^2 -1 \right]\right\}^{1/2}} ,\\
\label{diffcseqb}
&\simeq& \frac{(\lambdad/\lamJ)^2}{\diffncs \left[ 1 -
\left(\lambdad/\lamJ\right)^2\right]},
\end{eqnarray}
\beq
where 
\begin{eqnarray}
\label{diffcsdefeq}
\diffncs \equiv \tniod
\end{eqnarray}
($=\czero^2 \tnio$ dimensionally) is the neutral {\em pressure-driven
diffusion coefficient}. In obtaining equation (\ref{diffcseqb}) we have used
the fact that $\tniod^2 \ll 1$. For the representative model cloud used in this
paper, $\diffncs = 0.226$ ($\simeq \diff/16$). We note that, at $\lambdad=\lamJ$,
$\tdamp = \infty$ (see Fig. 6$b$, curve labeled ``PD'').
At this wavelength, the restoring pressure forces in this mode are exactly
balanced by self-gravitational forces, and the system is on the verge of
gravitational instability ($\omegad=0$).
Thus the Jeans instability manifests itself at $\lamJ$ even in the presence of
a magnetic field, as originally recognized by Langer (1979). Ambipolar diffusion 
allows this to happen, but, because of neutral-ion collisions, the growth time of 
the instability is longer than that of the nonmagnetic Jeans instability (compare eq. 
[\ref{tadtimescaleeq}] below with eq. [\ref{tffgrowtheq}]). For $\lambdad > \lamJ$, 
density perturbations in the neutrals grow exponentially in time as a result of 
gravitational contraction of neutrals through essentially stationary ions attached 
to rigid magnetic field lines. The growth time for this instability can be obtained 
easily from equation (\ref{soundeq}) under the conditions $\lambdad > \lamJ$ and 
$4\tniod^2\left[1 - \left(\lamJ/\lambdad\right)^2 \right] \, \ll \, 1$: 
\begin{eqnarray}
\label{tadtimescaleeq}
\tgrowth = \frac{(1/\tniod)}{1 - \left(\lamJ/\lambdad \right)^2}
=\frac{\vffo}{1 - \left(\lamJ/\lambdad\right)^2}, 
\end{eqnarray}
where $\vffo = 1/\tniod$ ($=\tffo/\tnio$ = 1/0.226 = 4.42) is the collapse
retardation factor, discussed in the penultimate paragraph of \S~2.3. It
follows from equation (\ref{tadtimescaleeq}) that, in the limit $\lambdad \gg
\lamJ$, $\tgrowth \rightarrow \vffo$. The growth time of this 
{\it ambipolar-diffusion--induced fragmentation} is shown in Figure $6c$ (the 
part of the curve labeled by ``AD,fr"). Dimensionally, the growth time for this
mode is $\vffo \tffo=\tffo^2/\tnio$. It is the same as the {\em nonlinear}
solution found analytically by Mouschovias (1979; see also 1983, 1987a, b, 1989) 
for the timescale of formation and evolution of protostellar cores (due to
gravitationally-driven ambipolar diffusion).  Numerical simulations
(including the effects of grains, UV ionisation, and magnetic braking) of the
formation of protostellar cores in magnetically supported molecular clouds
have also found that the evolution occurs on this timescale (Fiedler \&
Mouschovias 1993; Ciolek \& Mouschovias 1994, 1995; Basu \& Mouschovias 1994,
1995a, b). The same timescale was found in the one-dimensional similarity 
solution of Scott (1984).
It is clear from Figure $6c$ that, as predicted, $\tgrowth$ would tend to 
$1/\tniod =4.42$ for $\lambdad \gg \lamJ$; see the inflection point in the
curve (labeled ``AD,fr"). However, $\tgrowth$ falls below this would-be asymptotic 
value because, for $\lambdad \simgt \lamms$, where $\lamms$ is the 
{\em magnetic Jeans wavelength} (= 25.6; see eq. [\ref{lammsdefeq}] below),
gravitational forces on the neutrals, transmitted to the ions by neutral-ion
collisions, exceed the restoring magnetic forces on the ions, and the ions and
magnetic field are no longer able to remain stationary; the mode becomes a
gravitational (Jeans) instability against the magnetic field, as originally
found by Chandrasekhar \& Fermi (1953);
\footnote{The nonlinear equivalent of this instant in the development of the 
ambipolar-diffusion--induced, gravitationally-driven fragmentation is the 
instant at which a fragment's mass-to-flux ratio reaches its critical value 
and dynamical contraction ensues with the magnetic field essentially frozen 
in the matter.}
see Figure $6c$, part of curve beyond the inflecton point, labeled ``ff+".
The approximations $\vixd \simeq 0$, $\Bzd \simeq 0$ (which were used in 
deriving eq. [\ref{soundeq}]) are no longer valid, and equation 
(\ref{tadtimescaleeq}) no longer describes this mode. The proper expressions
are derived below.

The second mode resulting from the bifurcation of the modified sound waves at
$\lamcs$ is described by the negative root of equation (\ref{soundeq}). This
is a neutral collisional-decay mode (curve labeled ``n,coll" in Fig. $6b$), 
with damping time
\leteq
\begin{eqnarray}
\label{neutcollmodeeqa}
\tdamp &=& \frac{2 \tniod}{1 + \left\{1 - 4
\tniod^2\left[\left(\lamJ/\lambdad\right)^2 - 1 \right] \right\}^{1/2}} \\
\label{neutcollmodeeq}
&\simeq& \frac{\tniod}{1 - \tniod^2 \left[
\left(\lamJ/\lambdad\right)^2-1\right]}.
\end{eqnarray}
\beq
In equation (\ref{neutcollmodeeq}) we have again used the fact that, for the
typical model cloud, $\tniod^2 \ll 1$.

For $\lambdad \gg \lamJ$, $\tdamp \rightarrow \tniod$. This limit is never
attained by this mode, however, because the collisional coupling between the
neutrals and the ions becomes more effective with increasing $\lambdad$, and
the ions (and, hence, the magnetic field lines) begin to move with the
neutrals. As a result, the neutral collisional-decay mode and the (ion)
ambipolar-diffusion mode combine and merge to form wave modes, in a way
similar to that for the neutral {\Alf} waves in \S~3.1.2 (compare Figs. $4b$
and $6b$). This mode coupling occurs at $\lambdad=2.72$ (= $\lammsn$; see
below), which is only slightly greater than the neutral acoustic-wave cutoff
$\lamcs=2.59$ for the typical model cloud (see Fig. $6b$). The solution of
the dispersion relation (eq. [\ref{longdispeq}]) for these modes in the 
limit $|\omegad| \ll 1/\tinod$ is
\begin{eqnarray}
\label{nmagsoniceq}
\omegad = \pm \vmsod \kd \left[1-\frac{1}{\vmsod^2 \kd^2} -\frac{\left(
\vanod^2 \tniod \kd\right)^2}{16 \vmsod^2} \right]^{1/2} - \frac{i}{4}
\vanod^2 \tniod \kd^2 ,
\end{eqnarray}
where 
\begin{eqnarray}
\label{vmsdefeq}
\vmsod = \left(\vanod^2 + 1 \right)^{1/2}
\end{eqnarray}
[or, in dimensional form, $\vmso =\{\vano^2 + \czero^2\}^{1/2}$] is the
{\em magnetosonic speed in the neutrals}. For the representative model cloud used in
this paper, $\vmsod = 4.07 =1.03 \vanod$. Examination of equation
(\ref{nmagsoniceq}) reveals that $\omegad_{\rm r} > 0$ (i.e., $\vph > 0$) if
$\lammsn < \lambdad < \lamms$, where 
\begin{eqnarray}
\label{lammsndefeq}
\lammsn = \frac{\pi \vanod^2 \tniod}{\vmsod} =
\left(\frac{\vanod}{\vmsod}\right) \lamn
\end{eqnarray}
is the {\em neutral magnetosonic cutoff wavelength}, and
\begin{eqnarray}
\label{lammsdefeq}
\lamms = 2 \pi \vmsod 
\end{eqnarray}
(or, in dimensional form, $ \lammsdim =2 \pi \vmso \tffo$) is the {\em
magnetic Jeans wavelength}. For the typical model cloud, $\lammsn = 2.72$ and 
$\lamms=25.6$ (hence, $\lammsndim = 0.264~\rm{pc}$, and $\lammsdim=2.49~
\rm{pc}$). [Note: in deriving equations (\ref{lammsndefeq})
and (\ref{lammsdefeq}), we have used the fact that, for the conditions of
interest, $\tniod^2 \ll1$ and $1/\vmsod^2 \ll 1$.] Inserting equations
(\ref{lammsndefeq}) and (\ref{lammsdefeq}) in equation (\ref{nmagsoniceq}),
the phase velocity for magnetosonic waves in the neutrals is found to be
\begin{eqnarray}
\label{nmswaveeq}
\vph = \pm \vmsod \left[ 1 - \left(\frac{\lambdad}{\lamms}\right)^2 -
\left(\frac{\lammsn}{\lambdad}\right)^2 \right]^{1/2}, ~~~~ \lammsn \leq
\lambdad  \leq \lamms
\end{eqnarray}
(see Fig. $6a$, curve labeled ``n,ms"). The waves are weakly damped by
ambipolar diffusion (see Fig. $6b$, curve labeled ``n,ms"); $\tdamp = 
2 \lambdad^2/4 \pi^2 \diff$ (see eqs. [\ref{diffndefeq}] and
[\ref{nmagsoniceq}]).

For $\lambdad \simgt \lamms$, gravitational forces on the neutrals overwhelm
the restoring magnetic forces, and the neutral magnetosonic modes become
gravitationally suppressed ($\vph=0$), just like the thermal Jeans modes
become suppressed for $\lambdad \geq \lamJ$ (see discussion in \S~3.1.1). For
$\lambdad \simeq \lamms$, each (neutral) magnetosonic mode bifurcates into an
ambipolar-diffusion mode (with damping timescale $\tdamp = \lambdad^2/4 \pi^2
\diff$ for $\lambdad \gg \lamms$) and a conjugate Jeans mode (with damping
timescale $\tdamp =1$ for $\lambdad \gg \lamms$); see Figure $6b$, curves
labeled ``AD" and ``ff$-$", respectively.

The conjugate Jeans (or classical cosmological) mode was discussed earlier in
connection with equation (\ref{jeanmodeeq}). In this mode the neutrals, ions,
and magnetic field lines are well coupled, but the gravitational forces
prevent them from oscillating; motions are damped monotonically. The damping
time, corresponding to the negative root of equation (\ref{nmagsoniceq}), for
$\lambdad \geq \lamms$ is given by
\begin{eqnarray}
\label{cosmodeeq}
\tdamp = \left\{ \left[1 - \left(\frac{\lamms}{\lambdad}\right)^2 +
\left(\frac{2 \pi^2\diff}{\lambdad^2}\right)^2 \right]^{1/2} + \frac{2
\pi^2\diff}{\lambdad^2} \right\}^{-1}.
\end{eqnarray}
It is clear that, in the limit $\lambdad \rightarrow \infty$, $\tdamp
\rightarrow 1$ (i.e., $\tau_{\rm d}=\tffo$); see Figure $6b$, curve labeled
``ff$-$".

Finally, the gravitational instability mode (see eq. [\ref{tadtimescaleeq}])
also changes behaviour at $\lamms$, as discussed above. 
For $\lambdad > \lamms$, the neutrals, plasma, and magnetic field lines are
well coupled and behave like a single fluid; self-gravity overwhelms 
restoring magnetic and thermal-pressure forces, and the mode behaves as a 
classical Jeans instability, in which density perturbations grow 
exponentially in time with a timescale 
\begin{eqnarray}
\label{magjeanmodeeq}
\tgrowth  = \left\{ \left[1 - \left(\frac{\lamms}{\lambdad}\right)^2 +
\left(\frac{2 \pi^2\diff}{\lambdad^2}\right)^2 \right]^{1/2} + \frac{2
\pi^2\diff}{\lambdad^2} \right\}^{-1},
\end{eqnarray}
which is equal to the damping time of the conjugate Jeans mode (see eq. 
[\ref{cosmodeeq}]). 
From this equation we see that $\tgrowth \rightarrow 1$ (i.e.,
$\tau_{\rm{gr}}=\tffo$)
for $\lambdad \rightarrow \infty$, in agreement with the long-wavelength
behaviour of $\tgrowth$ exhibited in Figure $6c$.

\subsubsection{Longitudinal Modes for $\kvec \perp \Bveco$: Eigenvectors}

The main features of the eigenvectors are as follows.

\newcounter{tempb}
\begin{list}
{(\alph{tempb})}{\usecounter{tempb}}
\item{The dominant component of the ion magnetosonic and the ion 
     ambipolar-diffusion modes is $\vixd$ (see Fig. $7b$, curves labeled ``i,ms" 
     and ``AD"). It does not vanish at $\lambdad = \lami$; in fact, it hardly 
     changes from 1, because the ion-AD mode maintains $\vixd$ large beyond $\lami$. 
     Only when the ion-AD mode induces motions in the neutrals does $\vixd$ begin to 
     decrease, as the {\it magnitude} of $\vnxd$ increases as $\lambdad \rightarrow 
     \lammsn$ (see Figs. $7b$ and $7a$, curves labeled ``i,ms" and ``AD"). The 
     $z$-component of the magnetic field $\Bzd$ in Figure $7c$ actually does not 
     vanish. It is equal to $\vixd/\vims$ (see eq. [\ref{vybythetazeroeq}]), but 
     $\vims \simeq \vaio = 2.81 \times 10^{3}$.}  \\
     
\item{The velocity $\vixd$ is also the dominant component of the eigenvector of the 
     ion collisional-decay mode at all wavelengths (see Fig. $7b$, curve labeled 
     ``i,coll").} \\
     
\item{The eigenvector of the neutral acoustic mode has significant components 
     $\vnxd$ and $\rhond$. At $\lamcs$, beyond which neutral sound waves do not 
     exist and at which the acoustic mode bifurcates into the neutral 
     collisional-decay mode and the pressure-diffusion mode (see Fig. $6b$), the 
     collisional-decay mode is responsible for the increase in $|\vnxd|$ as $|\rhond|$ 
     decreases to zero. Because motion is induced in the ions as $\lambdad$ approaches 
     $\lamcs$ (see Fig. $7b$, curve labeled ``n,coll"), $|\vnxd|$ does not reach unity.} \\
     
\item{The most significant component of the PD mode is $\rhond$ (see Fig. $7d$); 
     $\rhond$ increases as $\lambdad$ increases from $\lamcs$ to $\lamJ$, at which 
     wavelength $\rhond$ reaches a maximum. Beyond $\lamJ$, ambipolar-diffusion--induced
     fragmentation sets in. At exactly $\lamJ$, $\rhond$ is the only nonvanishing 
     component of the eigenvector of the AD,fr mode (see Figs. $7a$ - $7d$). As 
     $\lambdad$ increases beyond $\lamJ$, the field lines begin to be compressed as 
     the neutrals begin to couple to the ions, and $\Bzd$ increases (see Fig. $7c$) 
     -- at the expense of $\rhond$ -- while $\vixd$ and $\vnxd$ are negligible. As 
     $\lambdad$ approaches $\lamms$, the ambipolar-diffusion--fragmentation mode 
     induces significant velocities $\vnxd$ and $\vixd$, while $\rhond$ remains large
     (see Figs. $7a$, $7b$ and $7d$). As discussed in relation to Figure $6c$, the AD,fr mode 
     turns into the Jeans free-fall mode (ff+) beyond $\lambdad \simeq \lamms$, as is 
     clearly shown in Figures $7a$ and $7b$ (see curves labeled ``AD,fr" and ``ff+"). 
     The bifurcation of the neutral magnetosonic mode into the AD and conjugate Jeans 
     (or, cosmological, ff-) modes beyond $\lambdad \simeq \lamms$, discussed in 
     relation to Figure $6b$, is also seen clearly in Figures $7a$ and $7b$.    
     [Note: If we had plotted only the real part of the eigenvector, we would have 
     found, for the neutral magnetosonic mode, that $\rm{Re}\{\rhond\}$ decreases 
     discontinuously to zero, then increases smoothly, reaches a maximum, and then 
     vanishes again as the magnetosonic waves are suppressed by gravity.]} \\

\item{Slightly beyond $\lammsn$, magnetosonic waves exist in the neutrals but, in addition to 
     maintaining $\vnxd$ large, they cause significant motion in the ions as well 
     (see Figs. $7a$ and $7b$, curves labeled ``n,ms"). The quantities $\Bzd$ and 
     $\rhond$ are also nonnegligible. Note that at $\lamJ$, $\vnxd$ and $\rhond$ 
     decrease while $\vixd$ increases.} 
\end{list}

\subsubsection{Transverse Modes for $\kvec \perp \Bveco$}

Examination of equation (\ref{yinducteqc}) reveals that there is one trivial
mode with $\omegad=0$. The remaining equations governing the transverse modes
with motions in the $y$-direction (eqs. [\ref{yneutforceqc}] and
[\ref{yionforceqc}]) are identical with those in the $z$-direction (eqs.
[\ref{zneutforceqc}] and [\ref{zionforceqc}]). Solving this simple system, we
find that
\leteq
\begin{eqnarray}
\omegad=0 ,
\end{eqnarray}
and
\begin{eqnarray}
\label{paralleleq}
\omegad = -i\left(\frac{1}{\tinod} + \frac{1}{\tniod}\right) \simeq
-\frac{i}{\tinod},
\end{eqnarray}
\beq
both of which are independent of wavelength. (Note: there are four modes in
all, two corresponding to the first solution, and two corresponding to the
second.) In the first mode the ions and neutrals move together with
$\vnyd=\viyd$ (or $\vnzd=\vizd$); hence, there are no frictional forces
between the two species, and $\tdamp = \infty$. The second mode consists of
oppositely flowing streams of ions and neutrals. The momentum of each species
decays by collisions with the other; hence, ion-neutral and neutral-ion
collisions occur in {\em parallel}, and the net damping time for this mode is
the {\em harmonic mean} of $\tinod$ and $\tniod$ (see eq. [\ref{paralleleq}]).
For the typical model cloud, the ion fluid has a much smaller inertia than the
neutral fluid and, therefore, the streaming velocity of the ions is much
greater than that of the neutrals. As a result, the motion of the ions is
collisionally damped on the timescale $\tdamp \simeq \tinod$.

\subsection{Propagation at Angles $0^{\circ} \, < \, \theta \, < \, 90^{\circ}$ 
with respect to $\Bveco$}

Equations (\ref{contineqd}) - (\ref{zinducteq}) reveal that, for $0^\circ \leq
\theta \leq 90^\circ$, the equations for the modes with motions in the
$y$-direction are uncoupled from those with motions in the $(x,z)$-plane.
Motions in the $x$- and $z$-direction, however, are coupled. The modes with
$\vnyd \neq 0 \neq \viyd$ (governed by eqs.
[\ref{yneutforceq}] - [\ref{yinducteq}]) are purely transverse; those with
nonvanishing velocities in the $(x,z)$-plane (governed by eqs.
[\ref{contineqd}] - [\ref{xinducteq}] and [\ref{zneutforceq}] - [\ref{zinducteq}])
are neither purely longitudinal nor purely transverse.

\subsubsection{Transverse Modes}

Figures $8a$ and $8b$ display the magnitude of the phase velocities and
the damping timescales of the three transverse modes for propagation at an
angle $\theta=45^\circ$ with respect to $\Bveco$. Because they are
incompressible, none of these modes can become gravitationally unstable.
Eigenvectors are displayed in Figure 9.
\begin{figure}
\includegraphics[width=180mm]{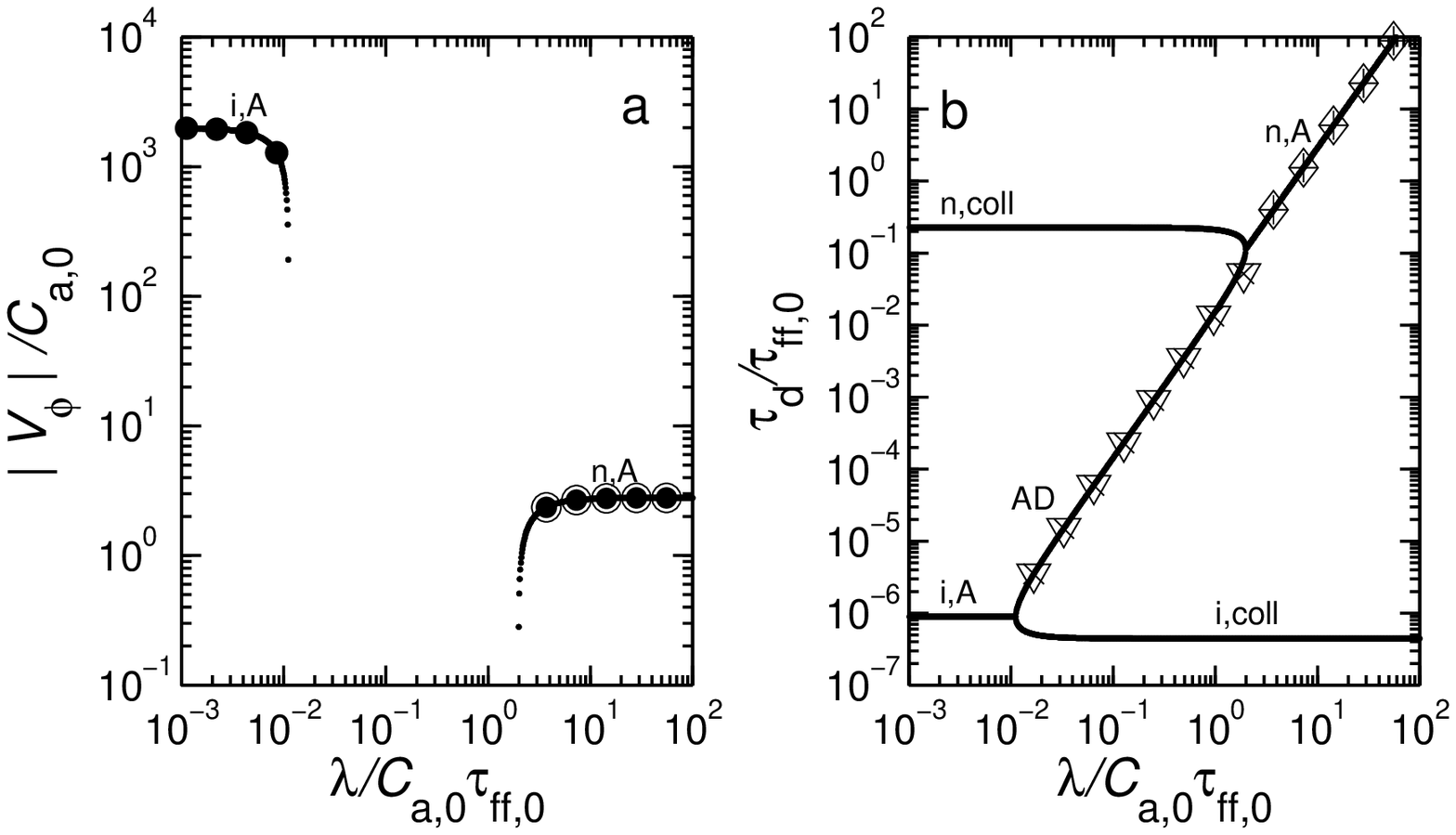}
\caption{{\it Eigenvalues of transverse modes as functions of wavelength, at an
angle of propagation $\theta=45^\circ$ with respect to $\Bveco$}. All
normalisations and labels are as in Fig. 2 and Table 1. There are three different 
modes. ({\it a}) Absolute value of phase velocity, $|\vphd|$. Phase speeds (large
black circles) obtained from eq. (\ref{ionAlfeq}) are shown, in addition
to those calculated from eq. (\ref{neutAlfeq}) (open circles with interior black 
circles). ({\it b}) Damping timescales $\tdampd$. Also depicted are values 
(downward-facing triangles with $\times$'s  inside) obtained from eq. 
(\ref{newdampeq}), and values (diamonds with interior crosses) obtained
from eq. (\ref{neutAlfdiffeq}).
}
\end{figure}
\begin{figure}
\includegraphics[width=180mm]{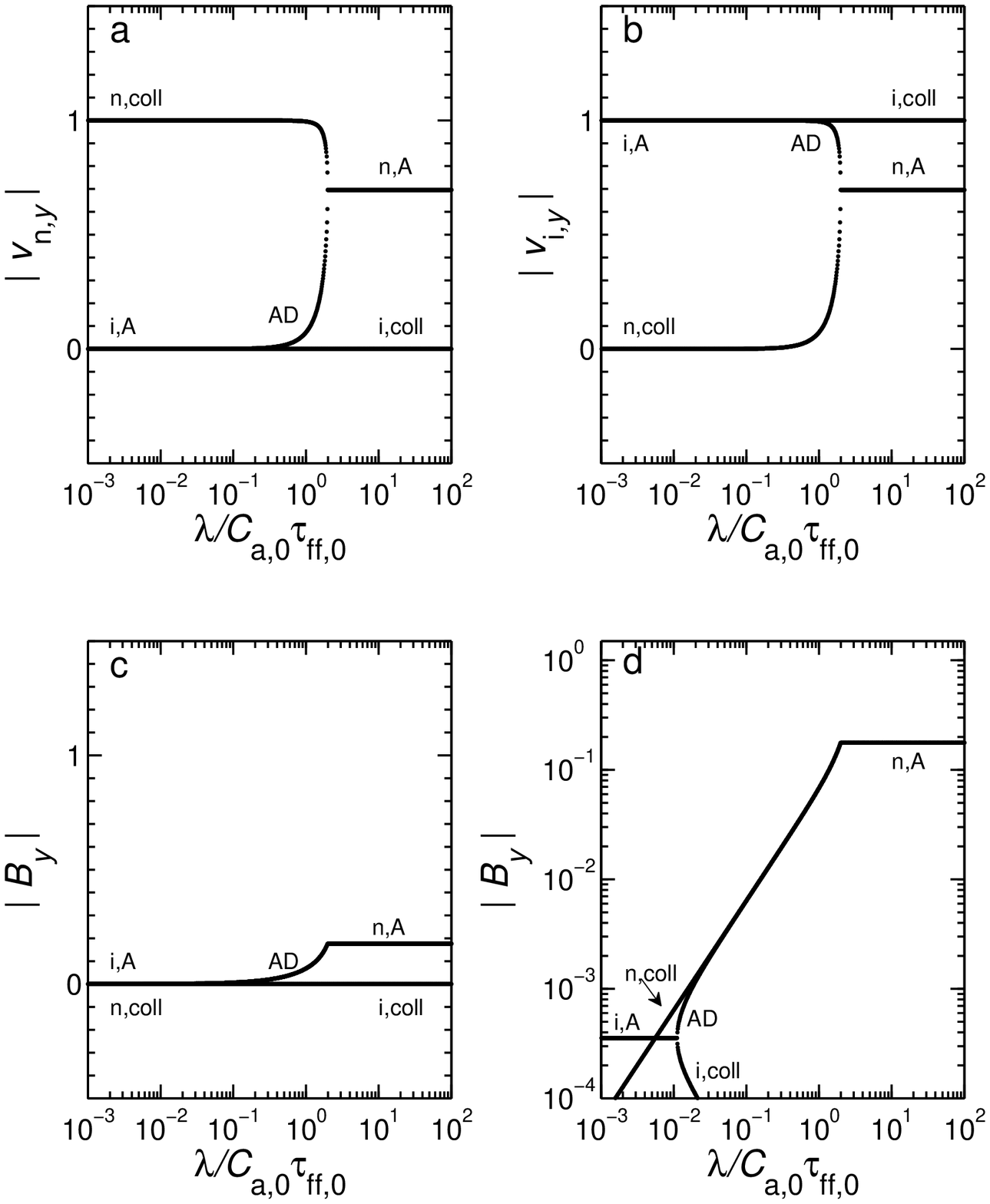}
\caption{{\it Magnitudes of eigenvectors of transverse modes for
$\theta=45^\circ$ as functions of wavelength.} All labels and normalisations
are as in Fig. 3 and Table 1. As in Fig. 8, there are three modes.
({\it a}) Transverse ($y$-) component of the neutral velocity, $|\vny|$. ({\it
b}) Transverse ($y$-) component of the ion velocity, $|\viy|$. ({\it c, d})
Transverse ($y$-) component of the magnetic field, $|\By|$, shown as in 
Figs. $5c$ and $5d$.}
\end{figure}
\begin{figure}
\includegraphics[width=180mm]{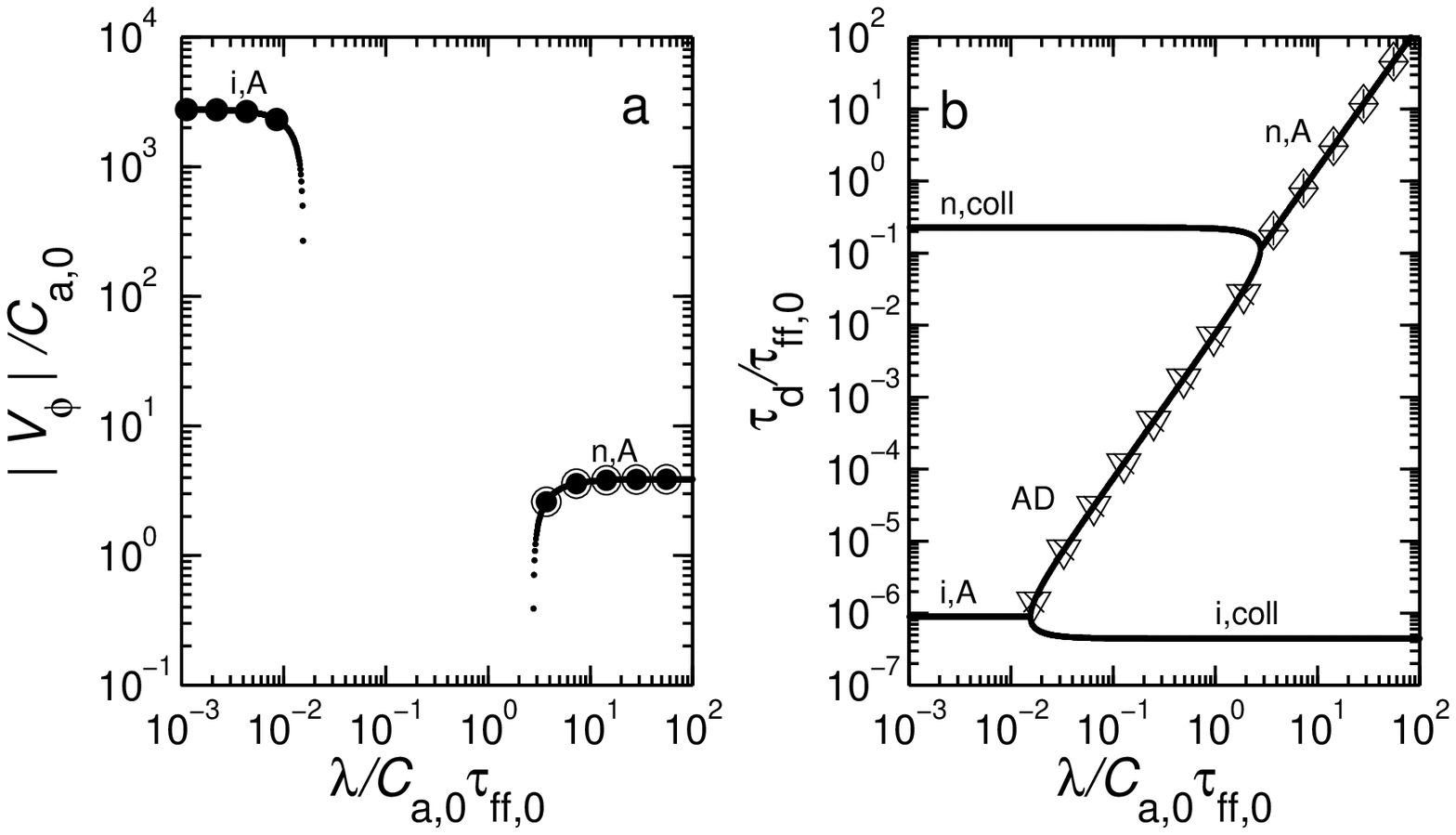}
\caption{{\it Eigenvalues of transverse modes as functions of wavelength, at an
angle of propagation $\theta=10^\circ$ with respect to $\Bveco$}. All
normalisations, labels, and symbols are as in Fig. 8 and Table 1. There are three 
different modes. ({\it a}) Absolute value of phase velocity, $|\vphd|$. 
({\it b}) Damping timescales $\tdampd$.}
\end{figure}
\begin{figure}
\includegraphics[width=180mm]{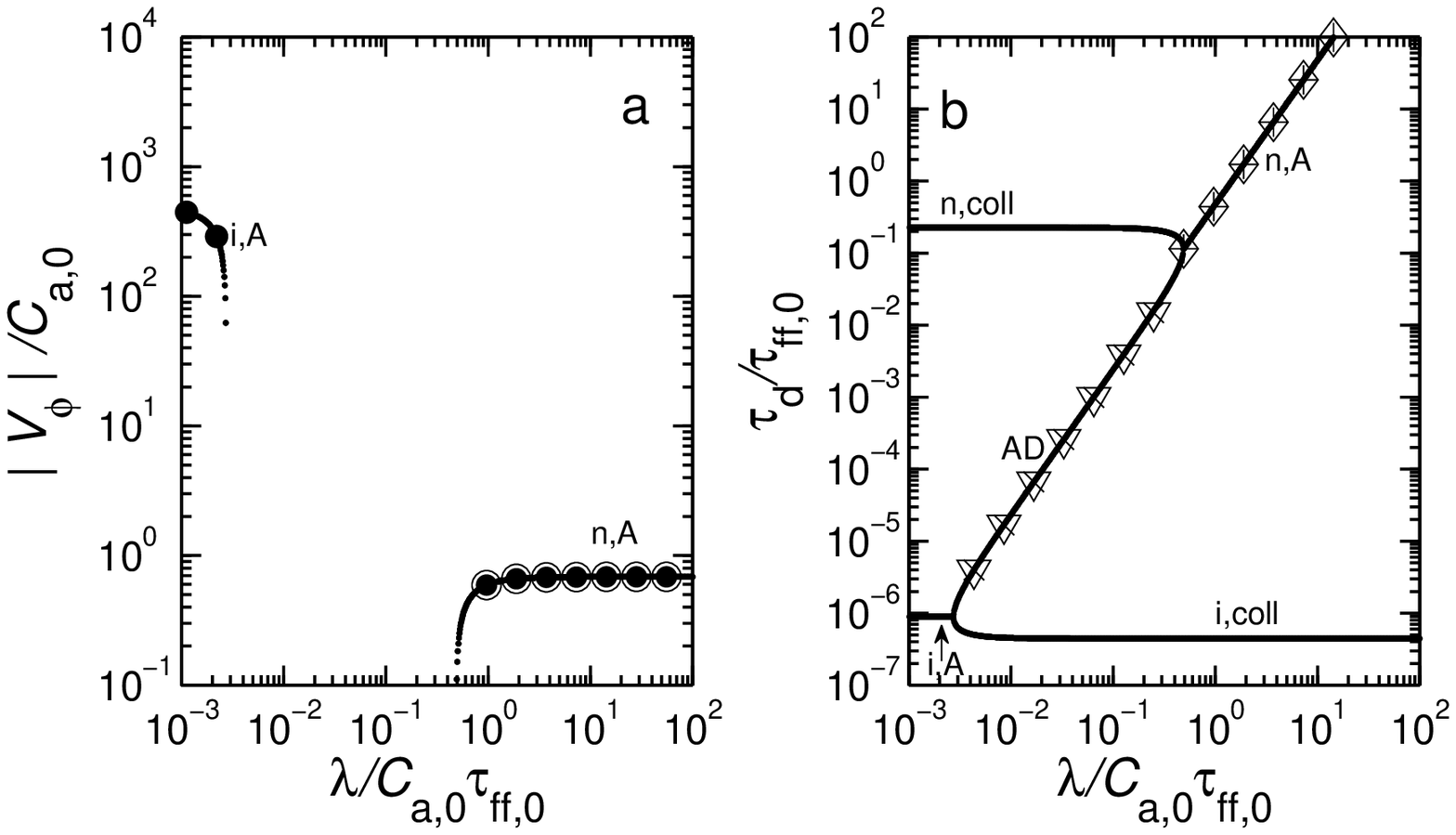}
\caption{{\it Eigenvalues of transverse modes as functions of wavelength, at an
angle of propagation $\theta=80^\circ$ with respect to $\Bveco$}. All
normalisations, labels, and symbols are as in Fig. 8 and Table 1. There are three 
different modes. ({\it a}) Absolute value of phase velocity, $|\vphd|$. 
({\it b}) Damping timescales $\tdampd$.}
\end{figure}

The dispersion relation for ion {\Alf} waves, obtained from equations
(\ref{yneutforceq}) - (\ref{yinducteq}), has solutions given by
equation (\ref{ionalfveneq}), with $\vaiod \cos \theta$ replacing $\vaiod$ in
that expression. Thus, the two ion {\Alf} waves propagate with 
\begin{eqnarray}
\label{ionAlfeq}
\vph = \pm \vaiod \cos \theta \left[1 - \left(\frac{\lambdad}{\lami \cos
\theta}\right)^2\right]^{1/2} 
\end{eqnarray}
(see Fig. $8a$, curve labeled ``i,A"). For $\lambdad/$($\lami \cos \theta$) 
$\ll 1$, $\vph \simeq \pm \vaiod \cos \theta=
\pm 1.99 \times 10^3$ for the typical model cloud. The $\pm$ sign denotes
degeneracy with respect to the direction of propagation. The damping timescale
for these waves is $2 \tinod$ (see Fig. $8b$, curve labeled ``i,A").

Collisions with the neutrals cut off the propagation of these waves for
$\lambdad \geq \lami \cos \theta = 1.11 \times 10^{-2}$ (see Fig $8a$). For
wavelengths greater than this value, each mode bifurcates into an ion
collisional-decay mode and an (ion) ambipolar-diffusion mode. The damping
timescale of the former mode is given by equation (\ref{negrooteq}) and is
shown in Figure $8b$ (curve labeled ``i,A"); it is identical with $\tdamp$
of the normal {\Alf} waves propagating along $\Bveco$, except for the fact
that $\lami \cos \theta$ replaces $\lami$ in that expression (see 
discussion following eq. [\ref{ionalfveneq}]). For very large
$\lambdad$, $\tdamp = \tinod$ (see Fig. $8b$, curve labeled ``i,coll"). For
the ambipolar-diffusion mode (Fig. $8b$, curve labeled ``AD"), the damping
timescale is the same as that given by equation (\ref{posrooteq}), but again
$\lami \cos \theta$ replaces $\lami$; hence, for long wavelengths, it follows
from equation (\ref{iADdiffeq}) that 
\begin{eqnarray}
\label{newdampeq}
\tdamp \simeq \frac{\lambdad^2}{4 \pi^2\diff \cos^2 \theta}. 
\end{eqnarray}

At small wavelengths the third mode is a low-frequency neutral 
collisional-decay mode, with $\viyd \simeq 0$ (see Fig. $9b$, curve labeled 
``n,coll"). The purely imaginary frequency of this mode is given by 
equation (\ref{neutdampeq}), $\omegad = -i/\tniod$; hence the
damping timescale is $\tdamp=\tniod$ ($=0.226$ for the typical model cloud;
see Fig. $8b$, curve labeled ``n,coll"). At longer wavelengths, the motions of
the neutrals and the ions become better coupled, and the neutral collisional-decay 
mode merges with the ion ambipolar-diffusion mode (see Fig. $8b$), 
as discussed in \S\S~3.1.2 and \S~3.2.1. For longer wavelengths, {\Alf} waves 
in the neutrals can be sustained, as seen in Figure $8a$. The frequencies 
of these modes are given by equations (\ref{neutalfveneqa}) and 
(\ref{neutalfveneqb}), but with $\vaiod \cos \theta$ and 
$\vanod \cos \theta$ replacing $\vaiod$ and $\vanod$, respectively. The
lower cutoff wavelength, below which these waves cannot propagate, is $\lamn
\cos \theta = 1.98$. They have phase velocity
\begin{eqnarray}
\label{neutAlfeq}
\vph = \pm \vanod \cos \theta \left[ 1 -
\left(\frac{\lamn \cos \theta}{\lambdad}\right)^2\right]^{1/2}  ,
\end{eqnarray}
and damping timescale 
\begin{eqnarray}
\label{neutAlfdiffeq}
\tdamp = \frac{2 \lambdad^2}{4 \pi^2 \diff \cos^2 \theta}  .
\end{eqnarray}
As $\lambdad \rightarrow \infty$, $|\vph| \rightarrow \vanod
\cos \theta =2.79$.

The components of the eigenvectors of these modes displayed in Figures $9a$ 
- $9d$ are qualitatively similar to those shown in Figures $5a$ - $5d$ (for 
transverse modes propagating along $\Bveco$), but with (small) quantitative 
decrease in the values of the critical wavelengths. We therefore do not discuss 
them further for economy of space.
 
In Figures $10a$ and $10b$ we display $|\vph|$ and $\tdamp$,  respectively,
for the three transverse modes propagating at an angle $\theta=10^\circ$ with
respect to the magnetic field. The qualitative solutions of the dispersion
relations for these modes are the same as for those described above for
propagation at an angle $\theta=45^\circ$. Quantitatively, the differences in
Figures $10a$ - $10b$ and $8a$ - $8b$ stem from the different values of the
numerical factors $\cos \theta$ and $\sin \theta$. Similarly, the magnitude of
the phase velocities and the damping timescales  for the transverse modes
propagating at an angle $\theta=80^\circ$ with respect to $\Bveco$ are shown
in Figures $11a$ and $11b$. They again differ from those for $\theta=45^\circ$
because of the factors $\cos \theta$ and $\sin \theta$. Although qualitatively the
modes at $\theta=10^\circ$ and $\theta=80^\circ$ are the same, comparison of
Figures 10 and 11 reveals that the quantitative differences in the magnitude
of the phase velocities and the cutoff wavelengths are substantial. Thus 
{\em the critical wavelengths that determine which modes can or cannot propagate 
in molecular clouds are expected to be very different depending on their 
direction of propagation with respect to the mean magnetic field} $\Bveco$.

\subsubsection{Modes with Motions in the $(x,z)$-Plane}

\begin{figure}
\includegraphics[width=180mm]{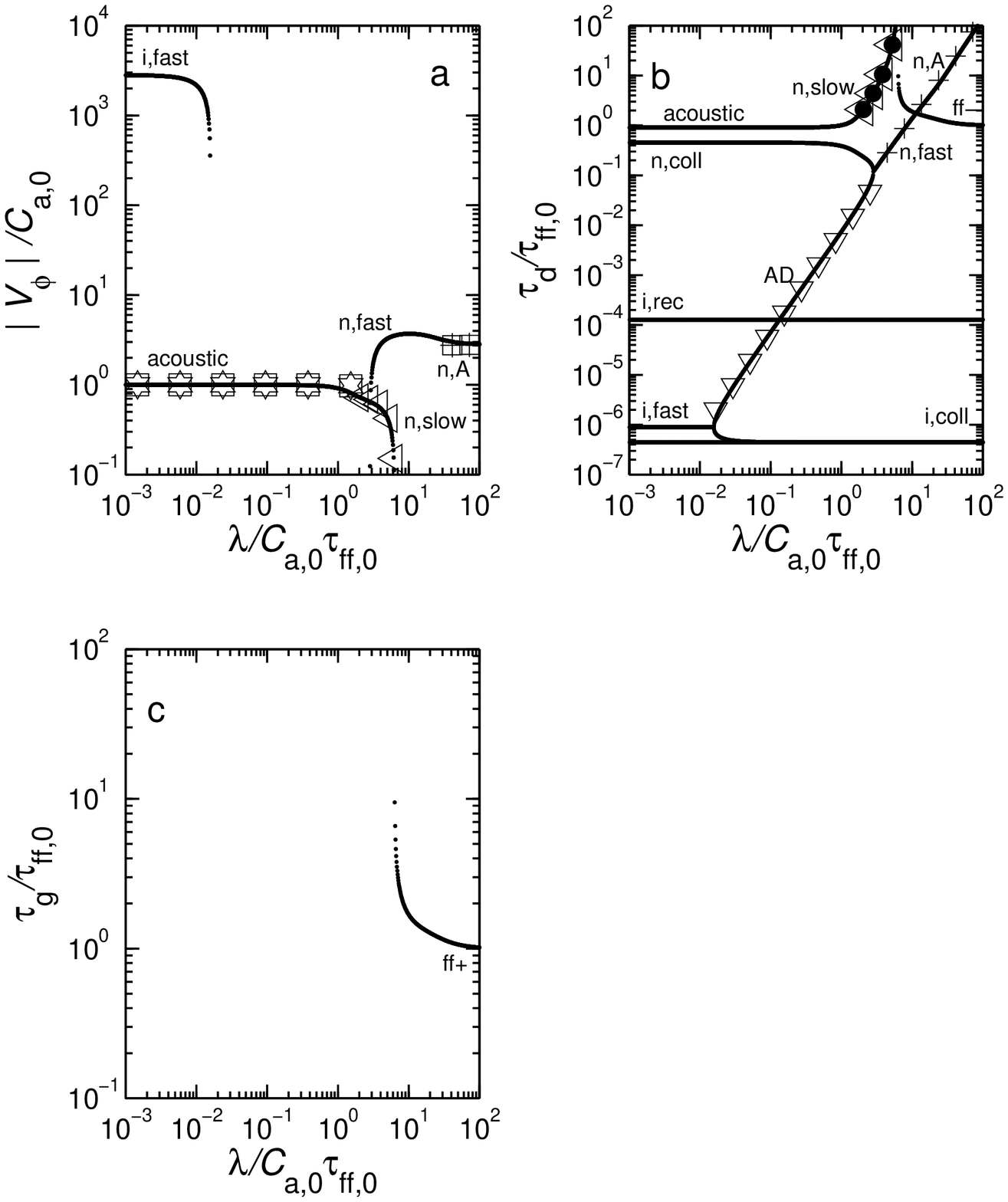}
\caption{{\it Eigenvalues of longitudinal modes as functions of wavelength, at
an angle of propagation $\theta=45^\circ$ with respect to $\Bveco$}. All
normalisations and labels are as in Fig. 2 and Table 1. There are seven different 
modes. ({\it a}) Absolute value of phase velocity, $|\vphd|$. Boxes with 
interior six-pointed stars denote phase velocity values derived from eq. 
(\ref{acousticthetaeq}). Phase speeds calculated by using eq. (\ref{jeanmodetheteq}) 
are shown as left-facing triangles, and phase speeds obtained from eq. 
(\ref{xzalfspeedeq}) are shown as boxes with interior crosses. 
({\it b}) Damping timescales $\tdampd$. Also displayed are values obtained from eq. 
(\ref{iADdiffeq}) (downward-facing triangles) and from eq. (\ref{jeandifftheteq}) 
(left-facing triangles with interior black circles). Values obtained  
from eq. (\ref{nADdiffeq}) are also displayed as crosses. ({\it c}) Growth timescale 
$\tgrowthd$.}
\end{figure}
Figures $12a$, $12b$, and $12c$ exhibit, respectively, the magnitude of the
phase velocity $\vph$, the damping timescale $\tdamp$, and the growth time
$\tgrowth$ as functions of $\lambdad$ for the specific case of propagation at
$\theta=45^\circ$ with respect to $\Bveco$ for modes with motions in the 
$(x,z)$-plane. Eigenvectors are displayed in
Figures $13a$ - $13g$. There are 7 modes in all displayed in these figures.

As in the preceding sections, one of the ion modes is a collisional-decay
mode (see Figs. $12$ and $13$, curves labeled ``i,coll"), with the ions
streaming through a fixed background of neutral particles
($\vnxd=\vnzd=0$); for this case $\Bzd=0$, and the ions move with
$\vixd=\vizd$. The frequency is given by equation (\ref{iondampeq}); hence,
the damping timescale is $\tdamp=\tinod$ ($=4.45 \times 10^{-7}$ for the
typical model cloud). This is the horizontal line labeled by ``i,coll" in
Figure $12b$. 

There also exist two high-frequency ion wave modes. These waves are {\em ion
fast modes}. For these modes, $\vixd/\vizd =-\sin\!\theta/\cos\!\theta = -1$
at $\theta=45^\circ$ (see Figs. $13c$ and $13d$, curves labeled ``i,fast";
in these Figures the ``i,fast" curves coincide with the ``i,coll" curves).
Since the ion {\Alf} speed and the magnetosonic speed
(in this typical case) are essentially the same, the dispersion relation
describing them is the same as equation (\ref{ionalfveneq}); it is again the 
case that the waves are cut off at $\lambdad \geq \lami$ ($= 1.57 \times
10^{-2}$ ; see Fig. $12a$). For wavelengths $> \lami$, each ion mode
bifurcates into an ion collisional-decay mode and an ion ambipolar-diffusion
mode (see Fig. $12b$). The decay timescales for these two modes are the same
as those previously discussed for the cases $\theta=0^\circ$ and $90^\circ$
(see \S~3.1.2 and \S~3.2.1).

The fourth mode is the ion dissociative-decay mode, discussed previously in
\S\S~3.1.1 and 3.2.1. In this mode, nonpropagating density perturbations in
the ions decay by dissociative recombinations of molecular ions and electrons.
The decay time is $\tdamp = \left(2 \rhoiod\alphdrd \right)^{-1}$ ($=1.28
\times 10^{-4}$ for the typical model cloud); this is the horizontal line 
labeled by ``i,rec" in Figure $12b$.

At small wavelengths there are two low-frequency acoustic-wave modes in the
neutrals and one neutral collisional-decay mode. The dispersion relation for
the acoustic modes, which are predominantly polarised in the $x$-direction, is
most easily found by first finding $\vixd$ in terms of $\vnxd$ and $\omegad$.
In the limit $|\omegad| \ll 1/\tinod$ and $\omegad \Bzd \simeq 0$, one finds
from equations (\ref{zneutforceq}), (\ref{zionforceq}), (\ref{zinducteq}),
and (\ref{xionforceq}) that
\begin{eqnarray}
\label{vixdeq}
\vixd \simeq \vnxd \cos^2 \theta \left[ \frac{\omegad +
\left(i/\tniod\right)}{\omegad + \left(i \cos^2 \theta/\tniod \right)}\right]
.
\end{eqnarray}
For $|\omegad| \simgt 1/\tniod$, the term in brackets in equation
(\ref{vixdeq}) is essentially unity, and $\vixd \simeq \vnxd \cos^2\! \theta$.
The factor $\cos^2 \!\theta$ is a measure of the opposition presented to the
neutrals by the ions, which are attached to magnetic field lines.  If
$\theta=0^\circ$ the ions are effectively inertialess, and the neutrals sweep
them up easily, so that $\vixd=\vnxd$. However, if $\theta=90^\circ$, 
$\vixd \simeq 0$, i.e., the ions are held in place by the magnetic field as
the neutrals move through them. In this case, the ions present the stiffest
opposition to the neutral motion. For $|\omegad| \simgt 1/\tniod$, we 
insert equation (\ref{vixdeq}) in equation (\ref{xneutforceq}) to find the
solution
\begin{eqnarray}
\label{soundtheteq}
\omegad = 
\pm \kd \left\{1 - \frac{1}{\kd^2}\left[1 + \left(\frac{\sin^2 \theta}{2
\tniod}\right)^2 \right] \right\}^{1/2} - i \frac{\sin^2 \theta}{2 \tniod}  .
\end{eqnarray}
[Note that for $\theta =0^\circ$ or $90^\circ$, equation (\ref{soundtheteq})
reduces to equations (\ref{jeanmodeeq}) or (\ref{soundeq}), respectively.] For
\leteq
\begin{eqnarray}
\label{lamdampeq}
\lambdad \simlt \frac{\lamcs}{\Asnfac(\theta)}~~,
\end{eqnarray}
where
\begin{eqnarray}
\label{Asndefeq}
\Asnfac(\theta) \equiv \left[\frac{\sin^{4}\theta + (2\tniod)^{2}}{1 + (2\tniod)^{2}}\right]^{1/2}~~,
\end{eqnarray}
\beq
these modes are sound waves, with 
\begin{eqnarray}
\label{acousticthetaeq}
\vph = \pm \left[1 - \left(\frac{\lambdad \Asnfac(\theta)}{\lamcs}\right)^{2}\right]^{1/2} 
\end{eqnarray}
(see Fig. $12a$) and
$\tdamp =2 \tniod/\sin^2 \theta=0.904$ for $\theta = 45^\circ$ (see horizontal
line labeled ``acoustic" in Fig. $12b$). $\Asnfac(\theta)$ is the acoustic-wave
angular (or, stiffness) parameter: for $\theta = 90^{\circ}$ equations
(\ref{lamdampeq}) and (\ref{acousticthetaeq}) yield our earlier result, 
that sound waves propagate only for $\lambdad \leq \lamcs$, while for 
$\theta =0^{\circ}$ we recover our other earlier result that $\lamJ$ is the
upper cutoff wavelength for these waves. 

The sound-wave upper cutoff wavelength (\ref{lamdampeq}) is relevant
only if the waves do not transition into neutral slow modes
at larger wavelengths (see discussion below). It turns out that this
depends on the angle $\theta$ of propagation: the upper cutoff equation is
applicable only to sound waves propagating at $\theta > \thetamax$ (for
the typical model cloud, $\thetamax = 62.9^{\circ}$ --- see below). For 
sound waves propagating at such angles, there is a mode bifurcation at the
cutoff (\ref{lamdampeq}). The dispersion relation (\ref{soundtheteq}) 
reveals that at larger $\lambdad$ one of the modes becomes a {\em neutral 
collisional-decay mode}, with damping timescale 
$\tdamp \rightarrow \tniod/\sin^{2}\theta$, and the other mode becomes a 
{\em pressure-driven diffusion mode} with damping timescale
\begin{eqnarray}
\label{pressurethetaeq}
\tdamp = \frac{(\lambdad/\lamJ)^{2}\sin^{2}\theta}{\diffncs [1 - (\lambdad/\lamJ)^{2}]} 
\, \, \, \, \, \, \, \, \, \, {\rm for} \, \, \, \lambdad \, \leq \, \lamJ ~.
\end{eqnarray}
This diffusion timescale is the same as that of
equation (\ref{diffcseq}), except that it is multiplied by $\sin^{2}\theta$, 
which reflects the reduced effectiveness of neutral-ion collisions in slowing 
down the neutrals at angles $\theta < 90^{\circ}$ with respect to the magnetic 
field. At $\lambdad = \lamJ$ the timescale (\ref{pressurethetaeq})
becomes infinite. For $\lambdad > \lamJ$ there is a {\em neutral
ambipolar-diffusion--induced gravitational fragmentation mode} at the
angle $\theta$ ($> \thetamax$), having a growth time 
\begin{eqnarray}
\label{gravfragthetaeq}
\tgrowthd = \frac{\sin^{2}\theta/\tniod}{1 - \left(\lamJ/\lambdad\right)^{2}}
= \frac{\vffo \sin^{2}\theta}{1- \left(\lamJ/\lambdad\right)^{2}}~~. 
\end{eqnarray} 
%
\begin{figure}
\includegraphics[width=180mm]{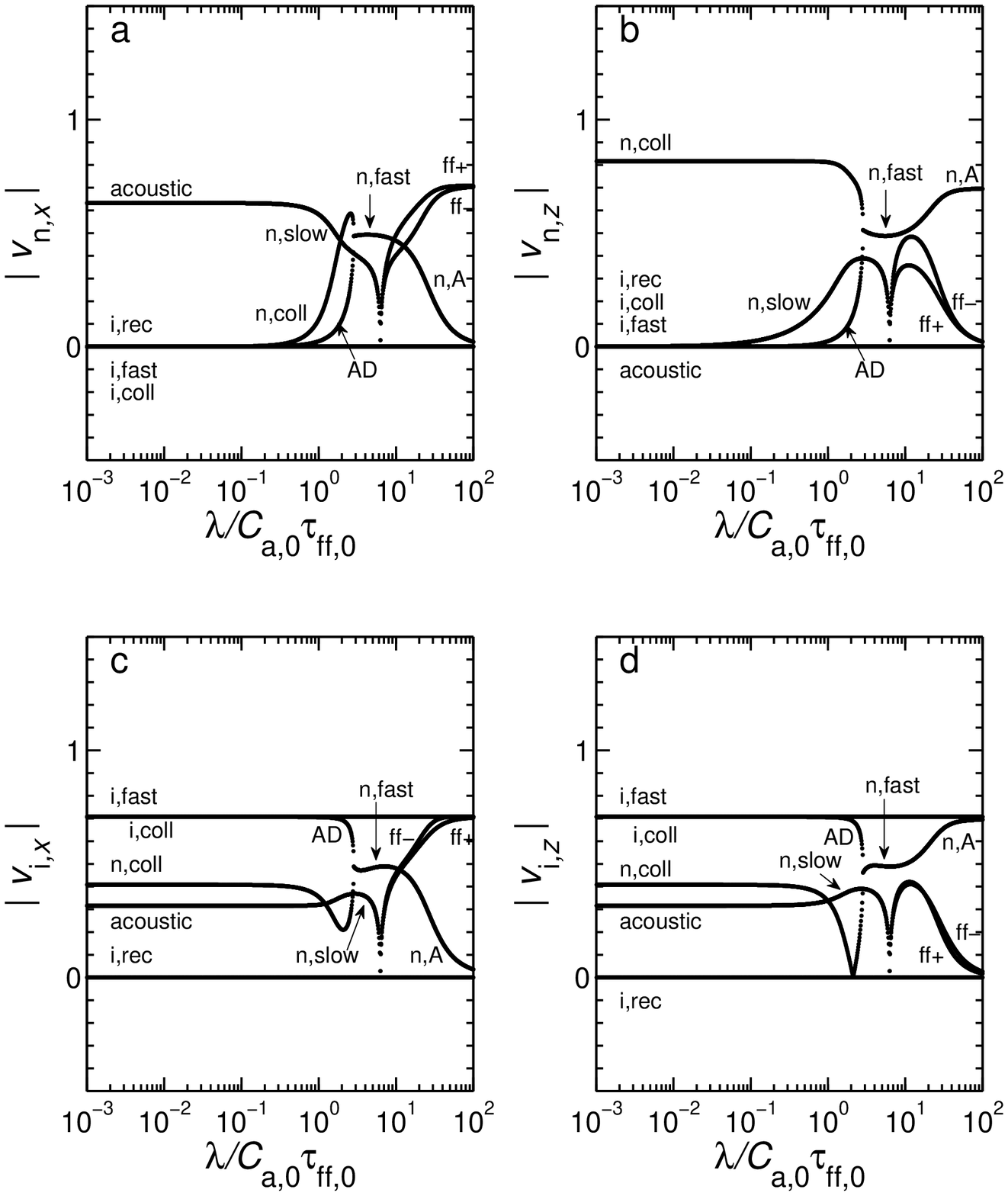}
\caption{{\it Magnitudes of eigenvectors of modes with motions in the (x,z)-plane
for $\theta=45^\circ$ as functions of wavelength.} All labels and
normalisations are the same as in Fig. 3 and Table 1. As in Fig. 12, there are 
seven modes. ({\it a}) Longitudinal ($x$-)component of the neutral velocity,
$|\vnx|$; ({\it b}) $|\vnz|$. ({\it c}) Longitudinal ($x$-)component of the ion
velocity, $|\vix|$; ({\it d}) $|\viz|$.} 
\end{figure}
\addtocounter{figure}{-1}
\begin{figure}
\includegraphics[width=180mm]{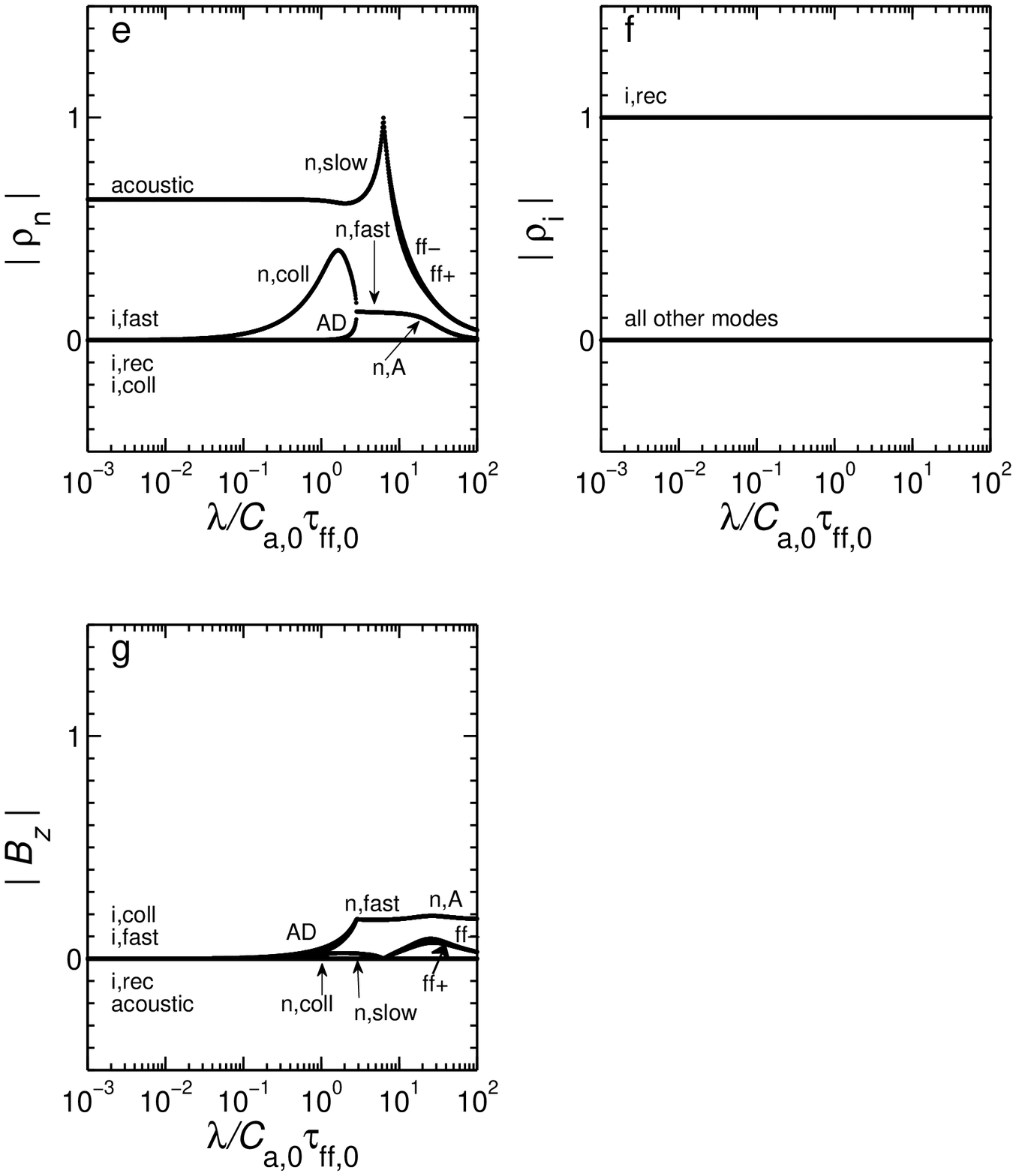}
\caption{{\it Cont.} ({\it e}) Neutral density $|\rhon|$. 
({\it f}) Ion density $|\rhoi|$. ({\it g}) Magnetic field component $|\Bz|$.}
\end{figure}
%
The fragmentation time (\ref{gravfragthetaeq}) is equal to the
growth time (\ref{tadtimescaleeq}) multiplied by $\sin^{2}\theta$,
again reflecting the reduced collisional resistance on the neutrals by 
magnetically-coupled ions when the angle of propagation is less than 
$90^{\circ}$ with respect to the field direction. For $\lambdad \gg \lamJ$,
equation (\ref{gravfragthetaeq}) yields 
$\tgrowth \rightarrow \sin^{2}\theta/\tniod = \vffo \sin^{2}\theta$.
However, similar to what occurs in the case for $\theta = 90^{\circ}$
(see \S~3.2.1 and Fig. $5c$), this limit will not be attained because
the approximation of stationary field lines breaks down when $\lambdad$
becomes $ \simgt \lamms$. When this happens, magnetic forces are
overwhelmed by self-gravitational forces and $\tgrowthd \rightarrow 1$ at 
these larger wavelengths.

From equation (\ref{soundtheteq}) one finds that collisional damping of the
motion of the neutrals by ions causes $|\vph|$ to become less than unity for
$\lambdad$ near the value given by the right-hand side of equation 
(\ref{lamdampeq}). However, for $0 < \theta < 90^\circ$, the sound waves 
are not always cut off at this wavelength. This is due to the fact that the
bracketed term in the expression for the $x$-component of the ion velocity
(eq. [\ref{vixdeq}]) is no longer essentially unity (because 
$|\omegad| \sim 1/\tniod$) at these wavelengths; thus, the dispersion 
relation (\ref{soundtheteq}) is no longer valid. Physically, this is a result of the
fact that the ions move readily with the neutrals in the $x$-direction at
these frequencies; this means that the waves suffer less damping, because the
frictional force on the neutrals is reduced when the ions and the neutrals
move together. (We note that, for $\theta < 90^\circ$, equation [\ref{vixdeq}]
yields $\vixd \simeq \vnxd$ in the limit $|\omegad| \ll 1/\tniod$.)

For $\lambdad$ sufficiently large, such that 
$|\omegad| \simlt \cos^{2}\theta/\tniod$, the frequency of the neutral waves is
\begin{eqnarray}
\label{jeantheteq}
\omegad \simeq
\pm \kd \cos \theta
\left\{\left(1 - \frac{1}{\kd^2}\right)
\left[1 - \left(\frac{\tniod \kd \sin\theta \tan\theta}{2}\right)^{2}
\left(1 - \frac{1}{\kd^2}\right)\right]\right\}^{1/2}
-\frac{i}{2}\tniod \left(\kd^2-1\right)\sin^{2}\theta  ~~.
\end{eqnarray}
Hence, for these conditions, the phase velocity is
\begin{eqnarray}
\label{jeanmodetheteq}
\vph = \pm \cos \theta \left\{\left(1 -
\frac{\lambdad^2}{\lamJ^2}\right)
\left[1 - \left(\frac{\lamcs\sin\theta\tan\theta}{4\lambdad}\right)^{2} 
\left(1 + 4 \tniod^{2}\right)
\left(1 - \frac{\lambdad^2}{\lamJ^2}\right)\right] \right\}^{1/2},
\end{eqnarray}
and
\begin{eqnarray}
\label{jeandifftheteq}
 \tdamp = \frac{2 (\lambdad/\lamJ)^{2}}{\diffncs
 \left[1-\left(\lambdad/\lamJ\right)^2\right]\sin^{2}\theta}.
\end{eqnarray}
These waves are {\em neutral slow modes}, modified by gravity, and cannot
propagate at wavelengths $\lambdad \geq \lamJ$ ($= 2 \pi$ for the typical
model cloud); 
at $\theta = 0^{\circ}$ the slow mode dispersion relation (\ref{jeantheteq})
is identical to the relation for the Jeans mode (eq. [\ref{jeanmodeeq}]).
From the low-frequency condition used to derive equation (\ref{jeantheteq})
it is found that slow modes exist only for wavelengths
\leteq
\begin{eqnarray}
\label{slowmodecutoffeq}
\lambdad \, \simgt \, \lamcs \Ssmfac(\theta)~,
\end{eqnarray}
where
\begin{eqnarray}
\label{Ssmdefeq}
\Ssmfac(\theta) \equiv \frac{\left[1 + (2 \tniod)^{2}\right]^{1/2}}{2 |\cos\theta|}~,
\end{eqnarray}
\beq
provided that
\begin{eqnarray}
\label{slowacousticeq}
\Asnfac(\theta) \Ssmfac(\theta) \, \simlt \, 1~.
\end{eqnarray}
The quantity $\Ssmfac(\theta)$ is the angular slow-mode factor.
The relation (\ref{slowacousticeq}) is derived from the requirement that
the slow modes arise from the acoustic wave modes. For this mode conversion
to occur, the minimum slow mode wavelength (\ref{slowmodecutoffeq}) must 
be less than or equal to the acoustic-wave upper cutoff wavelength 
(\ref{lamdampeq}); thus, the inequality (\ref{slowacousticeq}) follows.
This is equivalent to having $\theta \leq \thetamax$, where 
$\thetamax$ is defined by the condition 
\begin{eqnarray}
\label{slowangleeq}
\frac{\cos^{2}\thetamax}{\sin^{4}\thetamax + (2 \tniod)^{2}} =  \frac{1}{4}~~.
\end{eqnarray}
If $\theta \leq \thetamax$, slow modes emerge from the acoustic modes 
(without a bifurcation) and propagate for 
$\lamcs\Ssmfac(\theta) \simlt \lambdad \leq \lamJ$.
Otherwise, when $\theta > \thetamax$, the acoustic waves have an upper 
cutoff and mode bifurcation occurs at the wavelength
(\ref{lamdampeq}); in that case, there are no slow modes. 
\footnote{For clouds with perfect neutral-ion collisional coupling,
i.e., $\tniod = 0$, eq. (\ref{slowangleeq}) yields $\thetamax=65.5^{\circ}$.}

For the typical model cloud, with $\tniod = 0.226$, $\thetamax = 62.9^{\circ}$.  
At $\theta = 45^{\circ}$ the slow mode minimum wavelength
$\lamcs\Ssmfac(45^{\circ})=2.01$, and the transition from sound waves to slow 
modes can be seen in Figures $12a$ and $12b$ (curves labeled ``n,slow") to occur 
at this wavelength; Figures $13a$ and $13c$ show that, for these modes, 
$\vnxd \simeq \vixd$ at wavelengths greater than or equal to the transition 
wavelength. 
%
\begin{figure}
\includegraphics[width=180mm]{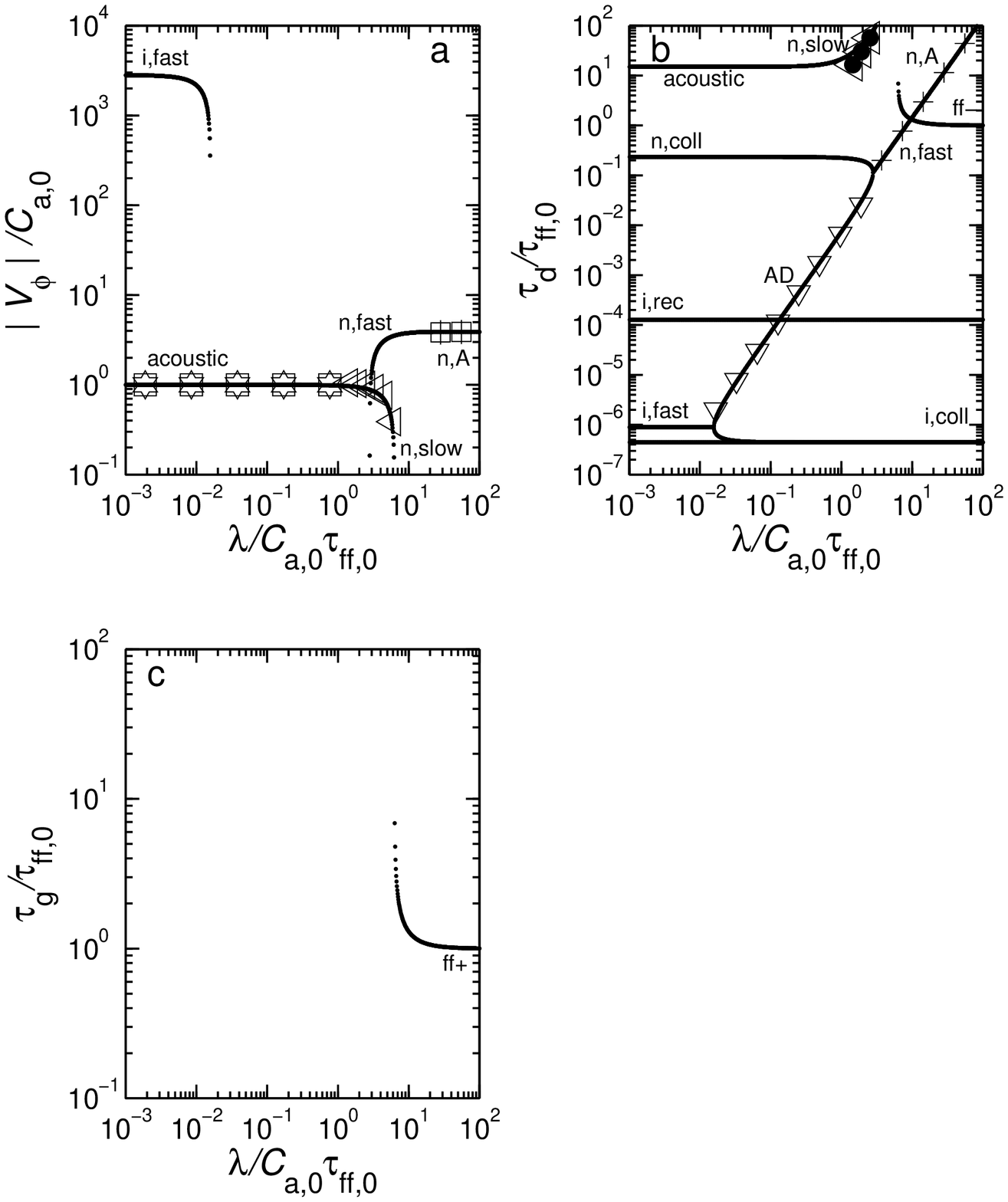}
\caption{{\it Eigenvalues of longitudinal modes as functions of wavelength, at
an angle of propagation $\theta=10^\circ$ with respect to $\Bveco$}. All
normalisations, labels, and symbols are as in Fig. 12 and Table 1. There are seven 
different modes. ({\it a}) Absolute value of phase velocity, $|\vphd|$. ({\it b}) 
Damping timescales $\tdampd$. ({\it c}) Growth timescale $\tgrowthd$.}
\end{figure}
\begin{figure}
\includegraphics[width=180mm]{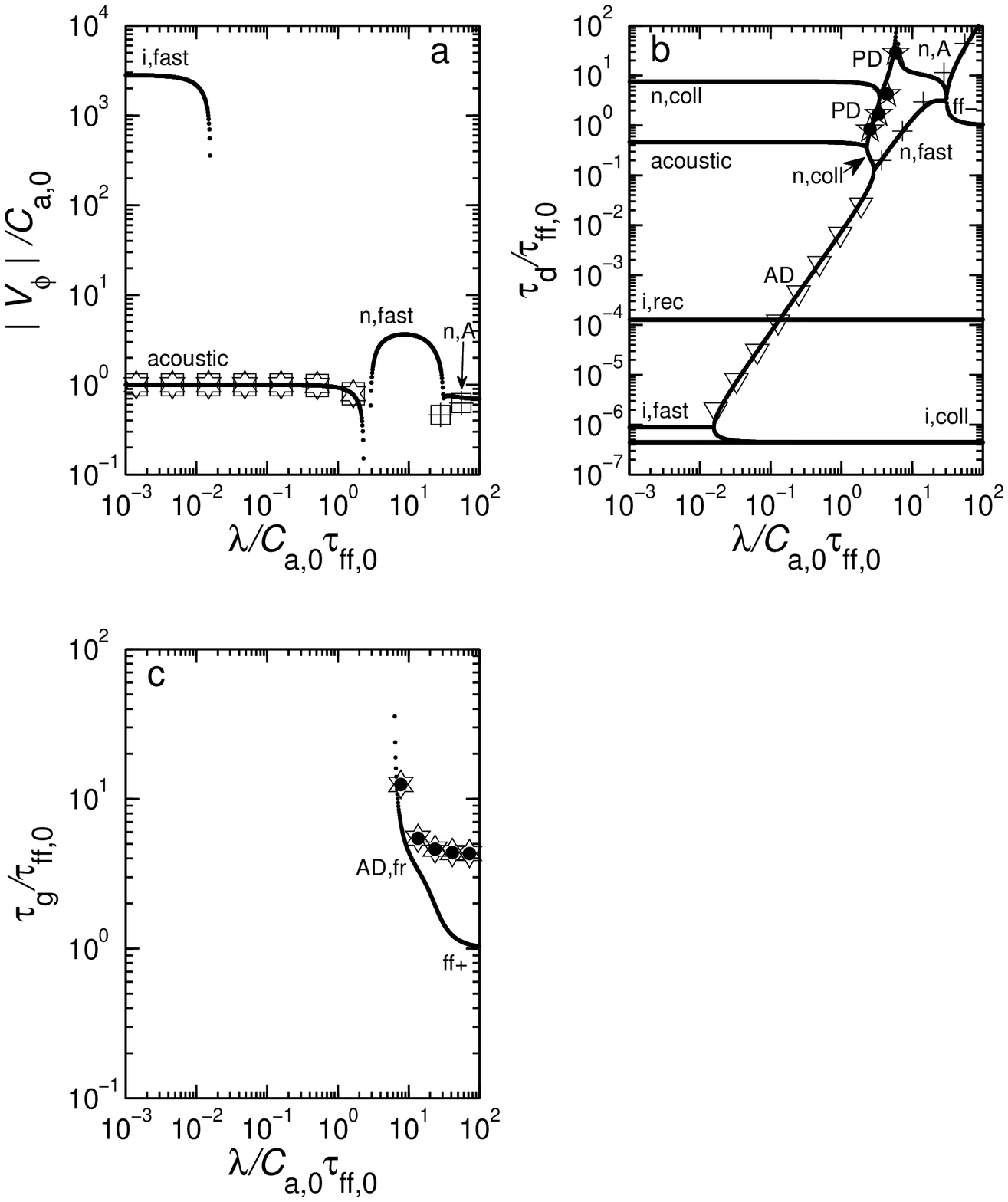}
\caption{{\it Eigenvalues of longitudinal modes as functions of wavelength, at
an angle of propagation $\theta=80^\circ$ with respect to $\Bveco$}. All
normalisations, labels, and symbols are as in Fig. 12 and Table 1. There are seven 
different modes. ({\it a}) Absolute value of phase velocity, $|\vphd|$. ({\it b}) 
Damping timescales $\tdampd$. Five-pointed stars with interior black circles 
represent values calculated by using eq. (\ref{pressurethetaeq}).
({\it c}) Growth timescale $\tgrowthd$. Values obtained from eq. 
(\ref{gravfragthetaeq}) are depicted as six-pointed stars with interior black circles.}
\end{figure}

Note that, as $\lambdad \rightarrow \lamJ$ from below, equation
(\ref{jeandifftheteq}) shows that the slow mode damping timescale 
$\tdamp \rightarrow \infty$ (see curve labeled by ``n,slow" in Fig. $12b$).
For $\lambdad > \lamJ$ there are again the two conjugate modes: the 
gravitational instability (or Jeans) mode
(see Fig. $12c$, curve labeled ``ff+") and the 
``cosmological" mode (see curve labeled ``ff$-$" in Fig. $12b$; although this 
curve ``crosses" the curve labeled ``n,fast" in Fig. $12b$, the two modes do 
not actually interact); as $\lambdad
\rightarrow \infty$, both the growth timescale of the unstable mode and the
damping timescale of the cosmological mode go to unity (i.e., in dimensional
form, $\tau_{\growth} = \tau_{\damp} = \tffo$), as seen in Figures $12b$ and
$12c$, respectively.

The other mode affecting the neutrals at short wavelengths is a {\it neutral
collisional-decay mode} (``n,coll"). The velocity of the neutrals at small 
$\lambdad$ is predominantly in the $z$-direction for this mode (see Figs. 
$13a$ and $13b$); the frequency is purely imaginary and given by
\begin{eqnarray}
\label{decaytheteq}
\omegad = - i \frac{\cos^2 \theta}{\tniod}.
\end{eqnarray}
Thus, $\tdamp = \tniod/\cos^2 \theta = 0.452$ (see Fig. $12b$, curve labeled
``n,coll"). For $\lambdad$
greater than the value given by the right-hand side of equation
(\ref{lamdampeq}), it is again the case that the motion of the ions and
magnetic field becomes better coupled to that of the neutrals. In this
wavelength regime $|\vnxd| \simeq |\vnzd| \simeq |\vixd| \simeq |\vizd|$ 
(see Figs. $13a$ - $13d$). This mode merges with the ion ambipolar-diffusion 
mode at $\lambdad=\lammsn$ (see Fig. $12b$), and, for $\lambdad > \lammsn$, 
{\em fast modes} are excited in the neutrals (curves labeled ``n,fast" in 
Figs. $12a, 12b$, and $13a$ - $13d$), degenerate with respect to the
direction of propagation. In these modes, the polarisation is given by
$\vnxd/\vnzd = -\sin\!\theta/\cos\!\theta$ ($=-1$ for $\theta=45^\circ$).
Hence, $\tilde{\vvec}_{\rm n} \cdot \Bveco =0$,  i.e., the fast modes are
polarised perpendicular to the magnetic field. The dispersion relation for
these modes is essentially the same as equation (\ref{nmagsoniceq}) because
of the fact that $\vanod^2 \gg 1$. They decay, as they propagate, on the 
ambipolar-diffusion timescale (see Fig. $12b$). For $\lambdad \geq \lamms$,
these waves tend to get suppressed by gravity.

For $\lambdad > \lamms$ thermal and magnetic restoring forces in the
$x$-direction are overwhelmed by gravitational forces, making $\vnxd \simeq
0$ (see Fig. $13a$). Hence the modes become essentially incompressible. Waves 
are still able to propagate for longer wavelengths, however, because of the 
{\em transverse} restoring magnetic tension force (i.e., $\Bzd \neq 0$; see 
Fig. $13g$). Solving equations (\ref{contineqd}) - (\ref{xionforceq}) and 
(\ref{zneutforceq}) - (\ref{zinducteq}) with the conditions $\vnxd=0=\rhond$ 
in the limit $|\omegad| \ll 1/\tinod$, we find
\begin{eqnarray}
\label{xzalfeq}
\omegad =
\pm \vanod \kd \cos \theta \left[1 - \left(\frac{\vanod \tniod \kd}
{2 \cos\!  \theta}\right)^2\right]^{1/2} -\frac{i}{2}\vanod^2 \tniod \kd^2 .
\end{eqnarray}
Hence, these modes are modified {\Alf} waves in the neutrals, with 
\begin{eqnarray}
\label{xzalfspeedeq}
\vph = \pm \vanod \cos \theta \left[ 1 -
\left(\frac{\lamn}{\lambdad \cos \theta} \right)^2 \right]^{1/2}
\end{eqnarray}
and $\tdamp = 2 \lambdad^2/4 \pi^2\diff$. In the limit $\lambdad \rightarrow
\infty$, $|\vph| \simeq \vanod \cos \theta = 2.79$, in agreement
with the long-wavelength behaviour of these modes shown in Figure $12a$ 
(curves labeled ``n,fast" and ``n,A"). {\it This is yet another example of a 
transition between wave modes without bifurcation.} 

Figures $14a$, $14b$, and $14c$ show $|\vph|$, $\tdamp$, and $\tgrowth$ as 
functions of $\lambdad$ for the seven different modes with motions in the
($x,z$)-plane propagating at an angle $\theta=10^\circ$ with respect to
$\Bveco$. Comparison with Figures $12a$ - $12c$ reveals that the
qualitative behaviour of the various modes as functions of wavelength is the
same as in the case of propagation at $\theta = 45^\circ$. The quantitative
differences stem from the numerical factors $\cos \theta$ and $\sin \theta$,
which become substantial for $\theta$ approaching $0^\circ$ or $90^\circ$. 
As Figure $14a$ shows clearly, {\it wave modes in the neutrals exist at all 
wavelengths and their decay times are very long} (see Fig. $14b$, curves labeled 
``acoustic", ``n,slow", and ``n,fast"). Figures $15a$, - $15c$ show the same 
quantities as Figures $14a$ - $14c$ but for propagation at $\theta=80^\circ$ 
with respect to the unperturbed magnetic field $\Bveco$.
There are no slow modes in Figure 15 (unlike the cases in Figs. 12 and 14) 
because $\theta > \thetamax$ for that angle of propagation in the typical
model. Instead, the sound waves are cut off at the maximum wavelength 
$\lambdad = \lamcs/\Asnfac(80^{\circ})= 2.66$, where there is a bifurcation.
At wavelengths greater than this maximum, the modes are a pressure-driven diffusion 
mode (``PD" curve) and a neutral collisional-decay mode (``n,coll"). There is
also an ambipolar diffusion-induced fragmentation mode seen in Figure $15c$
(``AD,fr" curve), which approaches the predicted limiting value (see eq. 
[\ref{gravfragthetaeq}]) of $\sin^{2}80^{\circ}/\tniod = 4.29$ at 
$\lambdad$ just below $\lamms$ (= 25.6 in the typical model cloud).

\vspace{-4ex}
\section{SUMMARY AND DISCUSSION}

We have obtained and analyzed the dispersion relations for MHD wave modes and 
instabilities for different directions of propagation with respect to the 
zeroth-order magnetic field $\Bveco$ in a two-fluid weakly ionised system, and 
we have applied the results to a typical interstellar molecular cloud. The system 
of equations has four dimensionless free parameters, $\tniod$, $\tinod$, $\vaiod$, 
and $\alphdrd$. They represent, respectively, the neutral-ion (momentum-exchange)
collision time and the ion-neutral collision time in units of the free-fall time 
of the zeroth-order state, the {\Alf} speed in the ions in units of the adiabatic 
speed of sound in the neutrals, and the dissociative recombination coefficient 
(see eq. [\ref{alphdrddefeq}]). (Because of ionisation equilibrium in the 
zeroth-order state, the dimensionless cosmic-ray ionisation rate $\zcrd$ is
expressible in terms of $\alphdrd$, $\tinod$ and $\tniod$.)

There are two distinct kinds of ambipolar diffusion, whose combined effect is 
unavoidable in typical molecular clouds and has crucial consequences on their 
evolution: 

\newcounter{tempc}
\begin{list}
{(\alph{tempc})}{\usecounter{tempc}}

\item{In the presence of hydromagnetic (HM) waves or turbulence, the tension 
of field lines (or the outward pressure due to compressed field lines) drives the motion 
of charged particles relative to the neutrals, with the tendency/consequence to straighten 
out the bent or tangled magnetic field lines (or to move compressed field lines apart, 
toward a more uniform configuration). The timescale of this process is proportional to the 
square of the wavelength of the HM waves (or the characteristic length of the field-line 
tangling, or the magnitude of the field gradient) -- see eqs. 
(\ref{nADdiffeq}) and (\ref{tADsummaryeq}). For lengthscales typical of molecular 
cloud cores ($\lesssim \, 0.1$ pc), it is much smaller than the free-fall time. This is
the {\em {magnetically-driven ambipolar diffusion}}. It is this kind of ambipolar
diffusion which is responsible for the observed large-scale ordering of 
polarisation vectors, indicating large-scale ordering of the magnetic field lines
in molecular clouds.} \\

\item{{\em Gravitationally-driven ambipolar diffusion} sets in with a short enough
timescale, but longer than the free-fall time, in the deep interiors of self-gravitating 
clouds, where the degree of ionisation is $x_{\rm i} \lesssim 10^{-7}$. Its onset can be 
spontaneous or initiated as a result of magnetically-driven ambipolar diffusion depriving 
a self-gravitating cloud of any support that most short-wavelength HM waves (or turbulence) 
may have provided against the cloud's self-gravity (Mouschovias 1987a). It (i) allows the 
clouds to fragment as neutral particles contract through almost stationary field lines (and 
the attached charged particles); and, consequently, (ii) increases the mass-to-flux ratio 
of the forming fragments (or cores). Once the critical mass-to-flux ratio (Mouschovias \& 
Spitzer 1976) of fragments is exceeded, dynamical contraction ensues, while the cloud 
envelopes remain magnetically supported, as found by numerical simulations starting 
with Fiedler \& Mouschovias (1993) and confirmed by numerous observations.} 
\end{list}

Hydromagnetic (HM) waves with phase velocity 
\begin{eqnarray}
|v_{\phi}| \simeq \vano = 0.96
\left(\frac{\Bo}{30~\mu\rm{G}}\right)\left(\frac{2 \times
10^3~\cc}{\nno}\right)^{1/2} \,\,\,\,\, \rm{km}~\rm{s}^{-1},
\end{eqnarray}
can propagate in all directions with respect to $\Bveco$, provided that
$\lambda \simgt \lamndim$, where
\begin{eqnarray}
\lamndim = \pi \vano \tnio = 0.22 \left(\frac{\Bo}{30~\mu\rm{G}}\right)
\left(\frac{2 \times 10^3~\cc}{\nno}\right)^{3/2}\left(\frac{2 \times
10^{-7}}{\xxio}\right) \,\,\,\,\, \rm{pc} .
\end{eqnarray}
The long-wavelength waves are long-lived; the {\em decay} time is the {\em magnetically-driven} 
ambipolar-diffusion timescale
\begin{eqnarray}
\label{tADsummaryeq}
\tau_{\damp} \simeq \frac{\lambda^2}{2 \pi^2 \vano^2 \tnio}
= 7.5 \times 10^5 
\left(\frac{\lambda}{1~\rm{pc}}\right)^2 \left(\frac{30~\mu{\rm G}}{\Bo}\right)^2 
\left(\frac{\nno}{2 \times 10^3~{\cc}}\right)^2 
\left(\frac{\xxio}{2 \times 10^{-7}}\right) \,\,\,\,\, \rm{yr ,}  
\end{eqnarray}
which is to be distinguished from the {\em growth} time of {\em gravitationally-driven} 
ambipolar-diffusion, relevant to fragmentation of molecular clouds into self-gravitating 
cores; namely, 
\begin{eqnarray}
\label{tADfrsummaryeq}
\tau_{\rm AD,fr} = \frac{{\tffo}^2}{\tau_{\rm ni,0}} \simeq 
1.1 \times 10^6 \, \left(\frac{\xxio}{10^{-7}}\right) \,\,\,\,\, \rm{yr} .
\end{eqnarray}
The (one-dimensional) free-fall timescale at the density $\nno = 2 \times 10^3
\, \cc$ as $\lambda \rightarrow \infty$ is $\tffo = (4 \pi G \rhono)^{-1/2} = 3.9
\times 10^5$ yr.
The nonlinear counterparts of these modes have been shown to explain quantitatively 
the observed highly supersonic but sub{\Alf}ic linewidths in molecular clouds, their 
cores, and even in OH and H$_{2}$O masers in which the strength of the magnetic field 
has been measured (Mouschovias \& Psaltis 1995; Mouschovias {\it et al.} 2006).

Most HM waves with $\lambda < \lamndim$ cannot propagate in the neutrals because 
they are damped rapidly by ambipolar diffusion. This means that there cannot be any
contribution from this wavelength regime to the spectrum of  HM
``turbulence" in molecular clouds (Mouschovias \& Psaltis 2011, in preparation),
which may provide clouds with a source of nonthermal pressure. This led
Mouschovias (1987a) to argue that the decay of HM waves  by ambipolar 
diffusion on wavelength scales $\simlt 0.1-0.3~\rm{pc}$ can initiate the
formation of protostellar cores in otherwise magnetically supported clouds
(see also Mouschovias 1991a). Damping of short-wavelength HM waves by ambipolar 
diffusion has also been proposed (Mouschovias 1987a, \S~2.2.5) as the cause of 
the observed narrowing and thermalization of linewidths with increasing column 
density, as observed, for example, by Baudry et al. (1981) in the cloud TMC 2.
Depending on the angle of propagation with respect to the unperturbed magnetic 
field $\Bveco$, however, we find that certain long-lived modes in the neutrals
exist at all wavelengths, while ion modes usually damp very rapidly even at 
short wavelengths. This may explain the observations by Li \& Houde (2008) in 
the M17 molecular cloud, which show neutral motions at small lengthscales but 
much smaller ion motions on the same scales. 

Gravitational instability is found to set in at $\lambda=\lambda_{\rm J,th}$ 
(the thermal Jeans wavelength) for all $\theta$, in agreement with 
Chandrasekhar \& Fermi (1953). For $\theta \simeq 90^\circ$, though, the 
timescale for the instability is $\simeq \vffo \tffo = \tffo^2/\tnio$, which
is the gravitationally-driven ambipolar-diffusion timescale. This is in 
agreement with the result for the ambipolar-diffusion--initiated 
formation and contraction of protostellar fragments (or cores) obtained from 
{\em non}linear calculations analytically by Mouschovias (1979, 1989) and Mouschovias 
\& Paleologou (1986), and numerically by Fiedler \& Mouschovias (1993) for a 
two-fluid system (neutral molecules and plasma), Ciolek \& Mouschovias (1994) 
for a four-fluid system (i.e., neutral molecules, plasma, negatively-charged, 
and neutral grains) , and Basu \& Mouschovias (1994, 1995a, b) for a two-fluid 
system including rotation and magnetic braking. It is also in agreement with 
the results of simulations by Tassis \& Mouschovias (2007a, b, c) and Kunz \& 
Mouschovias (2009, 2010) for a six-fluid system (neutral molecules, atomic and 
molecular ions, electrons, negatively-charged, positively-charged, and neutral 
grains).

Another recent success of the linear theory (applied to model molecular clouds 
flattened along the magnetic field) in predicting inherently nonlinear phenomena
was demonstrated by Kunz \& Mouschovias (2010). They obtained analytically the 
Core Mass Function (CMF) resulting from gravitationally-driven 
ambipolar-diffusion--induced fragmentation of molecular clouds. The predicted CMF 
is in excellent agreement with observations of more than 300 cores in Orion (Nutter 
\& Ward-Thompson 2007), not only at the high-mass end (for which many alternative 
models can obtain such agreement), but also at the turnover mass and low-mass end.

%
\begin{table}
\caption{\sc Critical Wavelengths}
\label{secondtable}
\begin{tabular}{llcl} \hline\hline
\mbox{\hspace{20em}}& \mbox{Dimensionless Expression \hspace{2em}}&
Model Value${}^{\dagger}$ \hspace{7em}&
Dimensional Expression
\\
\hline
ion {\Alf}-wave upper cutoff & $\lami = 4 \pi \vaiod \tinod$ & 0.0157 & 
$\lamidim
= 4 \pi \vaio \tino$ \\
ion magnetosonic-wave upper cutoff & $\lami$ & 0.0157 & $\lamidim$ \\
acoustic-wave upper cutoff &  
$\lamcs = 4 \pi \tniod/[1 + (2\tniod)^2]^{1/2}$ & 2.59 &
$\lamcsdim = 4 \pi \czero \tnio/[1 + (2\tniod)^2]^{1/2}$ \\
neutral {\Alf}-wave lower cutoff & $\lamn = \pi \vanod \tniod$ & 2.80 &
$\lamndim = \pi \vano \tnio$ \\
neutral magnetosonic-wave lower cutoff & $\lammsn =
(\vanod/\vmsod)\lamn$ &  2.72 &
$\lammsndim = (\vano/\vmso)\lamndim$ \\
thermal Jeans &  $\lamJ =  2 \pi $ & $2\pi$ & $\lamJdim =  2 \pi \czero \tffo$ \\
magnetic Jeans & $\lamms = 2 \pi \vmsod$ & 25.6 & $\lammsdim = 2 \pi \vmso \tffo$ \\
\hline
\hline
\end{tabular}
${}^{\dagger}$To convert to parsecs, multiply by the unit of length for the typical 
model cloud, $C_{\rm{a,0}} \tau_{\rm{ff,0}} = 9.72 \times 10^{-2}$ pc.
%
\end{table}
%
\begin{table}
\caption{\sc Modes and Wavelength Ranges in which They Exist}
\label{secondtable}
\begin{tabular}{lc} \hline\hline
\mbox{\hspace{30em}}&  Wavelength Range \\
\hline
{\sl Propagation Parallel to $\Bveco$ ($\theta = 0^{\circ}$)} &  \\
\hline
ion recombination mode  & All $\lambda$ \\
ion collisional-decay mode & All $\lambda$ \\
ion {\Alf} waves  & $\lambda \leq \lamidim$ \\
neutral collisional-decay mode & $\lambda \leq \lamndim$ \\
acoustic waves  & $\lambda \leq \lamJdim$ \\
ambipolar-diffusion mode & $\lamidim \leq \lambda \leq \lamndim$ \\
gravitational (Jeans) instability mode & $\lambda > \lamJdim$ \\
conjugate (``cosmological") Jeans mode & $\lambda > \lamJdim$ \\
neutral {\Alf} waves & $\lambda > \lamndim$ \\
\hline
{\sl Propagation Perpendicular to $\Bveco$ ($\theta = 90^{\circ}$)} & \\
\hline
ion recombination mode  & All $\lambda$ \\
transverse ion-neutral comoving mode & All $\lambda$ \\
transverse ion-neutral counterstreaming mode & All $\lambda$ \\
ion magnetosonic waves & $\lambda \leq \lamidim$ \\
ion collisional-decay mode & $\lambda > \lamidim$ \\
ambipolar-diffusion mode & $\lamidim \leq \lambda \leq \lammsndim$,
 ~~and ~~$\lambda > \lammsdim$ \\
acoustic waves & $\lambda \leq \lamcsdim$ \\
pressure-driven diffusion mode & $\lamcsdim < \lambda \leq \lamJdim$ \\ 
neutral ambipolar-diffusion--induced gravitational fragmentation mode & 
$\lamJdim < \lambda \leq \lammsdim$ \\
neutral magnetosonic waves & $\lammsndim \leq \lambda \leq \lammsdim$ \\
gravitational (magnetic Jeans) instability mode & $\lambda > \lammsdim$ \\
conjugate (``cosmological") magnetic Jeans mode & $\lambda > \lammsdim$ \\
\hline
{\sl Propagation at intermediate angles with respect to $\Bveco$
($0^{\circ} < \theta < 90^{\circ}$)} \\
\hline
ion recombination mode  & All $\lambda$ \\
ion collisional-decay mode & All $\lambda$ \\
ion {\Alf} waves  & $\lambda \leq \lamidim \cos \theta$ \\
transverse ambipolar-diffusion modes & 
$\lamidim \cos \theta <  \lambda \leq \lamndim \cos\theta$ \\
ion fast waves & $ \lambda \leq \lamidim$ \\
longitudinal ambipolar-diffusion modes & $\lamidim < \lambda \leq \lammsndim$ \\
transverse neutral collisional-decay mode & $\lambda \leq \lamndim \cos\theta$ \\
longitudinal neutral collisional-decay mode & $\lambda \leq \lammsndim$ \\
acoustic waves & $\lambda \leq \lamcsdim\Ssmfac(\theta)$ ~~~~~if ~$\theta \leq \thetamax$ ${}^{\dagger}$ \\
   & $\lambda \leq \lamcsdim/\Asnfac(\theta)$ ~if ~$\theta > \thetamax$ ${}^{\dagger}$\\
neutral slow waves &  $\lamcsdim \Ssmfac(\theta) < \lambda \leq \lamJdim$ 
~~only if ~$\theta \leq \thetamax$\\
pressure-driven diffusion mode & ~~~~~~$\lambda > \lamcsdim/\Asnfac(\theta)$ ~~~only if
~$\theta > \thetamax$ \\
neutral {\Alf} waves & $ \lambda > \lamndim \cos \theta$ \\
neutral ambipolar diffusion-induced gravitational fragmentation mode &
$\lamJdim < \lambda \leq \lammsdim$ ~~~only if ~$\theta > \thetamax$ \\
neutral fast waves & $\lammsndim < \lambda \leq \lammsdim$ \\
neutral modified {\Alf} waves & $\lambda > \lammsdim$ \\
gravitational instability mode & \hspace{4.5em} $\lambda > \lamJdim$ ~~~
if ~$\theta \leq \thetamax$ \\
& \hspace{4.5em} $\lambda > \lammsdim$ ~ if ~$\theta > \thetamax$ \\
conjugate (``cosmological") Jeans mode & \hspace{4,5em} $\lambda > \lamJdim$ ~~~ if ~$\theta \leq \thetamax$ \\
& \hspace{4.5em} $\lambda > \lammsdim$ ~ if ~$\theta > \thetamax$ \\
\hline
\hline \\
\vspace{-2ex}
${}^{\dagger}$ $\Asnfac(\theta)$, $\Ssmfac(\theta)$, and $\thetamax$ are 
defined in eqs. (\ref{Asndefeq}), (\ref{Ssmdefeq}), and (\ref{slowangleeq})
respectively. 
\end{tabular}
\end{table}
%
Table 2 lists all the critical wavelengths present in a two-fluid system, such
as a typical molecular cloud. The name of each critical wavelength, its defining
dimensionless expression and typical value, and the corresponding dimensional
expression are listed in columns 1 - 4, respectively.

Table 3 summarizes conveniently all the modes that can exist in a two-fluid
system (typical molecular cloud) for propagation parallel, perpendicular, and at
an arbitrary angle with respect to the unperturbed magnetic field. The name of
each mode is shown in the first column, and the wavelength range in which the
mode exists is shown in the second column. 

Although in a single-fluid system {\it linear} modes are independent of one 
another, in a multifluid system (such as a molecular cloud) this is not the 
case. A mode in one fluid can bifurcate due to interaction with 
the other fluid and give rise to two distinct daughter modes. Such interaction 
is also responsible for the reverse phenomenon of mode merging. 
Moreover, in a multifluid system, a wave mode in one fluid/species (e.g. a sound
wave in the neutrals) can transition into another wave mode (e.g., a slow MHD
wave in the neutrals) without bifurcation.
In other words, {\it linear waves in a multifluid system exhibit behaviour
that only nonlinear waves exhibit in a single-fluid system}. 

In a following paper we present a study of the free parameters, spanning 
the range of observationally allowed values, to examine their effect on
HM waves and instabilities in molecular clouds.

\vspace{+4ex}
This work was carried out in 1987 - 1989, when the bulk of 
this paper was also written. The paper was updated in 1994 - 1996, 
but never submitted for publication due to spatial dispersal of its three 
authors and their involvement in code development for nonlinear 
calculations. Its {\em Introduction} and some other parts have been 
updated to refer to relevant work published since then. The work was 
supported in part by the National Science Foundation under grant 
AST-93-20250, and also by the New York Center for Astrobiology under 
grant \# NNA09DA80A. TChM acknowledges the hospitality of the Alexander von
Humboldt Foundation and of the Max-Planck-Institut f\"{u}r Extraterrestrische
Physik during the writing of part of this paper in 1994. GC is supported by
the New York Center for Astrobiology (a member of the NASA Astrobiology
Institute) under grant \# NNA09DA80A.


\end{document}